\RequirePackage{fix-cm}
\documentclass[twocolumn,a4paper,natbib]{svjour3}        % twocolumn

\smartqed  % flush right qed marks, e.g. at end of proof

\usepackage{graphicx}
\usepackage{mathptmx}      % use Times fonts if available on your TeX system
\usepackage{amssymb,epsfig,color}
\usepackage{amsmath}
\usepackage{graphicx}
\usepackage{natbib}
\usepackage[hyphens]{url}

\usepackage[hyperindex,breaklinks=true, colorlinks, citecolor=blue]{hyperref}  % for active links (replace with one below if compilation fails!!)

\usepackage{flushend} % Package used to balance the columns on the last page
\def\reff@jnl#1{{\rm#1\/}}
\def\aj{\reff@jnl{AJ}}                  % Astronomical Journal
\def\araa{\reff@jnl{ARA\&A}}            % Annual Review of Astron and Astrophys
\def\apj{\reff@jnl{ApJ}}                % Astrophysical Journal
\def\apjl{\reff@jnl{ApJ}}               % Astrophysical Journal, Letters
\def\apjs{\reff@jnl{ApJS}}              % Astrophysical Journal, Supplement
\def\ao{\reff@jnl{Appl.Optics}}         % Applied Optics
\def\apss{\reff@jnl{Ap\&SS}}            % Astrophysics and Space Science
\def\aap{\reff@jnl{A\&A}}               % Astronomy and Astrophysics
\def\aapr{\reff@jnl{A\&A~Rev.}}         % Astronomy and Astrophysics Reviews
\def\aaps{\reff@jnl{A\&AS}}             % Astronomy and Astrophysics, Supplement
\def\azh{\reff@jnl{AZh}}                        % Astronomicheskii Zhurnal
\def\baas{\reff@jnl{BAAS}}              % Bulletin of the AAS
\def\jrasc{\reff@jnl{JRASC}}            % Journal of the RAS of Canada
\def\memras{\reff@jnl{MmRAS}}           % Memoirs of the RAS
\def\mnras{\reff@jnl{MNRAS}}            % Monthly Notices of the RAS
\def\pra{\reff@jnl{Phys.Rev.A}}         % Physical Review A: General Physics
\def\prb{\reff@jnl{Phys.Rev.B}}         % Physical Review B: Solid State
\def\prc{\reff@jnl{Phys.Rev.C}}         % Physical Review C
\def\prd{\reff@jnl{Phys.Rev.D}}         % Physical Review D
\def\prl{\reff@jnl{Phys.Rev.Lett}}      % Physical Review Letters
\def\pasa{\reff@jnl{PASA}}              % Publications of the ASA
\def\pasp{\reff@jnl{PASP}}              % Publications of the ASP
\def\pasj{\reff@jnl{PASJ}}              % Publications of the ASJ
\def\qjras{\reff@jnl{QJRAS}}            % Quarterly Journal of the RAS
\def\skytel{\reff@jnl{S\&T}}            % Sky and Telescope
\def\solphys{\reff@jnl{Solar~Phys.}}    % Solar Physics
\def\sovast{\reff@jnl{Soviet~Ast.}}     % Soviet Astronomy
 \def\ssr{\reff@jnl{Space~Sci.Rev.}}     % Space Science Reviews
\def\zap{\reff@jnl{ZAp}}                        % Zeitschrift fuer Astrophysik
\def\nat{\reff@jnl{Nature}}             % Nature 
\journalname{Astron Astrophys Rev (2017)}
\begin{document}

\title{The State-of-Play of Anomalous Microwave Emission (AME) Research%\thanks{Grants or other notes
}


\author{Clive\,Dickinson$^{1}$        \and
Y.\,Ali-Ha\"{i}moud$^{2}$ \and
A.\,Barr$^{1}$ \and
E.\,S.\,Battistelli$^{3}$ \and
A.\,Bell$^{4}$ \and
L.\,Bernstein$^{5}$ \and 
S.\,Casassus$^{6}$ \and
K.\,Cleary$^{7}$ \and
B.\,T.\,Draine$^{8}$ \and
R.\,G\'{e}nova-Santos$^{9,10}$ \and
S.\,E.\,Harper$^{1}$ \and
B.\,Hensley$^{7,11}$ \and
J.\,Hill-Valler$^{12}$ \and
Thiem\,Hoang$^{13}$ \and
F.\,P.\,Israel$^{14}$ \and
L.\,Jew$^{12}$ \and
A.\,Lazarian$^{15}$ \and
J.\,P.\,Leahy$^{1}$ \and
J.\,Leech$^{12}$ \and 
C.\,H.\,L\'{o}pez-Caraballo$^{16}$ \and
I.\,McDonald$^{1}$ \and
E.\,J.\,Murphy$^{17}$ \and
T.\,Onaka$^{4}$ \and
R.\,Paladini$^{18}$ \and
M.\,W.\,Peel$^{19,1}$ \and
Y.\,Perrott$^{20}$ \and
F.\,Poidevin$^{9,10}$ \and
A.\,C.\,S.\,Readhead$^{7}$ \and
J.-A.\,Rubi\~{n}o-Mart\'{i}n$^{9,10}$ \and
A.\,C.\,Taylor$^{12}$ \and
C.\,T.\,Tibbs$^{21}$ \and
M.\,Todorovi\'{c}$^{22}$ \and
Matias\,Vidal$^{6}$
}

\authorrunning{C. Dickinson et al.} % if too long for running head

\institute{C. Dickinson \at
              Jodrell Bank Centre for Astrophysics, Alan Turing Building, School of Physics and Astronomy, The University of Manchester, Oxford Road, Manchester, M13 9PL, U.K.\\
              Tel.: +44-161-275-4232\\
              Fax: +44-161-275-4247\
              \email{clive.dickinson@manchester.ac.uk}        \\ 
%          S. Author \at
%              second address}
\newline
$^{1}$Jodrell Bank Centre for Astrophysics, School of Physics and Astronomy, The University of Manchester, Oxford Road, Manchester, M13 9PL, Manchester, U.K. \\
$^{2}$Center for Cosmology and Particle Physics, Department of Physics, New York University, New York, NY 10003, U.S.A. \\
$^{3}$Sapienza, University of Rome, Italy \\
$^{4}$Department of Astronomy, Graduate School of Science, The University of Tokyo, Tokyo 113-0033, Japan \\
$^{5}$Spectral Sciences Inc., 4 Fourth Ave, Burlington, MA 01803-3304, U.S.A. \\
$^{6}$Universidad de Chile, Santiago, Chile \\
$^{7}$California Institute of Technology, Pasadena, CA 91125, U.S.A. \\
$^{8}$Princeton University Observatory, Peyton Hall, Princeton, NJ 08544-1001, U.S.A. \\
$^{9}$Instituto de Astrofis\'{i}ca de Canarias, 38200 La Laguna, Tenerife, Canary Islands, Spain \\
$^{10}$Departamento de Astrof\'{i}sica, Universidad de La Laguna (ULL), 38206 La Laguna, Tenerife, Spain \\
$^{11}$Jet Propulsion Laboratory, Pasadena, CA 91109, U.S.A. \\
$^{12}$Sub-department of Astrophysics, University of Oxford, Denys Wilkinson Building, Keble Road, Oxford OX1 3RH, U.K. \\
$^{13}$Korea Astronomy and Space Science Institute, Daejeon, 305-348, Korea \\
$^{14}$Sterrewacht, Leiden, The Netherlands \\
$^{15}$Department of Astronomy, University of Wisconsin-Madison, 475 Charter St., Madison, WI 53705, U.S.A. \\
$^{16}$Instituto de Astrof\'{i}sica and Centro de Astro-Ingenier\'ia, Facultad de F\'{i}sica, Pontificia Universidad Cat\'{o}lica de Chile, \\
Av. Vicu\~na Mackenna 4860, 7820436 Macul, Santiago, Chile \\
$^{17}$NRAO, U.S.A. \\
$^{18}$Infrared Processing and Analysis Center, California Institute of Technology, Pasadena, California, U.S.A. \\
$^{19}$Departamento de F\'{i}sica Matematica, Instituto de F\'{i}sica, Universidade de S\~{a}o Paulo, Rua do Mat\~{a}o 1371, S\~{a}o Paulo, Brazil \\
$^{20}$Cavendish Laboratory, University of Cambridge, Madingley Road, Cambridge CB3 0HA, U.K. \\
$^{21}$Directorate of Science, European Space Research and Technology Centre (ESA/ESTEC), Noordwijk, The Netherlands \\
$^{22}$University of South Wales,  Wales, U.K. \\
% Other people to include/ask/let them see early draft: 
% Bob Watson
% Doug Finkbeiner
% Ben Gold
% Angelica de Oliveira-Costa
% Anna Scaife
% Alex Lazarian
% Al Kogut
% Marc-Antoine Miville-Deschenes
% Guilaine Lagache
% Mathieu Remazeilles
% Nathalie Ysard
% Laurent Verstraete
% Peter Martin
% Adolf Witt
% Anthony Jones
% Vincent Guillet
% ADD MORE NAMES HERE BEFORE I EMAIL THEM ALL TO POSSIBLY CONTRIBUTE/BE CO-AUTHOR
}

\date{Received: date / Accepted: date}

\maketitle

%%%%%%%%%%%%%%%%%%%%%%%%%%%%%%%%%%%%%%%%%%%%%%%%%%%%%%%
%%%%%%%%%%%%%%%%%%%%%%%%%%%%%%%%%%%%%%%%%%%%%%%%%%%%%%%

\begin{abstract}
Anomalous Microwave Emission (AME) is a component of diffuse Galactic radiation observed at frequencies in the range $\approx 10$--60\,GHz. AME was first detected in 1996 and recognised as an additional component of emission in 1997. Since then, AME has been observed by a range of experiments and in a variety of environments. AME is spatially correlated with far-IR thermal dust emission but cannot be explained by synchrotron or free-free emission mechanisms, and is far in excess of the emission contributed by thermal dust emission with the power-law opacity consistent with the observed emission at sub-mm wavelengths. Polarization observations have shown that AME is very weakly polarized ($\lesssim 1$\,\%). The most natural explanation for AME is rotational emission from ultra-small dust grains (``spinning dust"), first postulated in 1957. Magnetic dipole radiation from thermal fluctuations in the magnetization of magnetic grain materials may also be contributing to the AME, particularly at higher frequencies ($\gtrsim 50$\,GHz). AME is also an important foreground for Cosmic Microwave Background analyses. This paper presents a review and the current state-of-play in AME research, which was discussed in an AME workshop held at ESTEC, The Netherlands, June 2016. 

%\PACS{PACS code1 \and PACS code2 \and more}
%\subclass{MSC code1 \and MSC code2 \and more}
\end{abstract}

%%%%%%%%%%%%%%%%%%%%%%%%%%%%%%%%%%%%%%%%%%%%%%%%%%%%%%%
%%%%%%%%%%%%%%%%%%%%%%%%%%%%%%%%%%%%%%%%%%%%%%%%%%%%%%%

\section{Introduction} 
\label{sec:introduction}

Anomalous Microwave Emission (AME) is a dust-correlated component of Galactic emission that has been detected by cosmic microwave background (CMB) experiments and other radio/microwave instruments at frequencies $\approx 10$--60\,GHz since the mid-1990s \citep{Kogut1996,Leitch1997}. It is thought to be due to electric dipole radiation from small spinning dust grains in the interstellar medium (ISM), although the picture is still not clear. The emission forms part of the diffuse Galactic foregrounds that contaminate CMB data in the frequency range \mbox{$\approx 20$--350\,GHz}, and hence knowledge of their spatial structure and spectral shape can be exploited during CMB component separation \citep{Dunkley2009a,Bennett2003b,Planck2015_X}. Since spinning dust emission depends critically on the dust grain size distribution, the type of dust, and the environmental conditions (e.g., density, temperature, interstellar radiation field), precise measurements of AME can also provide a new window into the ISM, complementing other multi-wavelength tracers.

The first mention of spinning dust grains in the literature was by \cite{Erickson1957}, who proposed this non-thermal emission as a contributor at high radio frequencies (GHz and above). The same basic mechanism of radio emission from rapidly spinning dust grains was also discussed by \cite{Hoyle1970} in the context of converting optical photons from stars into radio/microwave emission. \cite{Ferrara1994} further developed the theory, estimating the contribution from radio-emitting dust in spiral galaxies. These earlier works outlined the basic mechanisms of how small dust grains with finite electric dipole moments can be spun up to high rotational frequencies, thus producing radio emission. They also understood that such emission would predominantly arise at relatively high frequencies. However, it was not until the late 1990s, after when observations detected excess emission at frequencies \mbox{$\approx 10$--60\,GHz }(Sect.\,\ref{sec:observations}), that detailed predictions of spinning dust emission were made by \cite{Draine1998a,Draine1998b}. The field of AME research then became important, particularly since AME was known to be a significant CMB foreground (Sect.\,\ref{sec:cmb}). Magnetic dust emission (MDE) on the other hand had not been discussed in the literature until the seminal work of \cite{Draine1999} who proposed it as an alternative to spinning dust. 

In this article, we provide a comprehensive review the state-of-play of AME research. For a previous review, see \cite{Dickinson2013a} and articles within. Section\,\ref{sec:theory} provides an overview of the theory of spinning dust, magnetic dust, and other emission mechanisms that may be contributing to AME. Observations of AME are summarised in Sect.\,\ref{sec:observations}. Section\,\ref{sec:cmb} discusses AME as a CMB foreground while in Sect.\,\ref{sec:discussion} we discuss various methodologies and goals for future research. Concluding remarks are made in Sect.\,\ref{sec:conclusion}. This article is partially an outcome of the discussions at the AME workshop\footnote{\protect\url{http://www.cosmos.esa.int/web/ame-workshop-2016/schedule}} held  22--23\,June\,2016 at ESTEC (Noordwijk, The Netherlands). Previous AME workshops were held at Manchester\footnote{\protect\url{http://www.jb.man.ac.uk/~cdickins/Man_AMEworkshop_July2012.html}}  in 2012 and at Caltech\footnote{\protect\url{https://wikis.astro.caltech.edu/wiki/projects/ameworkshop2013/AME_Workshop_2013.html}} in 2013.

%%%%%%%%%%%%%%%%%%%%%%%%%%%%%%%%%%%%%%%%%%%%%%%%%%%%%%%
%%%%%%%%%%%%%%%%%%%%%%%%%%%%%%%%%%%%%%%%%%%%%%%%%%%%%%%

\begin{figure}
\begin{center}
\includegraphics[width=0.35\textwidth,angle=0]{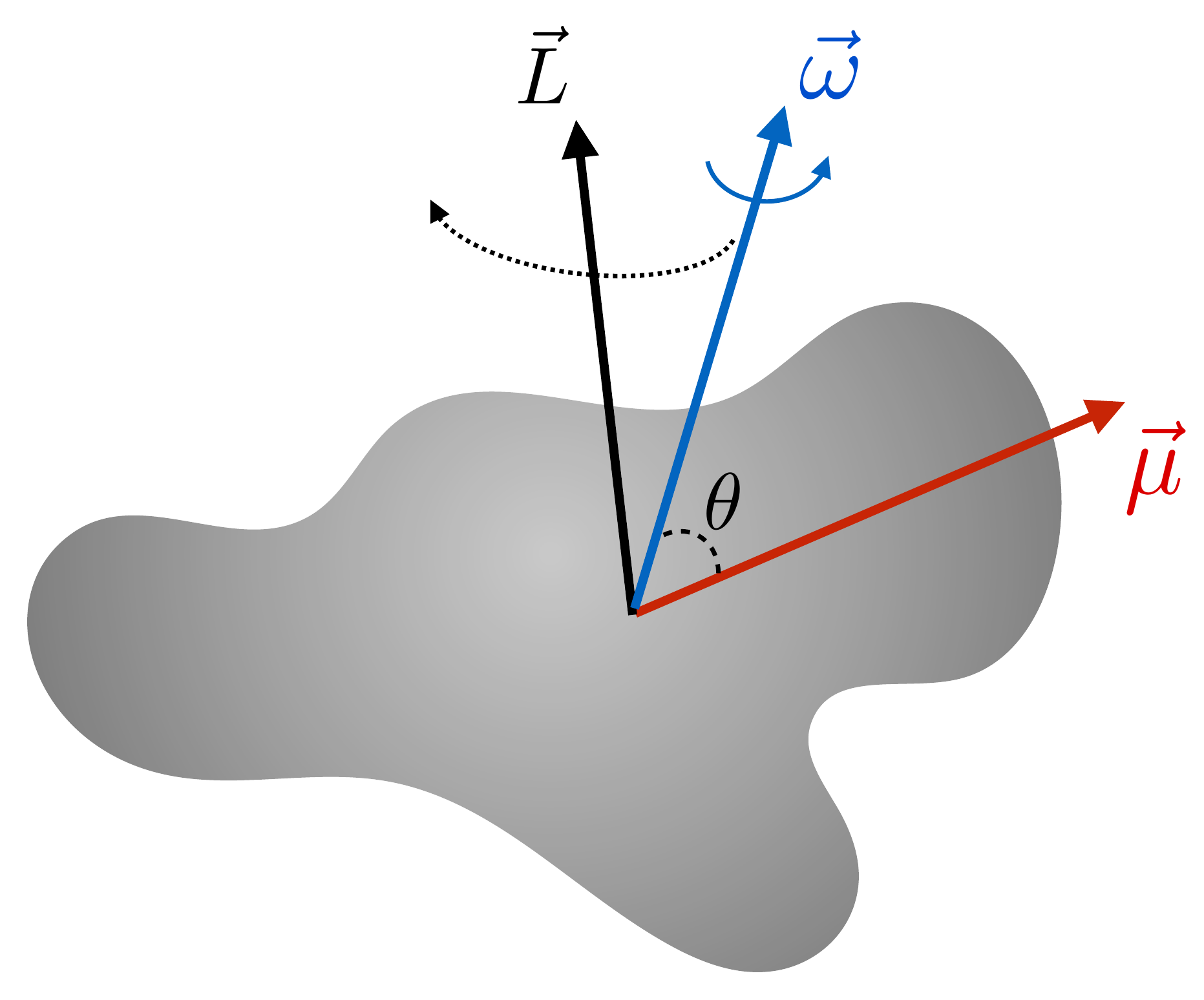}
\caption{Schematic spinning dust grain, with its permanent electric dipole moment $\vec{\mu}$, its instantaneous angular velocity $\vec{\omega}$, and its angular momentum $\vec{L}$, about which $\vec{\omega}$ precesses. }
\label{fig:dustgrain}
\end{center}
\end{figure}

\section{Models of Candidate AME Mechanisms} 
\label{sec:theory} 

\subsection{Spinning dust} 
\label{sec:theory_spinningdust}

\subsubsection{Basic Theory}
A dust grain with electric or magnetic dipole moment $\mu$ rotating with angular frequency $\omega$ will produce
emission according to the Larmor formula
\begin{equation}
P = \frac{2}{3} \frac{\omega^4 \mu^2 \sin^2\,\theta}{c^3}~,
\end{equation}
where $P$ is the total power emitted at frequency $\nu = \omega/2\pi$ and $\theta$ is the angle between $\omega$ and $\mu$. A schematic diagram of a single dust grain is shown in Fig.\,\ref{fig:dustgrain}. It is immediately evident
that the spinning dust emission spectrum will depend sensitively on the distribution of rotational frequencies attained by
the grains as well as their distribution of dipole moments. Indeed, much of the theoretical modelling efforts have
been toward accurate calculation of the distribution of rotation rates as a function of grain size and composition in
various interstellar environments.

A spherical grain of radius $a$ and mass density $\rho$ rotating thermally in gas of temperature $T$ will have a
rotational frequency of
\begin{equation}
\frac{\omega}{2\pi} = 21\,{\rm GHz}\,\left(\frac{T}{100\,{\rm K}}\right)^{1/2}\,\left(\frac{\rho}{3\,{\rm g}\,{\rm cm}^{-3}}\right)^{-1/2}
\,\left(\frac{a}{5\,{\rm \AA}}\right)^{-5/2}~.
\end{equation}
To emit appreciably in the 20--30\,GHz range as required to reproduce the observed AME, the grains must be very small, 
$a \lesssim 1\,$nm.

\setlength{\tabcolsep}{9pt}
\begin{table*}
\begin{center}
\caption{Key developments in the theory of spinning dust emission with associated references.}
\label{tab:spdust_developments}
\begin{tabular}{l l}
\hline
\noalign{\smallskip}
Development    &Reference   \\
\noalign{\smallskip}
\hline
\noalign{\smallskip}
First proposal for electric dipole radiation from spinning dust grains &\protect\cite{Erickson1957} \\
First full treatment of spinning dust grain theory  			&\protect\cite{Draine1998b} \\
Quantum suppression of dissipation and alignment			&\protect\cite{Lazarian2000} \\
Factor of two correction in IR damping coefficient    			&\protect\cite{Ali-Hamoud2009} \\
Fokker-Planck treatment of high-$\omega$ tail				&\protect\cite{Ali-Hamoud2009} \\
Quantum mechanical treatment of long-wavelength tail of PAHs &\protect\cite{Ysard2010a} \\
Rotation around non-principal axis						&\protect\cite{Hoang2010,Silsbee2011} \\
Transient spin-up events								&\protect\cite{Hoang2010} \\
Effect of tri-axiality on rotational spectrum					&\protect\cite{Hoang2011} \\
Effects of transient heating on emission from triaxial grains	&\protect\cite{Hoang2011} \\
Magnetic dipole radiation from ferromagnetic spinning dust	&\protect\cite{Hoang2016c,Hensley2017} \\
Improved treatment of quantum suppression of dissipation and alignment &\protect\cite{Draine2016} \\ 
\noalign{\smallskip}
\hline
\end{tabular}
\end{center}
\end{table*}
\setlength{\tabcolsep}{6pt}

%%%%%%%%%%%%%%%%%%%%%%%%%%%%%%%%%%%%%%%%%%%%%%%%%%%%%%%

\subsubsection{Rotational Dynamics}

An interstellar dust grain is subject to a number of torques arising from its interactions with the ambient interstellar matter,
which can both excite and damp rotation. Collisions with ions and neutral atoms, photon emission, H$_2$ formation,
photoelectric emission, and interaction with the electric fields of passing ions (``plasma drag'') have all been identified
as contributing to grain rotation. The distribution of grain rotational velocities will generally be non-thermal resulting from the interplay of a number of different excitation and damping processes, including collisions with atoms and ions, and absorption and emission of radiation. Systematic torques (i.e., torques that do not have a time average of zero in grain coordinates), and impulsive torques (i.e., impacts that produce large fractional changes in the grain angular momentum) can be important.

\begin{figure}
\begin{center}
\includegraphics[width=0.5\textwidth,angle=0]{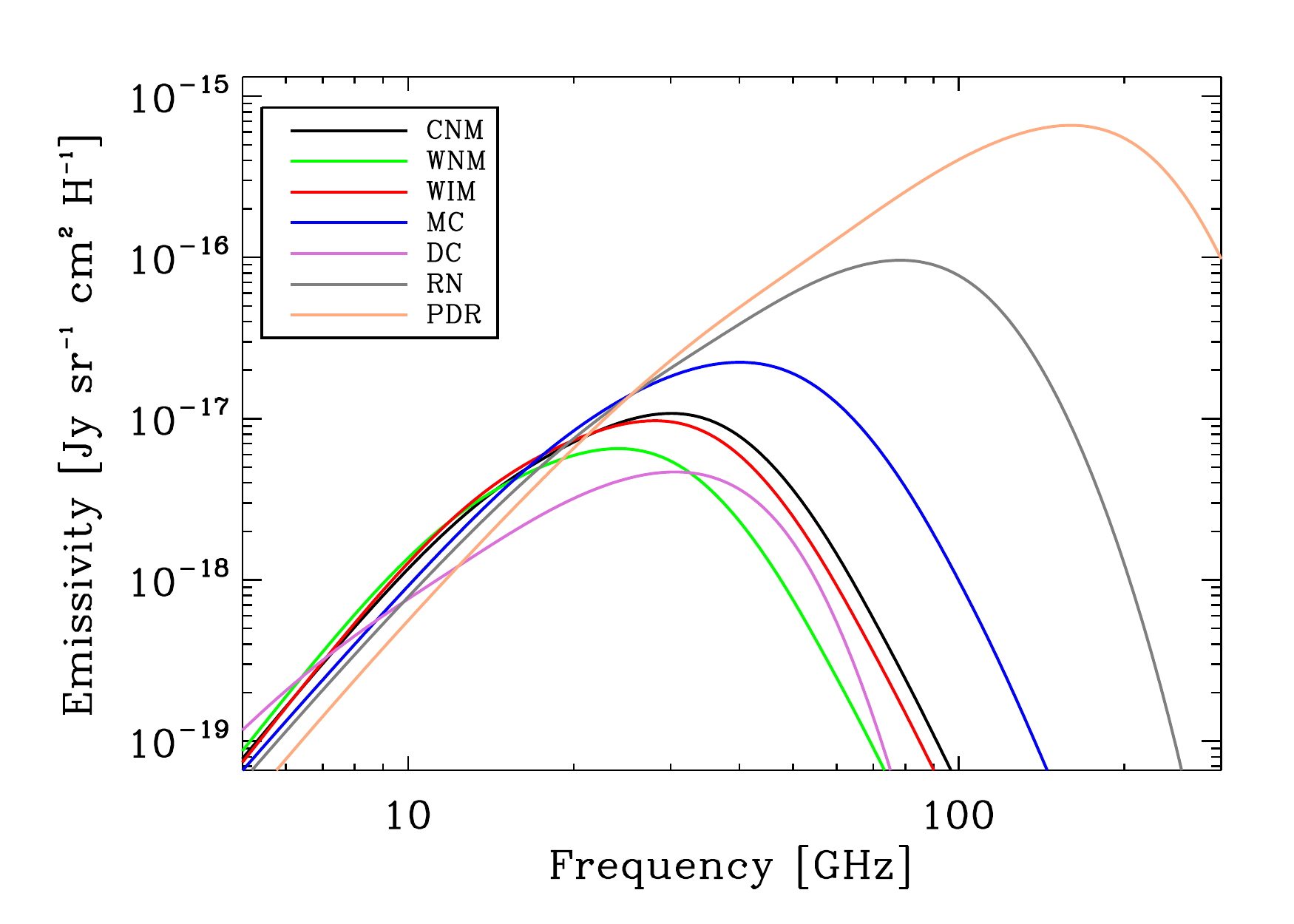}
\caption{Spinning dust emissivity curves as a function of frequency for different idealized phases of the interstellar medium (see text). The curves were produced using the {\sc spdust2} code, with parameters given in Table\,\protect{\ref{tab:spdust_params}, which can be compared with the original curves presented by \protect\cite{Draine1998b}}.}
\label{fig:spdust_ems}
\end{center}
\end{figure}

\setlength{\tabcolsep}{9pt}
\begin{table*}
\begin{center}
\caption{Environmental parameters for various idealized phases of the ISM, as was done by \protect\cite{Draine1998a} (see text). These parameters were used to produce the emissivity curves in Fig.~\protect\ref{fig:spdust_ems}.}
\label{tab:spdust_params}
\begin{tabular}{l|ccccccc}
\hline
\noalign{\smallskip}
 					&\multicolumn{7}{c}{Phase} \\
\noalign{\smallskip}
Parameter	 						&DC		&MC		&CNM	&WNM		&WIM		&RN		&PDR  	\\
\noalign{\smallskip}
\hline
\noalign{\smallskip}
$n_{\rm H}$ (cm$^{-3}$)				&$10^4$		&300		&30		&0.4		&0.1		&$10^3$ 	&$10^5$	\\				
$T$ (K)							&10			&20		&100		&6000	&8000	&100		&300		\\
$T_d$ (K)							&10			&20		&20		&20		&20		&40		&50		\\
$\chi$							&0.0001		&0.01	&1		&1		&1		&1000	&3000	\\
$y \equiv 2n({\rm H}_{2})/n_{\rm H}$		&0.999		&0.99	&0		&0		&0		&0.5		&0.5		\\
$x_{\rm H} \equiv n({\rm H}^{+})/n_{\rm H}$&0			&0		&0.0012	&0.1		&0.99	&0.001	&0.0001	\\
$x_{\rm M} \equiv n({\rm M}^{+})/n_{\rm H}$&$10^{-6}$	&0.0001	&0.0003	&0.0003	&0.001	&0.0002	&0.0002	\\
\noalign{\smallskip}
\hline
\end{tabular}
\end{center}
\end{table*}
\setlength{\tabcolsep}{6pt}

\citet{Draine1998b} presented the first comprehensive model of spinning dust emission taking most of these processes into 
account. For simplicity, they assumed $\langle\omega^4\rangle = 5/3\langle\omega^2\rangle^2$, consistent with a
Maxwellian distribution. Recognizing that ultrasmall grains could simultaneously furnish an explanation for AME and the
infrared emission bands, they focused their analysis on electric dipole emission from PAHs.

\citet{Ali-Hamoud2009} improved on the treatment of the rotational dynamics by employing the Fokker-Planck equation to 
compute the angular velocity distribution. Notably, they found significantly less power in the tails of the distribution, 
particularly toward high values of $\omega$, relative to a Maxwellian. Theoretical emissivity\footnote{The term ``emissivity" is usually defined as how well a body radiates relative to a blackbody. However, in astronomy it is used in a number of situations, including how much radiation is emitted per unit volume (or column density) as in spinning dust emissivity curves. Later, it will be used in the context of the emissivity law of dust grains (how well dust grains emit as a function of frequency) and also in terms of AME emissivities or correlation coefficients (AME brightness relative to various dust templates).} curves, as a function of frequency, can be calculated using the publicly available\footnote{\url{http://cosmo.nyu.edu/yacine/spdust/spdust.html}} {\sc spdust} code, written in the Interactive Data Language (IDL)\footnote{\url{http://www.harrisgeospatial.com/ProductsandTechnology/Software/IDL.aspx}}. \citet{Ysard2010a} presented a quantum mechanical analysis of the grain excitation and damping processes, finding good agreement with classical models. \citet{Hoang2010}
and \citet{Silsbee2011} further refined models of the rotational dynamics by considering asymmetric grains that rotate about a non-principal axis, incorporating this into the updated {\sc spdust2} code.

Both studies concluded that this grain ``wobbling'' could increase both the peak\footnote{The peak frequency strictly depends on whether flux density or brightness temperature is used (they are related by a factor of $1/\nu^2$). For AME, we will typically discuss the peak in flux density units at $\approx 30$\,GHz. In brightness temperature units the spinning dust spectrum does not have a clear peak; instead it has an inflection point, which turns over at $\approx 16$\,GHz.} frequency and the emissivity of the spinning dust emission spectrum. Additionally, \citet{Hoang2010} computed the angular velocity distribution via the Langevin equation, which, unlike the approach based on the Fokker-Planck equation, can account for impulsive torques. Subsequent improvements accounting for the effects of irregular grain shapes, stochastic heating, and emissivity enhancements due to compressible turbulence were made by \citet{Hoang2011}. Table\,\ref{tab:spdust_developments} lists the most important developments in spinning dust theory and the associated references.

Fig.\,\ref{fig:spdust_ems} shows spinning dust spectra for various phases (environments) of the interstellar medium: cold neutral medium (CNM), warm neutral medium (WNM), warm ionized medium (WIM), molecular clouds (MC), dark clouds (DC), reflection nebulae (RN), and photo-dissociation regions (PDR). The curves were produced using the {\sc spdust2} code using the same parameters used by \cite{Draine1998b} for idealized phases of the ISM; these are listed in Table\,\ref{tab:spdust_params} for seven representative environments. These include the gas density $n_{\rm H}$, gas temperature $T$, dust temperature $T_d$, strength of the interstellar radiation field $\chi$, and the fraction of molecular hydrogen $y$, ions of hydrogen $x_{\rm H}$, and heavier ions $x_{\rm M}$. Other inputs include the dust size distribution and electric dipole moments (amongst others). The spectra can be compared with the original curves presented by \cite{Draine1998b}, which are similar but are slightly different in detail, due to enhancements to the code already mentioned. One can see that a peaked shape spectrum is always produced, but with considerable variations in both emissivity (by almost 2 orders of magnitude) and peak frequency ($\approx 30$\,GHz to over 100\,GHz). The strongest signals and higher peak frequencies are typically produced by the densest environments, such as in PDRs and molecular clouds. 

The physics of spinning dust emission is thus very well-established, with numerous mechanisms affecting the rotational velocity distribution of interstellar grains having been worked out in detail \citep[see for example the review by][]{Ali-Haimoud2013}. The primary limitation with making a theoretical prediction of the spinning dust spectral energy distribution (SED) is not the current understanding of the various excitation and damping mechanisms, but rather by the unknown sizes, dipole moments, charges, and shapes of ultra-small interstellar grains.

\begin{figure}
\begin{center}
\includegraphics[width=0.45\textwidth,angle=0]{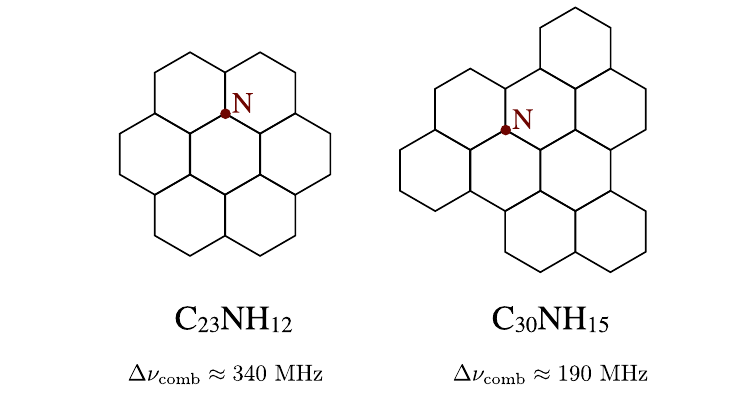}
\caption{Examples of two nitrogen-substituted symmetric PAHs that remain quasi-symmetric after substitution, i.e., their two smallest moments-of-inertia differ by less than a few per cent. Such molecules are the target of AME searches. $\Delta\nu_{\rm comb}$ indicates the approximate line spacing of their comb-like rotational spectrum. Figure reproduced from \protect\cite{Ali-Haimoud2015}.}
\label{fig:PAHlines}
\end{center}
\end{figure}

%%%%%%%%%%%%%%%%%%%%%%%%%%%%%%%%%%%%%%%%%%%%%%%%%%%%%%%

\subsubsection{Line Emission}
\label{sec:lines}

An interesting consequence of the spinning-PAH model is the possibility of rotational 
\emph{line} emission \citep{Ali-Haimoud2014}. The rotational quantum numbers 
of spinning PAHs are typically of order $\sim$100 \citep{Draine1998b}, large enough for 
a classical treatment to be accurate, but small enough for discrete rotational line 
transitions to be potentially distinguishable. The observability of individual transitions 
depends on two criteria. 

The first criterion is the diversity of PAH species carrying the emission: if a very large number 
of species are present in small abundances, their individual rotational spectra may be 
buried in the quasi-continuum total emission. \cite{Ali-Haimoud2014} argued that it is 
likely that a few select, highly stable PAH species are more resistant to the harsh ISM 
conditions, making them over-abundant. This ``grandPAH'' hypothesis, first put forward by 
\cite{Tielens2013}, was recently studied more quantitatively by \cite{Andrews2015}. These 
authors showed that available infrared data suggest that a limited number of compact, 
highly symmetric PAHs dominate the interstellar PAH family.

The second criterion for the observability of PAH lines is the degree of symmetry of the 
emitting molecules: large triaxial molecules have complex rotational spectra, with many 
weak individual transitions, making them impractical for spectroscopic identification. On 
the other hand, perfectly symmetric PAHs, such as coronene (C$_{24}$H$_{12}$) or 
circumcoronene (C$_{54}$H$_{18}$), which are likely to be among the ``grandPAHs'', have no permanent 
dipole moment hence no rotational emission. \cite{Ali-Haimoud2014} argued that Nitrogen 
substitution, a process likely to lead to large dipole moments in interstellar PAHs \citep{Hudgins2005}, 
breaks the symmetry of the moment-of-inertia matrix by a small enough amount that the rotational 
spectrum has the appearance of a ``comb'' of intense lines if observed with a resolution of 
$\sim$1\,MHz. Fig.\,\ref{fig:PAHlines} shows two possible symmetric PAHs that may be responsible for part of the AME and their approximate line spacing. The spacing of the ``teeth'' of the comb is inversely proportional to the moment of inertia of the carrier, hence providing a clear discriminative test of individual PAHs. The 
comb pattern moreover allows for blind searches with matched filtering.

\cite{Ali-Haimoud2015} performed an observational test of this idea by taking a 
spectrum of the Perseus molecular cloud at 25\,GHz, over a 3\,GHz bandwidth and with a 0.4\,MHz 
resolution. They matched-filtered the spectrum to search for comb patterns, though did not 
make a detection, and set upper limits on the abundance of individual PAHs assuming that the AME 
from Perseus is entirely due to spinning PAHs. A detection of PAH lines would not only be a smoking-gun signature of the spinning PAH hypothesis, but would also provide the long-awaited identification of specific planar PAHs.

%%%%%%%%%%%%%%%%%%%%%%%%%%%%%%%%%%%%%%%%%%%%%%%%%%%%%%%

\subsubsection{Non-PAH Carriers}

Each of the aforementioned studies focused on electric dipole emission
from PAHs, which are known to be a significant component of
interstellar dust due to their prominent mid-infrared features. \cite{Hensley2016} found no correlation between PAH emission fraction (presumably proportional to PAH abundance) and AME emission fraction (presumably proportional to AME carrier abundance). As a result, there has been interest in other potential carriers of spinning dust emission. The spinning dust theory developed for PAHs is equally applicable to nanoparticles of other compositions that have electric (or magnetic) dipole moments and thus most of the predictions based on models of spinning PAHs hold for other carriers.

\citet{Hoang2016a} and \citet{Hensley2017} considered 
spinning dust emission arising from the rotation of a {\it magnetic} dipole in interstellar iron grains. Both studies concluded
that such particles may constitute a portion of the observed AME, but that constraints on the solid-phase abundance of
interstellar iron preclude a population of such particles large enough to account for the entirety of the observed emission.
Rotational emission from nanosilicate grains was considered by \citet{Hoang2016b} and \citet{Hensley2017}, who determined
that such grains could account for the entirety of the observed emission without violating other observational constraints 
provided that the grains have a suitable electric dipole moment. 

%%%%%%%%%%%%%%%%%%%%%%%%%%%%%%%%%%%%%%%%%%%%%%%%%%%%%%%

\subsubsection{Polarization}

Electric (or magnetic) dipole emission from a single rotating grain is perfectly polarized. Thus, the spinning dust
emission spectrum could be highly polarized if the ultrasmall grains, which carry the emission, are substantially aligned.
However, the interstellar polarized extinction law is observed to drop rapidly in the UV with decreasing wavelength 
\citep[e.g.,][]{Martin1999}. Thus, on empirical grounds, it appears that the smallest grains are not systematically aligned, leading to a low level of polarization.

The physics of grain alignment is complex \citep[for a review see
e.g.,][]{Andersson2015}. Small grains could attain alignment via paramagnetic relaxation as
proposed by \citet{Davis1951}. However, paramagnetic relaxation may be
suppressed at high rotation frequencies. Further, \citet{Lazarian1999} argued that
thermal flipping prevents ultrasmall grains from achieving suprathermal
rotation, and so collisions with gas atoms would destroy this alignment on
short timescales. Nevertheless, the very smallest grains with extremely
rapid rotation rates could potentially align via ``resonance relaxation"
in which the rotational splitting of energy levels becomes important
\citep{Lazarian2000}. If ultrasmall grains are able to align in this way,
then spinning dust emission could be polarized at roughly the percent
level \citep{Lazarian2000}. \citet{Hoang2013} argued that the weak polarization in the 2175\,\AA\ 
feature observed along two sight-lines towards HD\,197770 and HD\,147933-4 could be explained with weakly-aligned PAHs, and computed that
the corresponding spinning dust emission from those PAHs would have a polarization fraction of  $\lesssim 1\,\%$ for 
\mbox{$\nu \gtrsim 20$\,GHz}. \citet{Hoang2016a} calculated that iron nanoparticles would be highly aligned with the interstellar
magnetic field due to their large magnetic susceptibilities, and that their rotational magnetic dipole radiation could be
polarized at the  40--50\% level. 

Most recently, \citet{Draine2016} argued that the quantization of
energy levels in ultrasmall grains would dramatically suppress the
conversion of rotational kinetic energy to vibrational energy,
thereby hindering all alignment processes dependent on this direct
conversion. They calculated that spinning dust emission at $\nu \gtrsim 10$\,GHz would be negligibly polarized with $P \lesssim 0.0001\,\%$
irrespective of the grain composition.

It should be noted that theoretical predictions of dust polarization are generally maximum values. The angle between the line-of-sight and the alignment axis, depolarization due to line-of-sight effects such as changing magnetic field direction, and contamination from other emission sources that may be polarized differently, may all contribute to a reduction of the polarization fraction in real observations. To mitigate some of these effects, it may be useful to compare the observed AME polarization (or upper bounds on it) to the observed thermal dust polarization in the same region-- regions with the highest observed polarization fractions for thermal dust emission would likely have the highest spinning dust polarization fractions.

A unique way to test physics of nanoparticle alignment is through the polarization of infrared PAH emission. Theoretical calculations by \cite{Hoang2017} found that, if PAHs are aligned with the magnetic field by paramagnetic resonance mechanism, their mid-IR emission can be polarized at a few percent for the conditions of reflection nebulae. The first detection of polarized PAH emission at $11.3\,\mu$m from the MWC\,1080 nebula was recently reported by \cite{Zhang2017}. The measured polarization of $1.9\pm0.2\,\%$ is much larger than the value predicted by the models with randomly oriented PAHs. PAH alignment with the magnetic field to $\sim 10\,\%$ can successfully reproduce this measurement. Therefore, further theoretical and observational studies in this direction are needed to achieve a convincing conclusion on the polarization of spinning dust. 

In summary, if quantum effects suppress dissipation in grains spinning at $\gtrsim 10$\,GHz, as suggested by \cite{Draine2016}, the polarization from spinning dust grains is likely to be very small ($\ll1\,\%$) and difficult to detect. Detailed further observations are required to confirm that this is the case.

%%%%%%%%%%%%%%%%%%%%%%%%%%%%%%%%%%%%%%%%%%%%%%%%%%%%%%%

\begin{figure}
\begin{center}
\includegraphics[width=0.35\textwidth,angle=0]{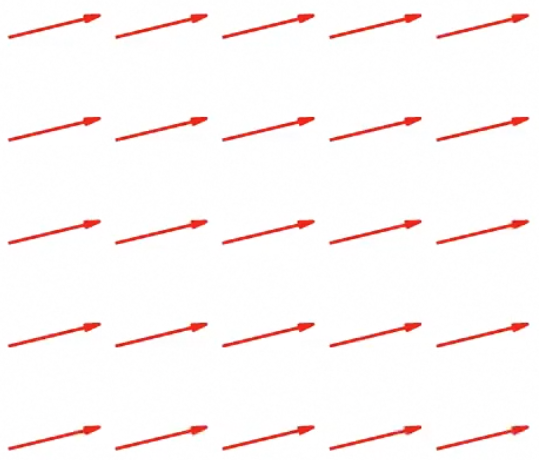}
\caption{Schematic representation of the electron spins within a ferromagnetic grain aligned along some preferred, energy-minimizing direction. Excitations cause the net magnetization to precess about this preferred direction, as illustrated in the animation (\textit{electronic version only}), eventually relaxing back into alignment with this direction. The precession of the magnetization gives rise to radiation.}
\label{fig:magnets}
\end{center}
\end{figure}

\begin{figure}
\begin{center}
\includegraphics[width=0.3\textwidth,angle=0]{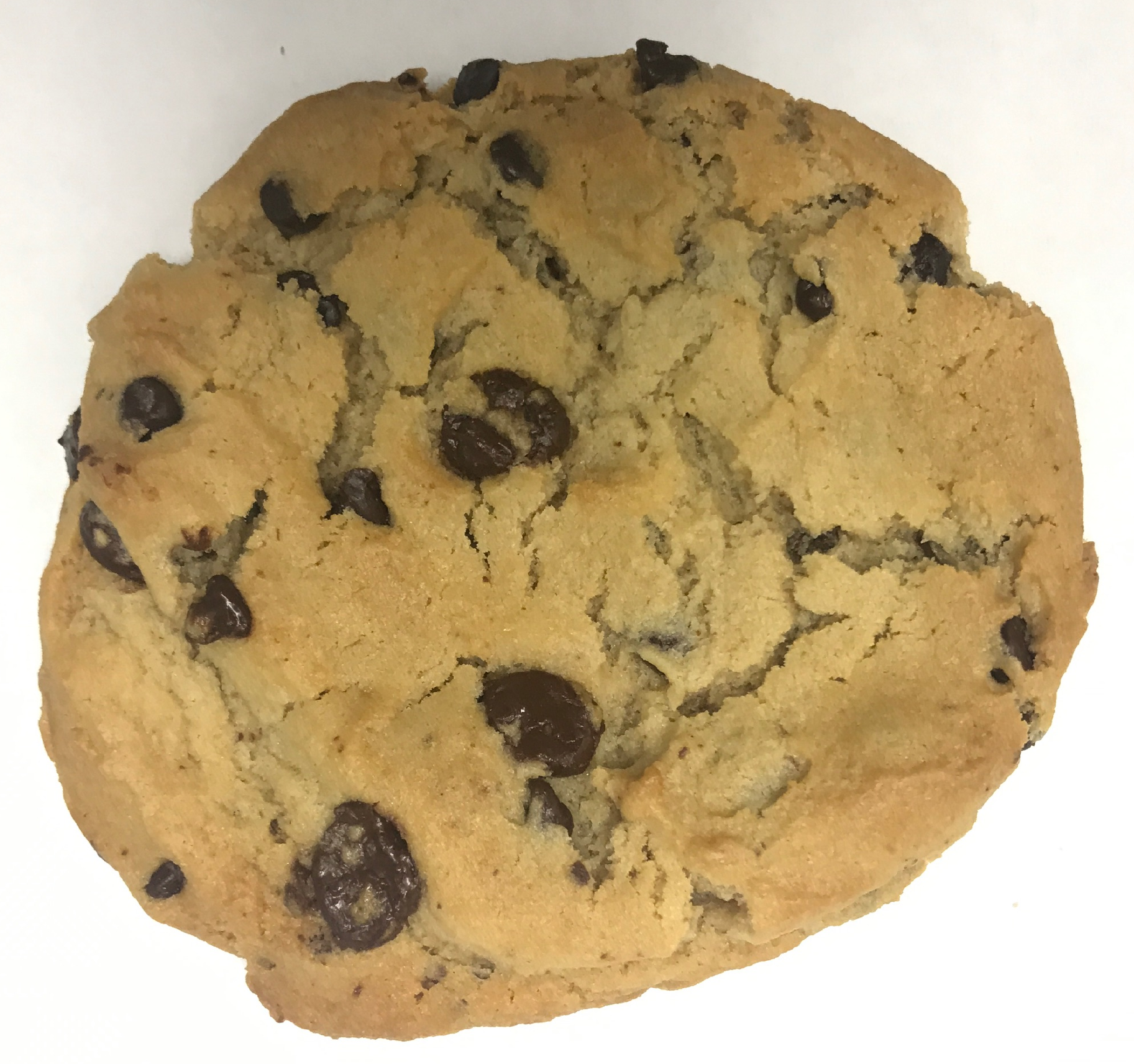}
\caption{Metallic iron inclusions may be found embedded in non-magnetic interstellar grains in much the same way that chocolate chips are embedded in chocolate chip cookies, as shown.}
\label{fig:cookie}
\end{center}
\end{figure}

\subsection{Magnetic dust (MDE)} 
\label{sec:theory_MD}

\subsubsection{Basic Theory}
Most of the interstellar iron resides in dust \citep{Jenkins2009}, and
iron inclusions (see Fig.\,\ref{fig:cookie}) have been observed in both interplanetary dust
\citep{Bradley1994} and putative interstellar grains collected in the Solar
System \citep{Westphal2014, Altobelli2016}. While the
chemical form of the iron in interstellar grains is unknown, it is plausible that some is in the form
of magnetic materials such as ferromagnetic metallic Fe or ferrimagnetic
magnetite (Fe$_3$O$_4$) or maghemite ($\gamma$-Fe$_2$O$_3$). In such
materials, the spins of unpaired electrons are spontaneously ordered,
giving rise to a net magnetization even in the absence of an applied field.
Thermal fluctuations can excite the magnetization away from its preferred,
energy-minimizing direction (see Fig.\,\ref{fig:magnets}). As the magnetization precesses and relaxes back
to the minimum energy state, radiation is emitted. Unlike spinning dust emission
(see Sect.\,\ref{sec:theory_spinningdust}), the emission is thermal and is not
associated with physical rotation of the grain.

\citet{Draine1999} put forward the first model of thermal magnetic dipole radiation
in the context of interstellar dust and as a possible explanation of the AME. They
modelled the magnetic response as a damped harmonic oscillator following
\citet{Morrish2001} and noted the
possible existence of a resonance feature in the absorption spectrum near 70\,GHz,
a magnetic analogue of a Fr\"{o}hlich resonance. They concluded that magnetic
materials exhibiting this resonance behaviour could possibly furnish an explanation for the
entirety of AME.

\citet{Draine2013} revisited the dynamics of the magnetic response by employing
the phenomenological Gilbert equation \citep{Gilbert2004}, which
explicitly accounts for the precession of the electron spins, to model
the time evolution of the magnetization, and resulting magnetic dipole
radiation. They found that magnetic grains can produce strong,
relatively grey emission from 
sub-millimetre to millimetre wavelengths and so could account for the apparent
excess emission observed in some low-metallicity dwarf galaxies and the Small Magellanic Cloud (SMC)
\citep{Draine2012}. Absorption resonances associated with the precession frequency
of the magnetization were also predicted between 1--30\,GHz, depending on the shape
of the grain. In particular, extremely elongated grains (e.g., spheroids with axial ratios
greater than 2:1) exhibit resonances at $\nu \gtrsim 20$\,GHz, which can more closely mimic
the observed AME spectrum.

The magnetic behaviour of materials in the microwave is still poorly
understood, and so direct laboratory measurements of the materials of
interest in this frequency range would be of great value. We outline
several key experimental tests of relevance to AME theory in
Section\,\ref{sec:future_theory}.

\subsubsection{Polarization}

The polarization properties of magnetic materials depend strongly on whether they are 
free-flying grains or whether they are inclusions in larger, non-magnetic grains.

\citet{Draine1999} and \citet{Draine2013} demonstrated that perfectly aligned free-flying iron nanoparticles
could achieve polarization fractions of $\simeq30\,\%$. \citet{Hoang2016a} calculated
the alignment efficiency of magnetic particles as a function of size, finding that large
grains ($\geq 1$\,nm) are poorly aligned whereas smaller grains attained high degrees
of alignment and could produce emission polarized at up to 10--30\,\% levels. \cite{Draine2016} showed that particles larger than $\sim 2$\,nm could be partially aligned; if ferromagnetic, thermal emission from such particles could be linearly polarized at the percent level.

In the case of magnetic inclusions within a larger non-magnetic grain, the polarized
emission depends both on the alignment of the grain and on the relative
importance of the magnetic dipole emission from the inclusions and the electric dipole
emission from the matrix of atoms, which are polarized orthogonally with respect to each other
\citep{Draine2013}. For randomly oriented magnetic inclusions in a silicate matrix, \citet{Draine2013}
found a drop in the polarization fraction beginning at $\simeq 10^3$\,GHz (300\,$\mu$m) and
extending to lower frequencies as the magnetic dipole emission becomes comparable to
the emission from the silicate material. At low frequencies ($\sim 10$\,GHz), the polarization
can even undergo a reversal, provided the magnetic Fe fraction is large enough. Thus, in the frequency range where both magnetic dipole emission and
electric dipole emission from the grain are important ($\simeq 10$--100\,GHz), the polarization
fraction of the emission is low ($\lesssim 5$\%). \cite{Hoang2016c} investigated the alignment efficiency of grains
with magnetic inclusions due to radiative torques, finding that the enhanced magnetic susceptibility due to the inclusions
enabled the grain to achieve nearly perfect alignment. Thus, the polarization fraction of the emission
from large grains with magnetic inclusions is likely limited only by the degree to which the magnetic dipole and electric dipole emission
processes are self-cancelling.

%%%%%%%%%%%%%%%%%%%%%%%%%%%%%%%%%%%%%%%%%%%%%%%%%%%%%%%

\subsection{Other emission mechanisms?}
\label{sec:theory_other}

Although spinning dust and magnetic dust have received the most attention as possible
explanations of AME, other mechanisms may still contribute a fraction, or even all,
of the observed emission. We now briefly review some of these possibilities.

Thermal emission from interstellar grains can be written as $\epsilon(\nu)B(\nu,T)$, where $\epsilon(\nu)$ is the emissivity ($=1$ for a black-body radiator) and $B(\nu,T)$ is the Planck function for a black-body at temperature $T$.  The emissivity $\epsilon(\nu)$ is sometimes approximated as a power-law, $\epsilon(\nu) \propto \nu^{\beta_d}$ \citep[see e.g.,][]{Draine_book,Martin2012}.  For spherical grains of radius $a$ and internal density $\rho$, the emissivity is directly related to the opacity, $\kappa(\nu)$ of the dust grains, via $\epsilon(\nu)=(4\rho a/3)\kappa(\nu)$.  The far-infrared opacity of nanoparticles of amorphous materials is notoriously difficult to both calculate theoretically and measure in the laboratory. As these frequencies are much lower
than the known resonances in the UV and optical, it is typically assumed that the opacity should
be decreasing as a power-law with $\kappa \propto \nu^{\beta_d}$ with $\beta_d = 2$. However,
amorphous materials exhibit a range of $\beta_d$ values depending on their composition and
temperature \citep[e.g.,][]{Agladze1996}. Current measurements show a range of values depending on environment and nature of the emission, but are typically in the range $\beta_d=1$--2, with an average value at high Galactic latitude of $\beta_d\approx 1.6$ \citep{PIP_XVII}. Given the current limited knowledge of the microwave
properties of amorphous materials, it is conceivable that some materials could have an absorption
resonance in the vicinity of 30\,GHz and thus explain AME via thermal dust emission. However, this seems unlikely and contrived, especially given how well the simple power-law model has worked so far, to explain the low-frequency Rayleigh-Jeans (R-J) tail of thermal dust at frequencies above $\approx 100$\,GHz \mbox{\citep[e.g., ][]{PIP_XVII}}. Furthermore, there are good theoretical reasons\footnote{According to the Kramer-Kronig relations, the real part of the dielectric constant $\chi$ of the interstellar medium is connected to an integral of the imaginary part \citep[e.g.,][]{Tielens_ISMbook,Draine_book} as $$ \mathrm{Re}\left[\chi(0)\right] -1 = \frac{2}{\pi} \int_{0}^{\infty} \mathrm{Im} \left[\chi(\nu)/\nu \right] d\nu ~.$$ The left-hand side is proportional to the total volume of interstellar dust \citep{Purcell1969}, and thus it must be finite. Then if $\mathrm{Im}(\chi) \propto \nu^{\gamma-1}$ for $\nu \rightarrow 0$, $\gamma$ must be larger than zero. This can be translated to the index of the emissivity, $\beta_d$, at low frequencies being larger than unity.} to assume that $\beta_d \ge 1$.

\cite{Jones2009} suggested that AME could arise from conformational changes in groups of atoms within amorphous, low-temperature grains. If this resonant tunnelling component is associated with sub-micron-sized grains, the AME should show a good correlation with the FIR emission. They argued that the Two Level System model \citep[TLS;][]{Phillips1973, Meny2007} of this phenomenon could reproduce the general shape of the AME spectrum. However, upper limits on AME polarization require that any grains with enhanced microwave emissivity be randomly-oriented, whereas observations by {\it Planck} show that a substantial fraction of the sub-mm emission from dust comes from grains that are aligned \citep{PIP_XIX}. Upper limits on AME polarization are thus not supportive of large grains as the origin of AME. \cite{Jones2009} pointed out an alternative possibility that resonant tunnelling may be associated with stochastically-heated very small grains, since they spend much of their time at low temperatures, where resonant tunnelling becomes important \citep{Meny2007}.  In this case a good correlation of AME with the mid-infrared excess emission (20--60\,$\mu$m) is expected. Low-temperature laboratory measurements of those materials at microwave frequencies are certainly needed to study the resonant tunnelling of these materials and draw a clear conclusion on this hypothesis.

Free-free emission from warm ionized gas could potentially contribute to excess emission above 10\,GHz. Optically thin free-free emission has a well-defined spectrum that is very close to a $\beta=-2.1$ ($\alpha=-0.1$) power-law with very little variation with frequency \citep{Draine_book}. However, at high gas densities, the gas becomes optically thick at lower frequencies and the spectrum acts like a blackbody with a R-J spectrum ($\beta=0$ or $\alpha=+2$) i.e., rising spectrum with frequency. This phenomena is well-known, for example, in ultracompact and hypercompact H{\sc ii} regions with densities of $\gtrsim 10^{6}$\,cm$^{-3}$ and higher, which can be optically thick up to frequencies of $\sim 15$\,GHz and higher \citep{Kurtz1994,Kurtz2002,Kurtz2005}; see also \cite{Dickinson2013b} for a review of AME from H{\sc ii} regions. However, this will only occur on small scales (typically arcsec) and along certain lines-of-sight, typically along the Galactic plane. For the majority of sight-lines across the sky, and at lower angular resolution, free-free emission is optically thin above 1\,GHz. As a guide, the Orion nebula (M42) is one of the brightest diffuse H{\sc ii} regions, with an angular size of $\approx 5$\,arcmin and a density of $\approx 10^{4}$\,cm$^{-3}$ and has a turnover frequency of $\approx 1$\,GHz. \cite{PIP_XXV} evaluated the contribution of UCH{\sc ii} regions to the flux density as seen by WMAP and {\it Planck} on scales of $1^{\circ}$ using IRAS colour ratios and high resolution 5\,GHz data from the CORNISH survey \citep{Purcell2013}. They found that, for most sources in the Galactic plane, the contribution from optically thick free-free emission is minimal. The two best examples of spinning dust, Perseus and Ophiuchus, do not have substantial ongoing high-mass star formation and high resolution observations do not reveal any bright compact sources. Nevertheless, care must be taken to consider optically thick free-free emission when observing compact and dense regions. 

Similarly, one can obtain a peaked synchrotron spectrum around the frequency of unit optical depth, but this requires extremely high brightness temperatures, $T_b \gg m_e c^2/k_B \sim 10^{10}$\,K. In fact, if the peak is at $\approx 30$\,GHz, either the magnetic field must be far lower than typical values in the ISM, or $T_b$ must be orders of magnitude higher. Given that the observed brightness of AME is of order mK at frequencies $\sim 30$\,GHz, this would have to be an artefact of extreme beam dilution; but high-resolution radio surveys show that the required population of compact sources peaking at 15--30\,GHz does not exist. Nevertheless, AME at high Galactic latitudes is everywhere superposed on the diffuse Galactic synchrotron emission, which at $\nu > 10$\,GHz can be fitted as a power-law with $\beta \approx -3$ or $\alpha \approx -1$ \citep[e.g.,][]{Strong2011}. 

\begin{figure}
\begin{center}
\includegraphics[width=0.45\textwidth,angle=0]{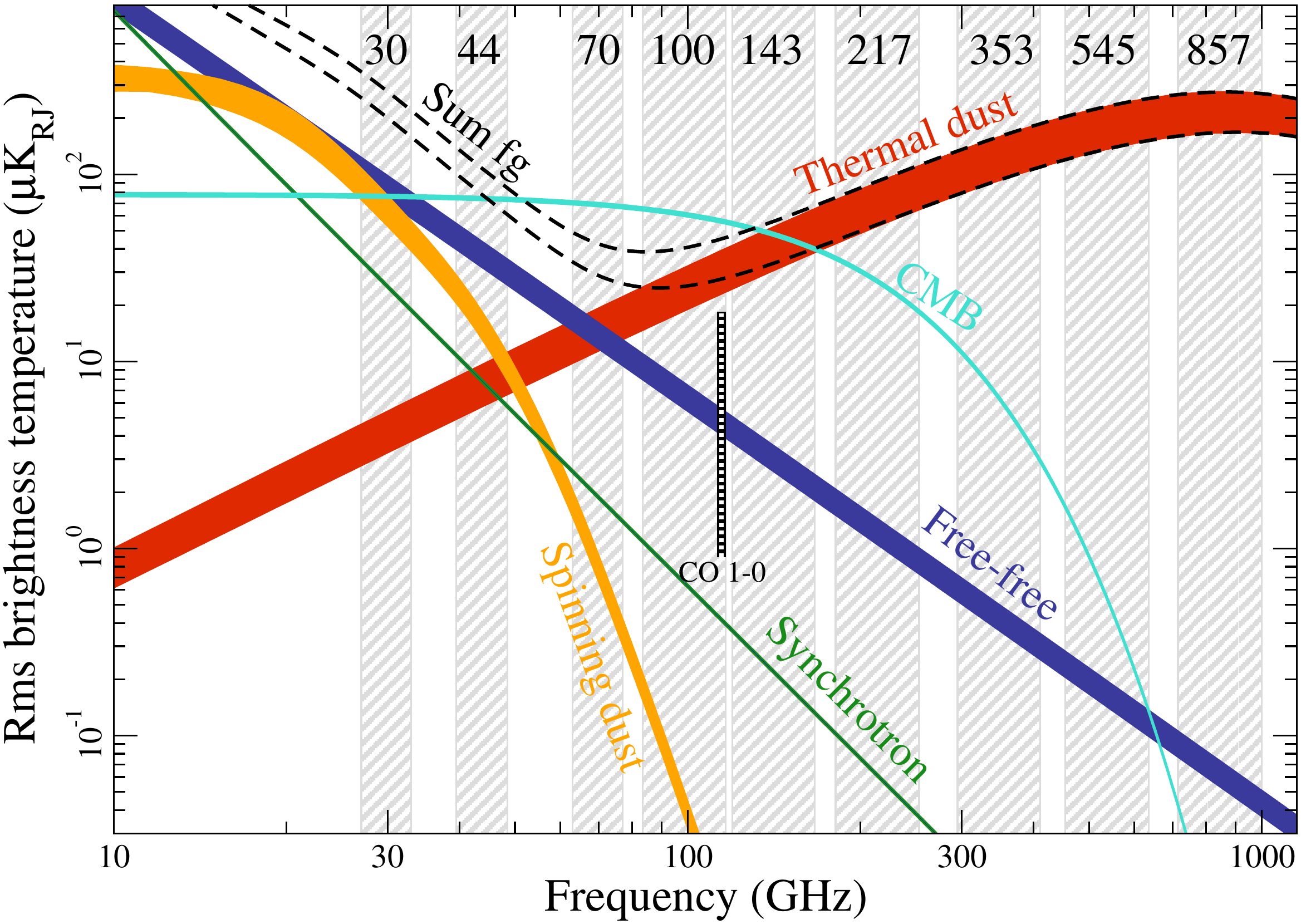}
\caption{Summary of the amplitude of intensity (temperature) foregrounds from the {\it Planck} component separation 2015 results; figure taken from \protect\cite{Planck2015_X}. The brightness temperature r.m.s. against frequency, on angular scales of 40\,arcmin, is plotted for each component. The width of the curves represents the variation when using $81\,\%$ and $93\,\%$ of the sky. }
\label{fig:planck_sed}
\end{center}
\end{figure}

In the frequency range 20--50\,GHz where AME contributes significantly to the sky maps made by CMB experiments, in particular {\it WMAP} and {\it Planck}, synchrotron and AME spectra are usually indistinguishable, since the low-frequency tail of the AME spectrum is just out of range (see Fig.\,\ref{fig:planck_sed}). This difficulty can clearly be seen in Fig.\,\ref{fig:planck_sed}, which presents a summary of the separation of diffuse Galactic components on $1^{\circ}$ scales covering 81--93\,\% of the sky \citep{Planck2015_X}. A contributing factor to the degeneracy is that the synchrotron spectrum is not expected to have spatially-uniform spectral index, nor to be exactly a power-law. These are both rather minor effects: although maps of Galactic spectral index \citep[e.g., ][]{Reich1988,Dickinson2009b} can sometimes show rather large variations in spectral index ($-2 > \beta > -3.5$)\footnote{Brightness temperature spectral indices ($T_{b} \propto \nu^{\beta}$) are related to flux density spectral indices ($S \propto \nu^{\alpha}$) by $\beta=\alpha-2$.}, these are largely due to (i) contamination by free-free emission, especially at low latitudes, and (ii) large uncertainties in the brightness near the minima of the all-sky synchrotron emission, due to uncertainties in the zero level. Residual variations in synchrotron $\beta$ where neither of these effects are important are around $\pm 0.2$ \citep[e.g.,][]{Davies1996}. The Galactic synchrotron spectrum steepens by about 0.4 between 1 and 10\,GHz, but the models of \cite{Strong2011} suggest a return to power-law behaviour above this knee. Although models of the high-frequency AME spectrum are much more strongly curved than synchrotron, this is not a very useful discriminator because both components are rapidly swamped by CMB and thermal dust emission above about 70\,GHz. This AME/synchrotron degeneracy accounts for the widely-divergent estimates of the fractional contribution of AME to the Galactic emission in the lowest-frequency {\it WMAP} and {\it Planck} bands. Nevertheless, the apparent lack of AME polarization (Sect.\,\ref{sec:obs_polarization}) suggests that synchrotron emission cannot account for the majority of AME.

In summary, other mechanisms including blackbody, synchrotron, free-free and various forms of thermal dust emission do not appear to be able to explain the majority of the AME. However, they should be considered carefully in case we are mis-understanding the emission mechanisms. Furthermore, they all contribute to the signal at some level, and therefore require accurate removal to accurately constrain AME.

%%%%%%%%%%%%%%%%%%%%%%%%%%%%%%%%%%%%%%%%%%%%%%%%%%%%%%%
%%%%%%%%%%%%%%%%%%%%%%%%%%%%%%%%%%%%%%%%%%%%%%%%%%%%%%%

\section{Observations} 
\label{sec:observations}

In this section we briefly review the observational status of AME research. We begin by discussing observational results in intensity (temperature) on large scales (Sect.\,\ref{sec:obs_intensity}), where we mainly focus on data from CMB experiments operating on angular scales typically $\gtrsim 1^{\circ}$. We then move on to targeted observations of specific regions in Sect.\,\ref{sec:obs_smallscales}, which are on angular scales typically $\lesssim 1^{\circ}$. We discuss extragalactic AME separately in Sect.\,\ref{sec:obs_extragalactic} and then review current polarization constraints in Sect.\,\ref{sec:obs_polarization}. Section\,\ref{sec:obs_summary} presents a summary of the observational constraints and discusses their interpretation.

%%%%%%%%%%%%%%%%%%%%%%%%%%%%%%%%%%%%%%%%%%%%%%%%%%%%%%%

\subsection{AME observations in intensity/temperature on large-scales} 
\label{sec:obs_intensity}

Until around the mid-1990s, sub-millimetre to centimetre radio emission from the Galaxy was thought to be understood, being a combination of (i) synchrotron radiation from
cosmic-ray electrons spiralling in the Galactic magnetic field, (ii) free-free (thermal bremsstrahlung) emission from electrons scattering in warm ionized ($T \sim 10^{4}$\,K) gas, and (iii) thermal (vibrational) emission from warm ($T\sim20$\,K) dust.

This simple picture has been complicated by mounting evidence for an additional component of emission at frequencies $\nu \sim $10--100\,GHz, spatially correlated with dust, but orders of magnitude stronger than any simple extrapolation of the thermal dust spectrum would predict. A widespread, dust-correlated component was detected in the {\it COBE}-DMR maps and attributed to a combination of thermal dust and free-free emission \citep{Kogut1996}. This interpretation could not be confirmed due to the lack of frequency bands and because full-sky H$\alpha$ maps were still not available at that time.

\begin{figure}
\begin{center}
\includegraphics[width=0.4\textwidth,angle=0]{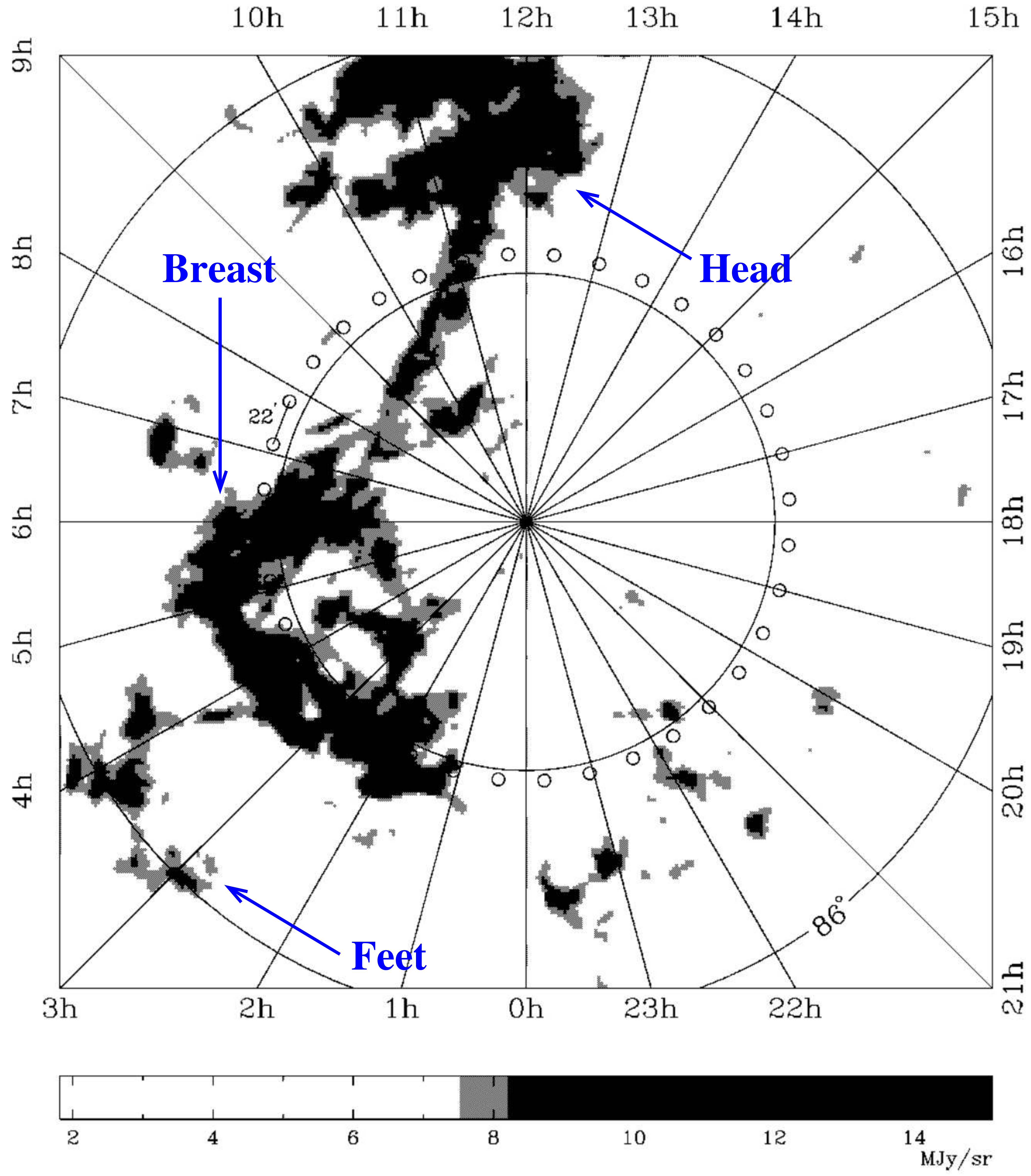}
\includegraphics[width=0.45\textwidth,angle=0]{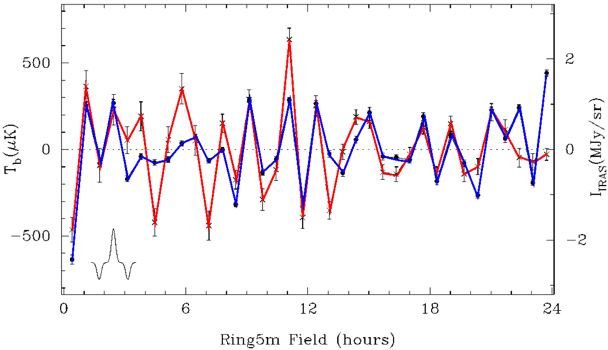}
\caption{{\it Top panel}: Far-IR $100\,\mu$m map of the NCP region. The major structure is sometimes referred to as ``the duck'' because of its similarity in shape. {\it Bottom panel}: 14.5\,GHz data from the RING5M experiment ({\it blue}) plotted against the $100\,\mu$m map, after convolving with the triple-beam of the experiment, shown in the bottom left corner. The remarkable correlation of AME with dust emission at far-IR wavelengths is evident by eye. Figures reproduced and adapted from \protect\cite{Leitch1997}.}
\label{fig:leitch}
\end{center}
\end{figure}

The dust-correlated component was first discovered to be anomalous (and hence the term ``anomalous microwave emission", or AME) by \cite{Leitch1997} at Caltech, who used the Owens Valley Radio Observatory (OVRO) 40-m and 5.5-m telescopes to observe the North Celestial Pole (NCP) region at 14.5\,GHz and 32\,GHz on angular scales 7--22\,arcmin. They detected foreground emission that was spatially correlated with the $100\,\mu$m IRAS maps (Fig.\,\ref{fig:leitch}) but with a microwave spectral index of $\beta \sim -2$, suggestive of free-free emission. Comparison with H$\alpha$ maps of the NCP region showed that the observed signal was at least 60 times stronger than predicted free-free levels. Leitch et al. concluded that if it were free-free, it could only be emission from very hot ($T_e \gtrsim 10^6$\,K) plasma. This explanation was suggested by the shock morphology of the NCP region but was subsequently shown to require implausibly high energy injection rates \citep{Draine1998a}.

Since then, ``anomalous'', dust-correlated emission has been seen in the Galaxy by numerous experiments aiming to detect CMB anisotropies. These include Saskatoon \citep{deOliveira-Costa1997}, 19\,GHz survey \citep{deOliveira-Costa1998}, Tenerife \citep{deOliveira-Costa1999,Mukherjee2001}, QMAP \mbox{\citep{deOliveira-Costa2000}}, ACME/SP94 \citep{Hamilton2001}, and Python V \citep{Mukherjee2003}, amongst others. These analyses largely relied on fitting of foreground templates to maps (Sect.\,\ref{sec:cmb}), where the AME was extracted using a far-IR dust template, such as the $100\,\mu$m IRAS map. By employing multiple spatial templates for the synchrotron (low frequency data), free-free (H$\alpha$), thermal dust (FIR/IR) and CMB (the CMB data themselves), the various foreground components can be separated based on their spatial morphology. The results showed a strong, dust-correlated component of Galactic emission, that could not be easily explained by synchrotron, free-free, thermal dust or CMB radiation. 

\begin{figure}
\begin{center}
\includegraphics[width=0.45\textwidth,angle=0]{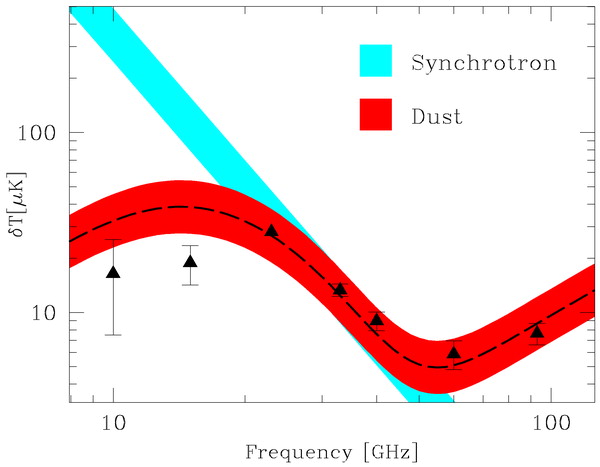}
\caption{Spectrum of foregrounds from WMAP data (22.8\,GHz and above) and Tenerife data (10 and 15\,GHz) from a template-fitting analysis at high latitudes ($|b|>20^{\circ}$). The dust-correlated foreground appears to turn over at frequencies $\approx 15$\,GHz, supporting the spinning dust origin for AME. Figure taken from \protect\cite{deOliveira-Costa2004}.}
\label{fig:tenerife}
\end{center}
\end{figure}

A joint analysis of  {\it COBE}-DMR and 19\,GHz data \citep{Banday2003} provided a high S/N detection of AME through cross-correlation of foreground templates. A joint analysis of Tenerife 10/15\,GHz data with {\it WMAP} \citep{deOliveira-Costa2002,deOliveira-Costa2004} showed the first evidence for a turnover at a frequency $\approx 15$\,GHz, supporting the spinning dust hypothesis. Fig.~\ref{fig:tenerife} shows the SED of diffuse emission in the high-latitude Tenerife ``strip", which clearly shows the preference for a flattening and turnover of the spectrum at a frequency $\approx 15$\,GHz due to the mysterious ``foreground X" (AME), which can be explained by spinning dust. 

First results from the WMAP team using 1-year data suggested that a harder (flatter spectrum, $\beta \approx -2.5$) component of synchrotron radiation could account for AME \citep{Bennett2003b}.  However, their interpretation was different in later releases, where a spinning dust component was considered \citep[e.g.,][]{Gold2011}. Several other results using WMAP data showed further evidence for AME \citep{Lagache2003,Finkbeiner2004,Dobler2008}. 

\begin{figure*}
\begin{center}
\includegraphics[width=1.0\textwidth,angle=0]{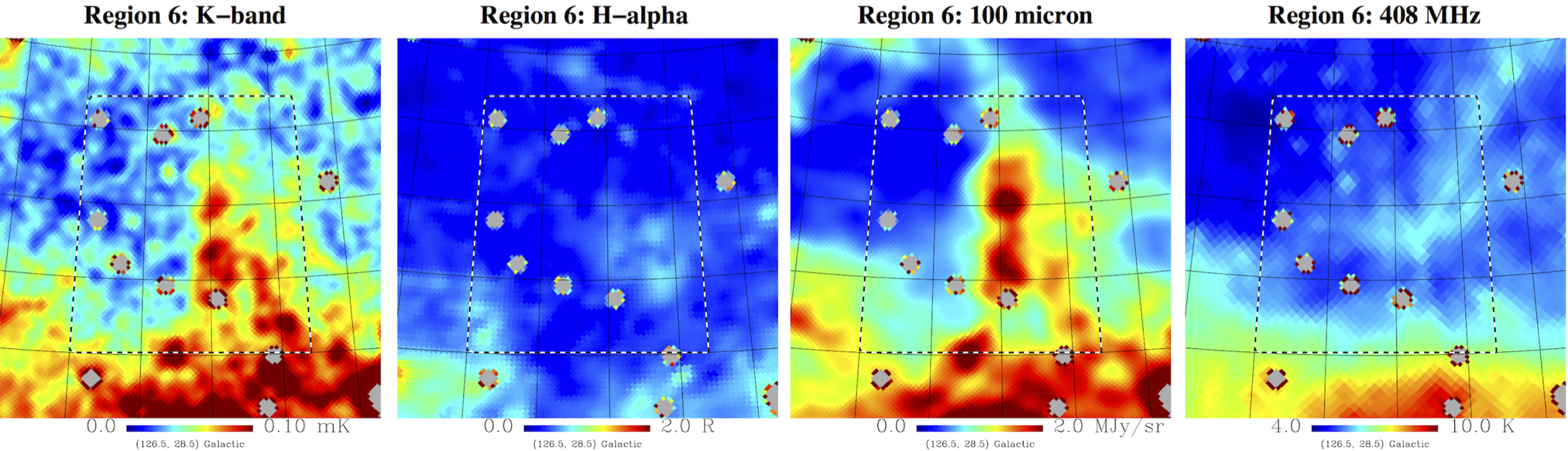}
\caption{Maps of a dust-dominated region centred at $(l,b)=(126^{\circ}\!\!.5,+28^{\circ}\!\!.5)$, near the NCP region. From {\it left} to {\it right} are the WMAP K-band (22.8\,GHz), H$\alpha$ to trace free-free emission, $100\,\mu$m to trace dust emission and 408\,MHz to trace synchrotron emission. All maps are smoothed to an angular resolution of $1^{\circ}$. There is a strong correlation of the 22.8\,GHz data with dust but not the other foregrounds, indicating AME. Figure reproduced from \protect\cite{Davies2006}.}
\label{fig:region6}
\end{center}
\end{figure*}

A comprehensive study of template fits to {\it WMAP} data was made by \cite{Davies2006}. They analysed 15 regions, chosen by hand to be dominated by one specific component of either synchrotron/free-free/dust. Fig.~\ref{fig:region6} shows one of the dust-dominated regions that corresponds to the NCP region where AME was first identified (Fig.\,\ref{fig:leitch}). The dust-correlated emission is easily discernible by eye while the other foregrounds (synchrotron traced by 408\,MHz data and free-free traced by H$\alpha$ data) do not correlate strongly with the K-band (22.8\,GHz) data. \cite{Davies2006} also found that the brightness of the AME per unit thermal dust was remarkably constant (sometimes, confusingly, referred to as ``emissivity"), with an average value of $\approx 10\,\mu$K at 30\,GHz per MJy/sr at a wavelength of $100\,\mu$m. At 22.8\,GHz the emissivity is $\approx 20\,\mu$K/(MJy/sr); see Table~\ref{tab:emissivities}. This ratio varied by up to a factor of $\approx 2.5$ across the sky when not including the Galactic plane, showing the apparent ubiquitous nature of AME, at least at high Galactic latitude. A more detailed work, covering 35 regions, was made by \cite{Ghosh2012}. \cite{Miville-Deschenes2008} used WMAP polarization data to constrain the synchrotron spectral index and therefore separate the Galactic components, again showing a strong dust-correlated AME component.

Other large-scale data have been combined with WMAP data to extend the frequency range and improve component separation. \cite{Lu2012} re-analysed archival CMB data at 8\,GHz and showed that an additional component of emission was required to explain the 23\,GHz data. COSMOSOMAS data at 13--17\,GHz also showed AME with a peaked spectrum around 22\,GHz \citep{Hildebrandt2007}. Data from the ARCADE2 experiment at 3, 8 and 10\,GHz gave a strong preference for an AME component that accounts for $40 \pm 10\,\%$ of the Galactic plane emission at 23\,GHz \citep{Kogut2011}.

With the release of {\it Planck} data \citep{Planck2015_I}, a more complete picture of the spectrum of diffuse Galactic emission became available, particularly at higher frequencies. As well as the targeted analysis of the Perseus and Ophiuchus clouds \mbox{\citep{PEP_XX}}, a separation of the interstellar medium components by ``inversion'' \citep{PEP_XXI}, which utilized multiple tracers including velocity-resolved line emission spectra thus allowing a 3-D separation to be made, showed the need for AME with an amplitude that meant that a substantial fraction ($25\pm5\,\%$) of the Galactic plane emission at 30\,GHz was due to AME. A separate study of the Galactic plane components as seen by {\it Planck} estimated that $\approx 45\,\%$ of the 30\,GHz emission could be due to AME \citep{PIP_XXIII}.

\setlength{\tabcolsep}{6pt}
\begin{table*}
\begin{center}
\caption{Selected AME emissivities (correlation coefficients) at $\approx 20$--30\,GHz measured with various reference dust templates and regions of sky.}
\label{tab:emissivities}
\begin{tabular}{lccccl}
\hline
\noalign{\smallskip}
Sky region				&Freq. (GHz)	&AME emissivity 		&Units 				&Template 				&Reference \\
\noalign{\smallskip}
\hline 
\noalign{\smallskip}
WMAP Kp2 mask ($85\%$\,sky) 		&22.8		&$21.8\pm1.0$			&$\mu$K/(MJy/sr)		&$100\,\mu$m	(IRIS)		&\cite{Davies2006}   		\\
$|b|>10^{\circ}$				&22.8		&$21\pm2$ 			&$\mu$K/(MJy/sr)		&$100\,\mu$m (IRIS)		&\cite{Planck2015_XXV}	\\
Perseus					&22.8		&$24\pm4$			&$\mu$K/(MJy/sr)		&$100\,\mu$m	(IRIS)		&\cite{PIP_XV}   		\\
$\rho$\,Oph\,W				&22.8		&$8.3\pm1.1$			&$\mu$K/(MJy/sr)		&$100\,\mu$m	(IRIS)		&\cite{PIP_XV}   		\\
32 source mean 			&22.8		&$32\pm4$			&$\mu$K/(MJy/sr)		&$100\,\mu$m	(IRIS)		&\cite{PIP_XXV}   		\\
Diffuse HII regions 			&33			&$4.65 \pm 0.40$		&$\mu$K/(MJy/sr)		&$100\,\mu$m	(IRIS)		&\cite{Todorovic2010}  	\\
\noalign{\smallskip}
\hline
\noalign{\smallskip}
$|b|>10^{\circ}$ 			&22.8		&$9.7\pm1.0$			&$10^{6}\,\mu$K		&$\tau_{353}$ ({\it{Planck}})		&\cite{Planck2015_XXV}	\\
Perseus					&28.4		&$12.3\pm1.2$			&$10^{6}\,\mu$K		&$\tau_{353}$ ({\it{Planck}})		&\cite{PIP_XV}	\\
$\rho$\,Oph\,W				&28.4		&$23.9\pm2.3$			&$10^{6}\,\mu$K		&$\tau_{353}$ ({\it{Planck}})		&\cite{PIP_XV}	\\
$26\,\%$ high latitude sky		&30			&$7.9\pm2.6$			&$10^{6}\,\mu$K		&$\tau_{353}$ ({\it{Planck}})		&\cite{Hensley2016}	\\
\noalign{\smallskip}
\hline
\noalign{\smallskip}
$|b|>10^{\circ}$				&22.8		&$70\pm7$ 			&$\mu$K/(MJy/sr)		&545\,GHz ({\it{Planck}})		&\cite{Planck2015_XXV}	\\
\noalign{\smallskip}
\hline
\noalign{\smallskip}
$26\,\%$ high latitude sky		&30			&$6240\pm1210$		&MJy/sr /(W/m$^2$/sr)	&({\it{Planck}}) (${\cal R}$)		&\cite{Hensley2016}	\\
\noalign{\smallskip}
\hline
\noalign{\smallskip}
$26\,\%$ high latitude sky		&30			&$271\pm89$			&$\mu$K/(MJy/sr)		&$12\,\mu$m (WISE)		&\cite{Hensley2016}	\\
\noalign{\smallskip}
\hline
\end{tabular}
\end{center}
\end{table*}
\setlength{\tabcolsep}{6pt}

The {\it Planck} 2015 results included full-sky maps of AME based on a parametric SED-fitting algorithm \citep{Planck2015_X}. The fit included two AME components based on models of spinning dust with the {\sc spdust2} code \citep{Ali-Hamoud2009,Silsbee2011}, one with a variable peak frequency and one with a fixed peak frequency (33.35\,GHz). Two components with different peak frequencies are needed to account for the broadening of the total AME, presumably due to the presence of multiple components along the line-of-sight (which would inevitably broaden the spectrum compared to a single component model). Even using the latest {\it Planck}, {\it WMAP}, and ancillary data, the separation is known to be far from perfect. Due to degeneracies between parameters, caused by lack of frequency coverage particularly in the range $\approx 5$--20\,GHz, the data cannot always distinguish between the continuum components. The primary limitation was having to effectively fix the synchrotron spectrum in each pixel, and thus not fully accounting for spectral variations across the sky. Indeed, careful inspection of the component maps shows clear examples of aliasing of power between the various components, such as AME signal leaking into the free-free solution; see \cite{Planck2015_XXV} for a detailed discussion. Nevertheless, the same close correlation with dust was observed and with comparable emissivities to previous works. The emissivity defined by the thermal dust optical depth at a wavelength of $250\,\mu$m was shown to be more constant than previous estimates based on FIR brightness that was affected by variations in dust temperature \citep{Tibbs2012b}. The analysis also revealed more tentative evidence for a variable spinning dust peak frequency, where some pixels prefer a higher peak frequency. The peak frequency is typically near 30\,GHz (in flux density), while some regions prefer a peak near $40$--50\,GHz. Such regions, including the California nebula, tend to be bright HII regions where the environment might lead to smaller and more rapidly spinning dust grains \citep{PIP_XV}. However, as already mentioned, this requires independent confirmation given the complexity of component separation, particularly in the Galactic plane where the simple synchrotron spectral model is likely to be insufficient. 

\cite{Hensley2016} examined the correlation of AME with the \textit{Planck} 353\,GHz optical depth $\tau_{353}$ and with total far-infrared radiance ${\cal R}$, also estimated from \textit{Planck} data and analyses. For fixed dust size distribution, spinning dust models predict that AME should closely follow the total dust column ($\propto \tau_{353}$), but be relatively insensitive to variations in the starlight heating rate ($\propto {\cal R}/\tau_{353}$). Surprisingly, AME is more strongly correlated with ${\cal R}$ than with $\tau_{353}$. The reason for this still remains unclear. \cite{Hensley2016} combined WISE 12$\mu$m observations with ${\cal R}$ from IRAS and \textit{Planck} to estimate the PAH abundance, which appears to show regional variations. They found that AME/${\cal R}$ showed no correlation with estimated PAH abundance, suggesting that AME may be dominated by a source other than PAHs.

Various maps of Galactic dust emission have been used as spatial templates to both trace and estimate the relative brightness of dust-correlated AME. These include maps from COBE-DIRBE 100--240$\,\mu$m \citep{Banday2003}, IRAS/IRIS 12--100$\,\mu$m \citep{Miville-Deschenes2005,Ysard2010b}, combination dust model such as those of \cite{Finkbeiner1999}, dust radiance \citep{Hensley2016}, and HI \citep{Lagache2003} amongst others. As discussed by \cite{Finkbeiner2004}, comparisons of the various emissivity values as a function of different tracers and models has been quite confusing. Results based solely on data are easier to compare, such as the brightness in $\mu$K (or Jy/sr) per unit of MJy/sr at $100\,\mu$m, but are difficult to relate to theory. For example, \cite{Tibbs2012b} demonstrated that in warmer environments, such as HII regions, the $100\,\mu$m brightness is not a useful reference for the dust column since it varies dramatically with dust temperature. This may explain why AME in HII regions typically has a much lower (less than half) $100\,\mu$m emissivity than diffuse high-latitude emission \citep{Dickinson2013b} (see Table~\ref{tab:emissivities}). \cite{Ysard2010b} found that by dividing the $12\,\mu$m IRAS map by the intensity of the interstellar radiation field ($G_0$), a better correlation with AME was obtained on large scales, a result which has also been seen on small scales \citep[e.g.,][]{Tibbs2011,Tibbs2012}; see Sect.\,\ref{sec:obs_smallscales}.

Alternatively, one can choose physical properties as the AME emissivity reference, such as the column density (e.g., Jy\,sr$^{-1}$\,cm$^{2}$), which can (for example) be estimated from the optical depth of thermal dust emission \citep{Planck2013_XI}, making it easier to compare with theoretical models (as in Fig.\,\ref{fig:spdust_ems}) and each other. As can be seen in Fig.\,\ref{fig:spdust_ems}, theoretical models yield values of $\sim 10^{-17}$\,Jy\,sr$^{-1}$\,cm$^{2}$ at $\approx 30$\,GHz, although there is considerable scatter and these models are only indicative. Nevertheless, the observed emissivities are of order this level \citep[e.g.,][]{PEP_XX,PIP_XV,Planck2015_XXV,Hensley2015}. As an example, the mean value at high latitudes ($|b|>15^{\circ}$) of $\tau_{353}/N_{\rm H}$ is $\approx 7.3\times10^{-27}$\,cm$^{2}$ \citep{Planck2013_XI} while typical observed AME emissivities are $\approx 8 \times 10^6\,\mu$K/$\tau_{353}$ (Table~\ref{tab:emissivities}). This corresponds to a brightness per unit column density of $\Delta T/N_{\rm H} \sim 6 \times 10^{-20}\,\mu$K\,cm$^{2}$ or $2 \times 10^{-18}$\,Jy\,sr$^{-1}$\,cm$^{2}$ at 30\,GHz, which is of the same order as the theory values. \cite{Hensley2015} discuss that the observed emissivities are typically a few times $10^{-18}$\,Jy\,sr$^{-1}$\,cm$^{2}$, slightly below the reference value, bringing theory and observation into better agreement. In a recent paper looking at the HI-$E(B-V)$ connection \mbox{\citep{Lenz2017}}, they derived $\tau_{545}/N_{\rm H} = 4.46 \times 10^{-26}$\,cm$^{2}$. Assuming an emissivity index $\beta=+1.6$ (therefore $\tau^{1.6}$)  gives $2.0 \times 10^{-26}$\,cm$^{2}$, which corresponds to an emissivity per $\tau_{353}$ of $1.5 \times 10^8$\,Jy/sr or 5.4\,K, which compares favourably to the observed values $\sim 8$\,K (Table\,\ref{tab:emissivities}). 

In Table~\ref{tab:emissivities} we list a few selected AME emissivities at $\approx 23$--33\,GHz from the literature using various reference dust templates. It can be seen that diffuse AME emissivities have typical values of $\approx 20\,\mu$K/(MJy/sr) at 22.8\,GHz relative to the $100\,\mu$m brightness (one of the most commonly used dust templates). At 30\,GHz values are typically $\approx 6$--$10\,\mu$K/(MJy/sr). A sample of diffuse HII regions have a lower $100\,\mu$m emissivity ($< 5\,\mu$K/(MJy/sr), presumably due to the higher dust temperatures that affects the $100\,\mu$m brightness; the lower value for $\rho$\,Oph\,W region is also likely to be due to the warmer dust temperature \mbox{\citep{PIP_XXV}}. The AME emissivity relative to the optical depth at 353\,GHz is $\approx 8 \times 10^6\,\mu$K/$\tau_{353}$ for diffuse high latitude emission but is higher in the Perseus molecular cloud and, in particular, the $\rho$\,Oph\,W molecular cloud by a factor of $\approx 3$. We also include the results from correlating with the {\it Planck} 545\,GHz brightness, IRIS/WISE 12$\,\mu$m brightness and the total dust radiance, ${\cal R}$, all of which have been found to be even more closely correlated with AME \citep{Planck2015_XXV,Hensley2016}. 

On the other hand, deriving reliable column densities is difficult. First, it is model-dependent, which can make it difficult to make comparisons between different tracers. Thermal dust optical depths have been estimated at several common observing wavelengths, including $100\,\mu$m, $250\,\mu$m, and more recently with \textit{Planck} data at 353\,GHz ($\lambda=850\,\mu$m). However, like many forms of AME emissivity, it is not straightforward to convert between the various estimates because one has to assume properties for the dust (e.g., temperature, emissivity index), while converting from a brightness at say $100\,\mu$m to an optical depth depends on empirical relations, which are known to be non-linear especially over the entire range of brightness/densities observed in the sky \citep{Planck2013_XI,Hensley2016}. Second, observations of compact objects with large beams can result in artificially low values of intensity due to dilution within the beam, resulting in lower effective column densities. An example of this would be the analysis using WMAP/{\it Planck} using $2^{\circ}$ diameter apertures \citep{PIP_XV}, resulting in effective optical depths at $250\,\mu$m\footnote{In table~3 of \protect\cite{PIP_XV}, the thermal dust optical depths at $250\,\mu$m are listed as multiplied by $10^{5}$, when in fact they have been multiplied by $10^{4}$. This results in all the $\tau_{250}$ values in their table~3 being a factor 10 too small.}  (1200\,GHz) of $\tau_{250} \sim 10^{-4}$, which corresponds to the optical depth at 353\,GHz of $\tau_{353} \sim 10^{-5}$, which is below what would be expected at low latitude sight-lines ($\tau_{353} \gtrsim 10^{-4}$ typically). Ultimately, this means that one must be careful when comparing AME emissivities, particularly when using column densities. Trying to convert various observable quantities over a range of environments can lead to large (up to a factor of several) unphysical variations in AME emissivity. Nevertheless, the observed AME emissivities do appear to be the same order of magnitude across various analyses and regions in the sky and the associated column densities ($N_{\rm H}$) inferred from spinning dust models are consistent with expectations. We will discuss this further at the end of the next section.

%%%%%%%%%%%%%%%%%%%%%%%%%%%%%%%%%%%%%%%%%%%%%%%%%%%%%%%

\begin{figure}
\begin{center}
\includegraphics[width=0.4\textwidth,angle=0]{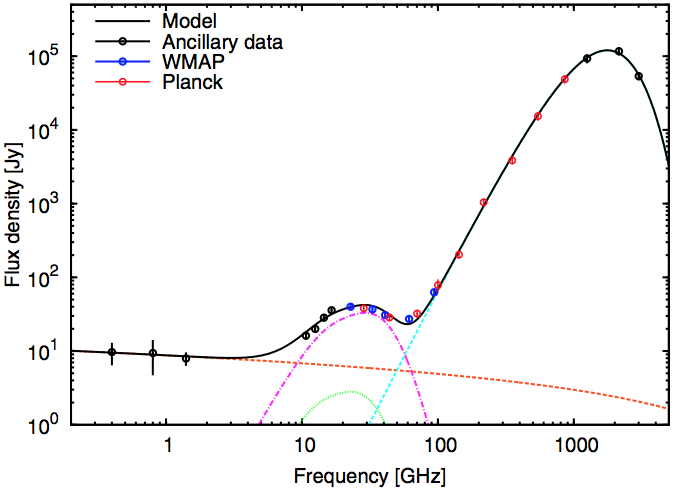}
\includegraphics[width=0.4\textwidth,angle=0]{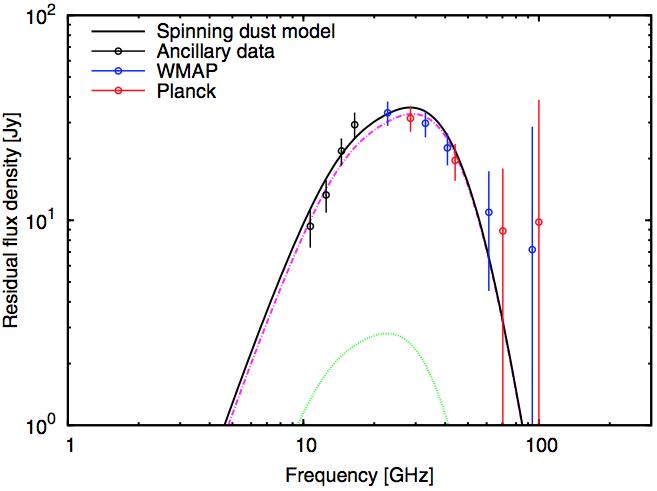}
\caption{The spectrum of G160.26--18.62 in the Perseus molecular cloud ({\it top}) and the residual spectrum showing the spinning dust component ({\it bottom}). The spectrum is fitted by components of free-free ({\it orange dashed line}), CMB (not visible), thermal dust ({\it dashed cyan line}) and spinning dust ({\it green dotted} and {\it magenta dot-dashed lines} for the atomic and molecular phases, respectively), which peaks at $\approx 30$\,GHz. The theoretical spectrum is a remarkably good fit to the data with parameters that are physically motivated. Reproduced from \protect\cite{PEP_XX}.}
\label{fig:perseus}
\end{center}
\end{figure}

\subsection{AME on small ($\lesssim 1^{\circ}$) scales} 
\label{sec:obs_smallscales}

On smaller angular scales, typically on scales of  $\lesssim 1^{\circ}$, a number of dedicated observations have been made to study AME in more detail in specific environments and clouds.

The Green Bank 140ft telescope was used to observe 10 dust clouds at 5, 8, and 10\,GHz \citep{Finkbeiner2002}. Using 1-D scans, a spectrum was estimated from 5 to 10\,GHz. They found 8 with negative (falling with increasing frequency) spectral indices and 2 showing a rising spectrum, indicative of spinning dust. The first was the dark cloud LDN1622, which was confirmed later with 31\,GHz data from the Cosmic Background Imager (CBI) \citep{Casassus2006}. The GBT 100-m telescope was also used to map the peak of the emission at 5 and 14\,GHz, constraining the free-free emission and showing a rising spectrum from 14 to 31\,GHz \citep{Harper2015}. The second source was the HII region LPH201.663+1.643. However, follow-up observations with the CBI at 31\,GHz showed no significant excess above the expected free-free level \citep{Dickinson2006}; unpublished GBT follow-up observations (Doug Finkbeiner, priv. comm.) were also not able to reproduce the initial result. The Green Bank Galactic Plane survey was used to study the diffuse emission from the Galactic plane at 8 and 14\,GHz \citep{Finkbeiner2004b}. When combined with lower frequency data at 2.3\,GHz and WMAP 1-year data, a rising spectrum between 8 and 14\,GHz provided strong evidence for AME.

\begin{figure*}
\begin{center}
\includegraphics[width=1.0\textwidth,angle=0]{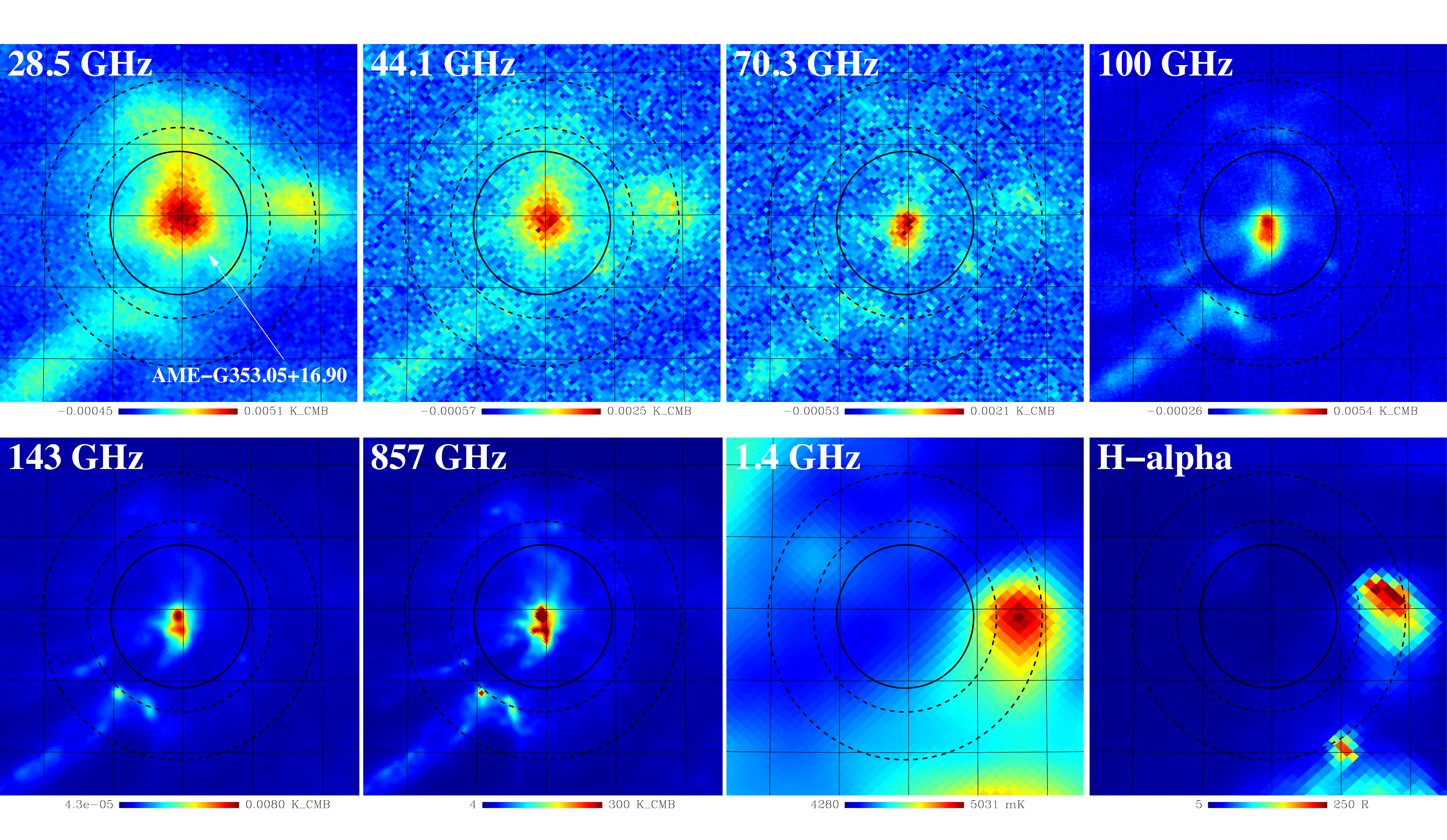}
\caption{Multi-frequency maps of the $\rho$\,Oph\,W molecular cloud region centred at $(l,b)=(353^{\circ}\!\!.05,+16^{\circ}\!\!.90)$. From the {\it left} to {\it right}, from the {\it top} row: 28.5, 44.1, 70.3, 100, 143 and 857\,GHz from {\it Planck}, 1.4\,GHz and H$\alpha$. The maps cover $5^{\circ}$ by $5^{\circ}$ and the graticule has a $1^{\circ}$ spacing. The strong AME at $\approx 20$--40\,GHz is evident. Note how the relatively weak HII region to the right of the main cloud is strong at low radio frequencies (1.4\,GHz) and in H$\alpha$ but is weak compared to AME at frequencies $\gtrsim 20$\,GHz. Figure reproduced from \protect\cite{PEP_XX}.}
\label{fig:roph_maps}
\end{center}
\end{figure*}

One of the most important AME detections was made towards the Perseus molecular cloud by the COSMOSOMAS experiment on angular scales of $\approx 1^{\circ}$. \cite{Watson2005} combined COSMOSOMAS data at 11, 13, 17 and 19\,GHz, with WMAP data to produce an AME spectrum that closely follows the prediction from spinning dust grains. This was the first time that a clear rise in the spectrum below the peak at $\approx 30$\,GHz had been seen, which is consistent with expectations for spinning dust. \mbox{\cite{PEP_XX}} made a definitive detection by combining these data with {\it Planck} data, providing more data points and showing that the R-J tail of thermal dust emission was essentially negligible at frequencies $\sim 30$\,GHz. More importantly, a physically-motivated spinning dust model, based on knowledge of the environment within the Perseus molecular cloud, provided an excellent fit to the data. Fig.\,\ref{fig:perseus} shows the spectrum of G160.26--18.62 on $1^{\circ}$ scales, with synchrotron/free-free/spinning dust/thermal dust components shown. The spinning dust spectrum is a very high S/N detection and is well-fitted by 2 distinct components (low and high density), based on a plausible physical model. 

Another important AME region is the $\rho$~Ophiuchus molecular cloud. The first detection was made at 31\,GHz with the CBI on arcmin scales, showing strong cm-emission associated with $\rho$~Oph~W PDR \citep{Casassus2008}. Fig.~\ref{fig:roph_maps} shows large-scale multi-frequency maps of the region, where AME is clearly evident. Spectral modelling showed that the spinning dust model could comfortably fit the data. A more detailed picture was made with the WMAP and {\it Planck} data, which showed that a plausible physical model could easily explain the emission \citep{PEP_XX}. Follow-up observations with ATCA (Casassus et al., in prep.) not only confirm the emission but, for the first time, shows a spatial shift of the spinning dust emission with frequency. This might be as expected from the varying dust properties across the PDR but requires detailed modelling. Nevertheless, this is potentially one of the strongest pieces of evidence for the spinning dust explanation. 

A number of close-packed, microwave interferometers, operating either at $\sim 15$ or $\sim 30$\,GHz, have been extensively used for AME research. In particular, the CBI at 26--36\,GHz \citep{Padin2002}, the Very Small Array at 26--36\,GHz (VSA; \citealt{Dickinson2004}), and the Arcminute Microkelvin Imager at 13--18\,GHz \citep[AMI;][]{Zwart2008} have been used. Although they were primarily designed for CMB studies, their compact configuration resulted in good brightness sensitivities and resolutions of a few arcmin, ideal for AME research. The Combined Array for Research in Millimeter-wave Astronomy at 27--35\,GHz (CARMA) has also been used.

Data from the CBI resulted in detections of AME from LDN1622 \citep{Casassus2006} and LDN1621 \citep{Dickinson2010}, $\rho$~Oph~W \citep{Casassus2008}, the H{\sc ii} region RCW175 \citep{Dickinson2009a,Tibbs2012}, the translucent cloud LDN1780 \citep{Vidal2011},  and the reflection nebula M78 \citep{Castellanos2011}. The CBI was used to refute the earlier claim of AME in the H{\sc ii} region LPH96 \citep{Dickinson2006}, which was shown to follow a normal optically-thin free-free spectrum. The CBI was also used to survey the brightest 6 H{\sc ii} regions in the southern sky \citep{Dickinson2007} and two bright star-forming regions \citep{Demetroullas2015}, both finding little evidence for AME at 31\,GHz. AME studies with the VSA revealed a flattening of the spectrum of the supernova remnant 3C396, which was interpreted as a possible signature of spinning dust. However, \cite{Cruciani2016} made follow-up observations with the Parkes 64-m telescope at 8--19\,GHz and combined it with unpublished 31.2\,GHz GBT data but found no evidence for AME. 

The VSA was used to make a survey of the Galactic plane $(l=27^{\circ}$--$46^{\circ}$, $|b|<4^{\circ}$) at 33\,GHz with an angular resolution of 13\,arcmin, finding an AME detection in RCW175 and statistical evidence of excess emission from the brightest sources in the sample \citep{Todorovic2010}. The VSA was also used to perform the first detailed morphological analysis of AME in the Perseus cloud at 7\,arcmin resolution, identifying five regions of AME \citep{Tibbs2010}. However, the total flux density of the AME in these five regions accounted for only $\sim 10\,\%$ of AME detected on degree angular scales by {\it Planck}, implying that the AME in Perseus is coming from a diffuse component of gas/dust and is not concentrated in the five compact regions. A similar result was found by \citet{PIP_XV} in their sample of potential new AME candidates. \cite{Tibbs2013b} used the GBT 100-m telescope at 1.4 and 5\,GHz to constrain the level of the free-free emission in Perseus on arcmin scales, confirming that the observed 33\,GHz VSA data were mostly due to AME.

\begin{figure}
\begin{center}
\includegraphics[width=0.45\textwidth,angle=0]{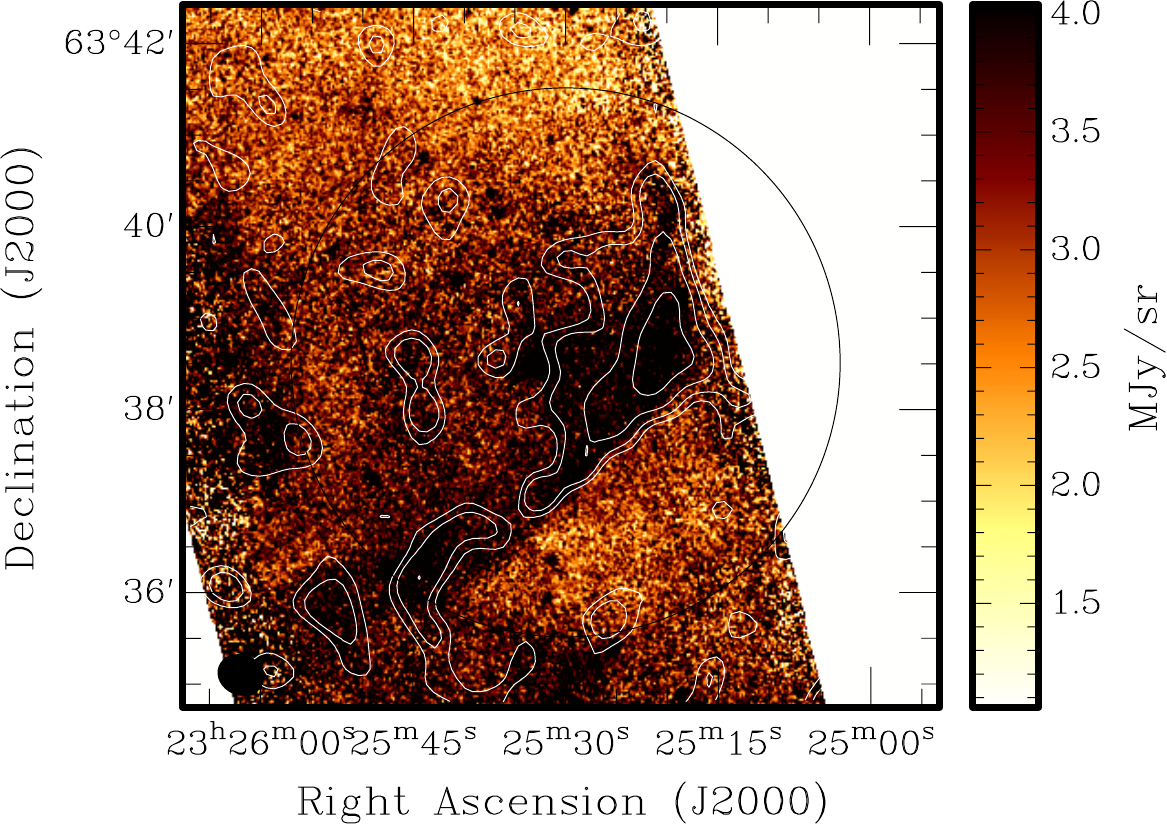}
\caption{AMI Large Array (AMI-LA) combined 16\,GHz data shown as white contours at 1, 2, 4, 8$\sigma$ of the local noise level. \textit{Spitzer} band 4 ($8\,\mu$m) map shown as greyscale in MJy/sr, saturated at both ends of the scale to emphasise the diffuse structure present. The correlation of the microwave emission with IR is evident. The AMI-LA primary beam (field-of-view) is shown as a circle and the synthesized beam (angular resolution FWHM) as a filled ellipse in the bottom left corner. Figure reproduced from \protect\cite{Scaife2010b}.}
\label{fig:L1246}
\end{center}
\end{figure}

\begin{figure*}
\begin{center}
\includegraphics[width=0.95\textwidth,angle=0]{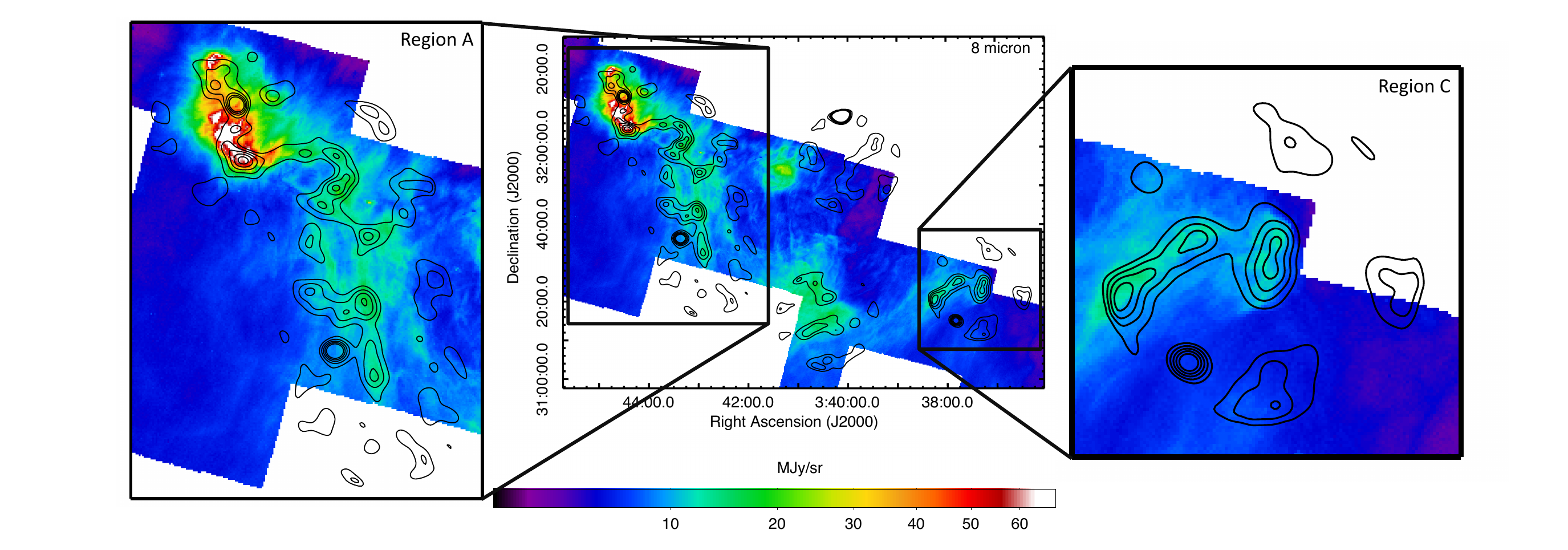}
\includegraphics[width=0.95\textwidth,angle=0]{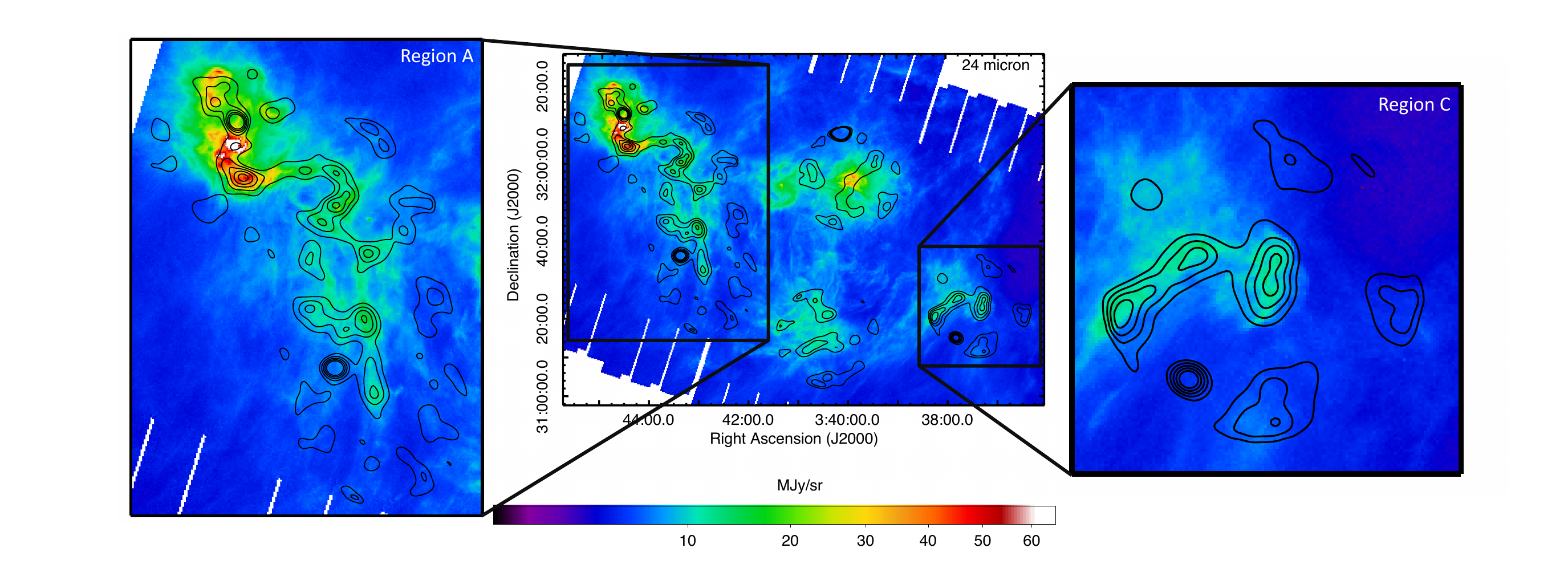}
\caption{\textit{Spitzer} colour maps of the G159.6--18.5 region of the Perseus molecular cloud at $8\,\mu$m (\textit{top}) and $24\,\mu$m (\textit{bottom}), overlaid with AMI 16\,GHz black linear contours at an angular resolution of $\approx 2$\,arcmin. There remains a strong correlation between AME and IR emission, with the tightest correlation being with the $24\,\mu$m band, suggesting that PAHs might not be responsible for the bulk of the AME in this region.}
\label{fig:AMI_Perseus}
\end{center}
\end{figure*}

Several interesting results have come from high angular resolution AMI data, operating in the unique frequency band of 13--18\,GHz. A survey of compact H{\sc ii} regions found essentially no evidence of excess emission \citep{Scaife2008}. Observations of a sample of Lynds clouds resulted in detections in LDN1111 \citep{Scaife2009a} and further detections in several more Lynds clouds \citep{Scaife2009b}. Of these, approximately one third of these were shown to be contaminated by optically thick free-free emission from young stellar objects \citep{Scaife2010b}. However, LDN1246 shows diffuse emission at 16\,GHz on arcmin scales that is closely correlated with $8\,\mu$m maps from {\it Spitzer} \citep{Scaife2010b}; Fig.\,\ref{fig:L1246} shows the striking correlation, which remains one of the best examples of AME on scales of about an arcmin.  \cite{Perrott2013} used AMI to resolve the structure of two {\it Planck} AME sources (G107.1+5.2 and G173.6+2.8; \citealt{PEP_XX}), finding a rising spectrum in G107.1+5.2 that is consistent with either AME or emission from an ultra-compact H{\sc ii} region, but with a much lower flux density than {\it Planck}, consistent with AME originating on larger angular scales. \cite{Perrott2013} found no evidence for AME in G173.6+2.8 on angular scales $\approx 2$--10\,arcmin, suggesting that the bulk of AME is very diffuse. Recently, a blind survey of Sunyaev-Zoldovich clusters has revealed by accident the first blind detection of AME on scales of an arcmin \citep{Perrott2017}.

\cite{Tibbs2013} used AMI to study the AME in the Perseus cloud in even more detail than the VSA, at a resolution of $\approx 2$\,arcmin, and found that the spatial correlation between AME and IR emission remained strong on these scales, as shown in Fig.\,\ref{fig:AMI_Perseus}. More interesting is that the correlation is visibly stronger with $24\,\mu$m emission than it is with $8\,\mu$m emission. This might indicate that AME is originating from a population of stochastically heated small interstellar dust grains rather than PAHs, in agreement with the conclusions from the more general analyses by \mbox{\cite{Hensley2016}} and \mbox{\cite{Hoang2016b}}. Although the AME-IR correlation persisted on small scales, the results also indicated that the AME intensity did not correlate with PAH abundance, but rather with the interstellar radiation field, which may be shaping the dust grain size distribution.

\begin{figure}
\begin{center}
\includegraphics[width=0.5\textwidth,angle=0]{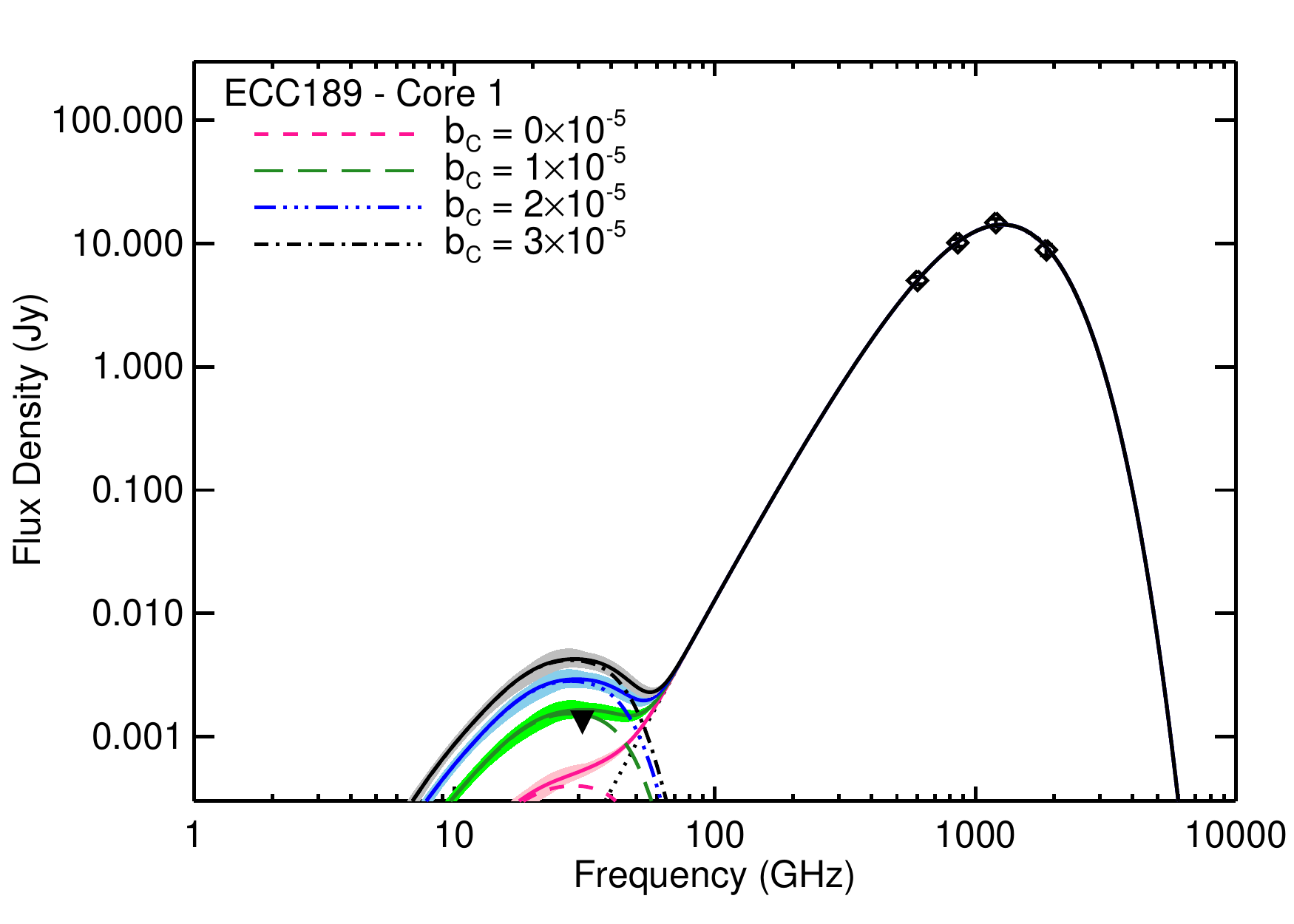}
\caption{SED for the cold core ECC189. The black line is the best-fitting thermal dust model while the coloured lines represent spinning dust models with different values for the total number of C atoms governing the number of very small grains responsible for spinning dust emission. The CARMA upper limit at 30\,GHz is shown, which yields a limit of $b_{\rm C} \le 1 \times 10^{-5}$. Figure taken from \protect\cite{Tibbs2016}.}
\label{fig:Tibbs_core_SED}
\end{center}
\end{figure}

Using CARMA data at 31\,GHz, \cite{Tibbs2015} performed the first search for AME in a sample of dense Galactic cold cores at 2\,arcmin angular resolution, finding less AME than expected. The nominal predictions from spinning dust models predicted detectable emission, providing constraints on the size distribution of dust grains and environmental conditions. Fig.~\ref{fig:Tibbs_core_SED} shows the SED of one of the sources (ECC189) with the CARMA upper limit at 30\,GHz providing an upper limit on the parameter $b_{\rm C}$, the total number of carbon atoms per H nucleus, which governs the number of small ($\lesssim 1$\,nm grains). As discussed by \cite{Tibbs2016}, it is possible to explain these CARMA observations in terms of AME by assuming that the smallest dust grains in the dense cores are coagulating, which decreases the expected level of AME. These observations were the first time that spinning dust modelling had been used to constrain the physical properties of interstellar dust grains such as the abundance of small grains \citep{Tibbs2016}. A similar attempt was made by comparing high-resolution (1--2.4\,arcmin) {\it Parkes} radio data with IR data \citep{Battistelli2015} where astrophysical information was extracted from AME fits of two components, allowing both concentrated and diffuse AME to be distinguished.

A survey of bright Galactic clouds in the {\it Planck} data has detected a large number of potential candidates  on $1^{\circ}$ scales \mbox{\citep{PIP_XV}}. Out of 98 targets, 42 were found to potentially have excess emission at frequencies $\approx 20$--40\,GHz, but the difficulty of detection due to overlapping sources, estimating source flux densities in the presence of bright backgrounds and optically thick sources means that some of these are likely to be false AME detections. Nevertheless, these sources tended to have similar properties (such as emissivity, peak frequency, or correlations with other datasets) to known AME sources. The most significant sources tended to be at high latitudes in regions of low free-free emission, often associated with dark clouds. Higher resolution, multi-frequency follow-up observations are needed to confirm and investigate these sources in more detail. A follow-up of some of these sources could be provided at 11, 13, 17, and 19\,GHz by the Multi-Frequency Instrument (MFI) of the QUIJOTE experiment \citep[see][for an update on the status of the project]{Rubino-Martin2017}. A study of the characterisation of AME toward the Taurus Molecular Cloud Complex is currently under investigation (Poidevin et al., in prep.) with $1^{\circ}$ FWHM resolution smoothed maps obtained at these frequencies.

As discussed in Sect.\,\ref{sec:obs_intensity}, reliable physical emissivities relating the intensity to the gas column density (e.g., in units Jy/sr cm$^{2}$) are difficult to accurately compare. Nevertheless, a trend has been noticed by several authors that appears to corroborate the spinning dust explanation. \cite{Vidal2011} compared the emissivities of several clouds observed with the CBI at an angular resolution of $\approx 4$--6\,arcmin (to reduce bias in the estimation of the average column density) and found an anti-correlation with the gas column density. A similar trend with $\tau_{250}$ (proxy for column density) was noticed by \cite{PIP_XV} and also with dust radiance by \cite{Hensley2016}. 

\begin{figure}
\begin{center}
\includegraphics[width=0.45\textwidth,angle=0]{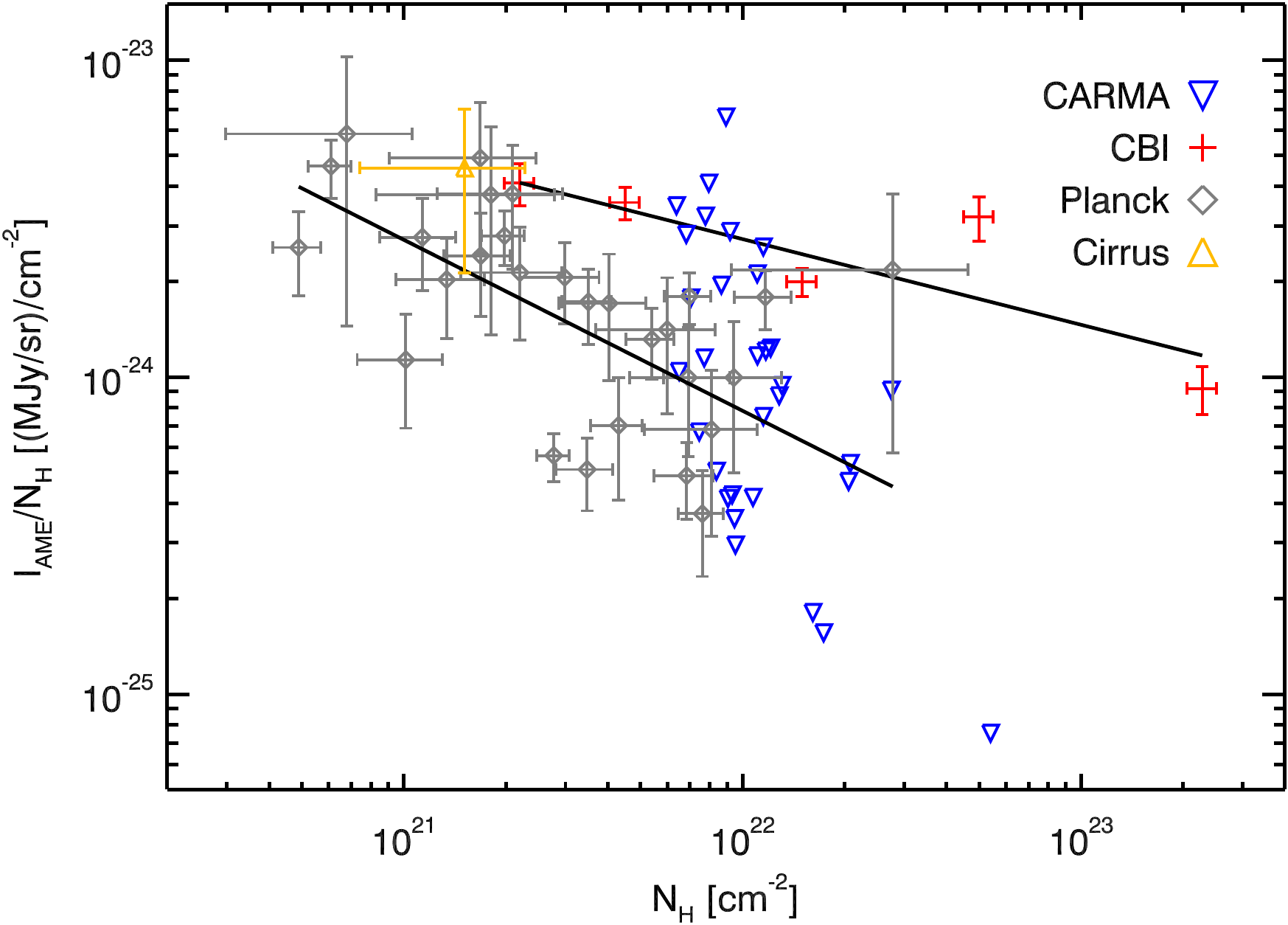}
\caption{AME emissivity at $\approx 30$\,GHz as function of column density for different Galactic objects. The {\it red} points represent observations of Galactic clouds with the CBI \protect\citep[see][]{Vidal2011}, the {\it grey} points are clouds measured with the {\it Planck} satellite \protect\citep{PIP_XV} and the {\it blue} points are 2$\sigma$ upper limits on AME emission from a sample of Galactic cold cores observed by CARMA \protect\citep{Tibbs2016}. The {\it yellow} point is an average value for cirrus clouds taken from the \protect\cite{Leitch1997} observations of AME at high Galactic latitudes.The lines represent the best-fitting power-laws to the {\it Planck} ($\alpha=-0.54\pm0.07$) and CBI ($\alpha=-0.27\pm0.04$) data, respectively. We did not attempt a fit to the CARMA points as these represent $2\sigma$ upper limits. The AME intensity is calculated as the mean intensity over the aperture for the integrated fluxes of the {\it Planck} and CARMA sources, while the CBI intensities correspond to the peak value of each source. The column densities for each point are estimated using thermal dust opacities and temperature fits. Systematic errors in this calculation can account for some of the scatter of the points.}
\label{fig:emm}
\end{center}
\end{figure}

In Fig.~\ref{fig:emm} we plot the AME emissivities for a range of objects, including the AME detections of \cite{PIP_XV} and upper limits from cold cores by \cite{Tibbs2015}. Note that the different colour points have been measured over different angular scales. For instance, the {\it Planck} estimates of $N_{\rm H}$ are mean values over a 2$^{\circ}$ diameter aperture, while the CBI points are peak values in a 4~arcmin beam. This produces systematic differences for the measured column density, with the {\it Planck} values being typically smaller than the CARMA and CBI ones, even though one might expect them to be in the same range or larger. So we caution that the $N_{\rm H}$ values on Fig.~\ref{fig:emm} should be strictly read as relative values, only comparable for points measured with the same instrument, and not as the true (average) column density for the cloud studied. Regardless of this, there is a systematic trend that is clearly visible, where for higher column density there is less AME emissivity. However, there is considerable scatter in the data. For example, fitting only the CBI data points gives\footnote{In the paper by \protect\cite{Vidal2011}, the best-fitting line is correct but the quoted value ($\alpha=0.54\pm0.1$) is incorrect; it should have been $-0.27\pm0.04$.} $\alpha=-0.27\pm0.04$, but when removing the $\rho$~Oph data point with higher emissivity (red data point at $N_{\rm H}=5\times10^{22}$) the slopes is $-0.34\pm0.05$. The steeper trend ($\alpha=-0.54\pm0.07$) found in the {\it Planck} data may be due to systematic errors either in the photometry or in the evaluation of $N_{\rm H}$. Taking this trend to be real, within the spinning dust framework, the observed behaviour can be explained by a change in the dust size distribution with column density. In denser clouds the smallest grains tend to aggregate into larger ones thus reducing the number of small grains available to produce spinning dust emission \citep[e.g.,][]{Tibbs2016}. Indeed, \cite{Draine1999} discussed this test as a way to discriminate between rotational emission from spinning dust and bulk magnetic dipole radiation from magnetic dust grains.

%%%%%%%%%%%%%%%%%%%%%%%%%%%%%%%%%%%%%%%%%%%%%%%%%%%%%%%

\begin{figure}
\begin{center}
\includegraphics[width=0.45\textwidth,angle=0]{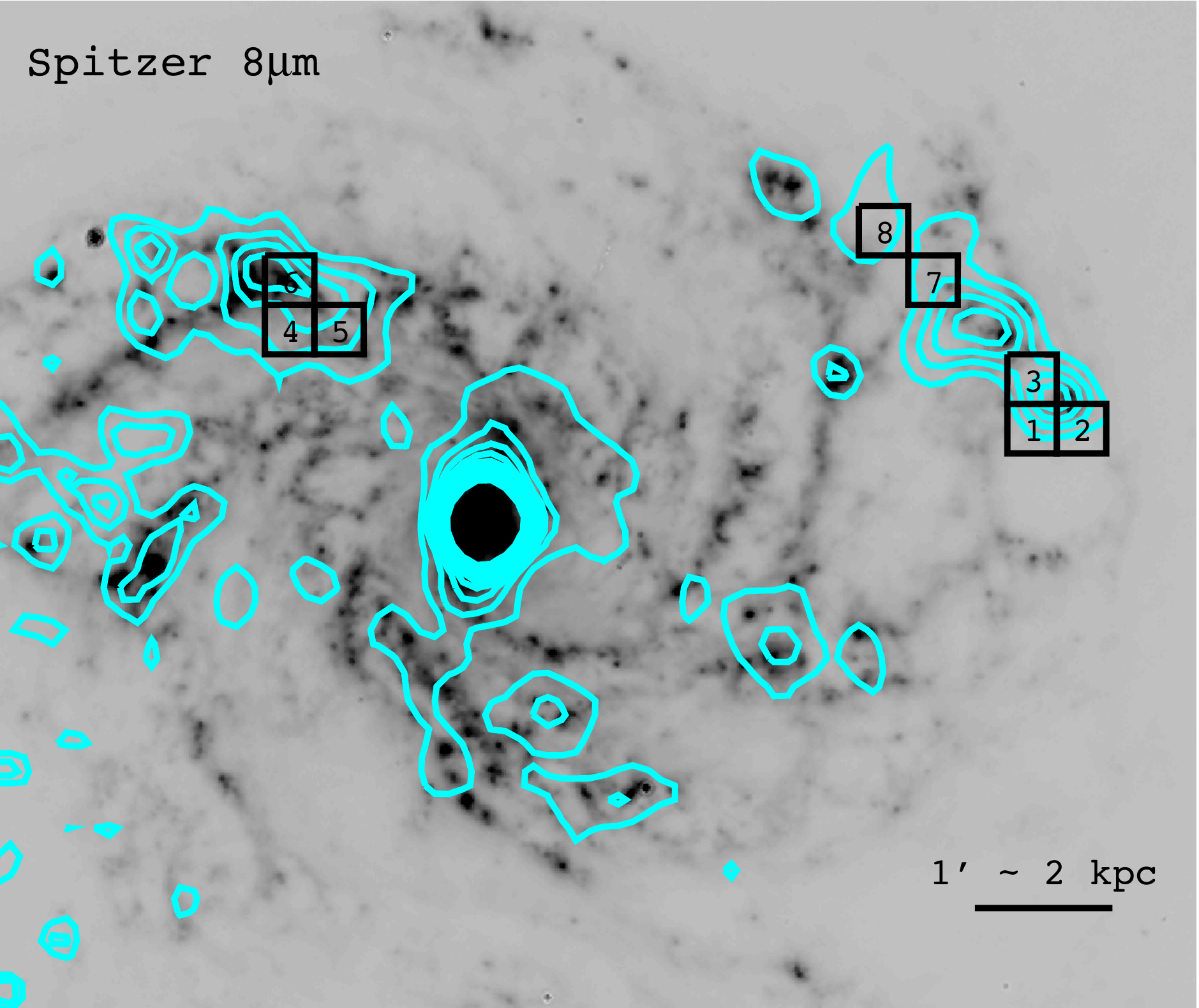}
\caption{From \protect\cite{Hensley2015}:  33\,GHz emission contours on a {\it Spitzer}/IRAC 8\,$\mu$m image of the face-on spiral galaxy NGC\,6946, where the ten contour levels are linearly spaced between the $3\sigma$ noise level of 0.048\,MJy/sr and 0.48\,MJy/sr. The eight regions with significant AME detections are shown in black boxes.  Boxes 1, 2, and 3 cover the location of Enuc\,4 in \protect\cite{Murphy2010}, where a very significant ($7\,\sigma$) AME detection was found.  
\label{fig:bms15}}
\end{center}
\end{figure}

\subsection{Extragalactic AME} 
\label{sec:obs_extragalactic}

The majority of AME detections have come from the interstellar matter and clouds in our Galaxy. Perhaps surprisingly, only a few detections from sources beyond our Galaxy have been made so far. This is partly due to the lack of high precision and high frequency ($\gtrsim 15$\,GHz) radio data with the required angular resolution. For example, WMAP and {\it Planck} data have very poor sensitivity to compact ($\lesssim 1$\,arcmin) sources. \cite{Peel2011a} studied the integrated spectrum of three nearby bright galaxies that were detected at all WMAP/{\it Planck} frequency channels: M82, NGC253 and NGC4945. Almost all
other galaxies are significantly weaker at these wavelengths causing flux density measurements between 10 and 300\,GHz to be mostly lacking, so that no conclusion on a possible excess can be drawn at this time. They found that the bulk of the emission can be explained by synchrotron, free-free, and thermal dust emissions. No significant contribution of AME was required with upper limits $\approx 1\,\%$ at 30\,GHz. These limits are marginally consistent with expectations based on the AME emissivity in our own Galaxy. The brightest (ultra)-luminous infrared galaxies Arp220, Mrk231, NGC3690, and NGC6240 do have available data; although they are an order of magnitude weaker, their entire radio to far-infrared spectrum is relatively well-sampled and well-determined. None of these four galaxies shows any indication of AME (Israel, in prep.).

\cite{Murphy2010} observed 10 star-forming regions in the nearby galaxy NGC6946 at Ka-band (33\,GHz) with the GBT. They found evidence for several regions having a marginal excess at 33\,GHz, with one region (Enuc\,4) being very significant ($7\,\sigma$). This region appears to emit $\approx 50\,\%$ AME at 33\,GHz, which was confirmed by follow-up observations with the AMI telescope at 15--18\,GHz \citep{Scaife2010a}. A spinning dust model for the spectrum was preferred over the free-free-only model. For the other star forming regions, the 33\,GHz flux appears to be dominated by free-free emission. 

Follow-up observations of NGC\,6946 with CARMA, at yet higher angular resolution (21\,arcsec), revealed additional regions having excess 33\,GHz emission attributed to AME \citep[][see Fig.\,\ref{fig:bms15}]{Hensley2015}.  The strength of the excess emission in these regions was found to be correlated with the total Infrared (IR; 8--$1000\,\mu$m) luminosity, lending credence to the interpretation as AME. However, no correlation with the dust mass fraction in PAHs was observed. These conclusions are consistent with later results indicating that the dust luminosity, rather than mass in PAHs, correlates more reliably with AME strength. Further, these results suggest that inferences from Galactic AME may extrapolate to extragalactic systems. In particular, measurements of IR luminosity lead to direct predictions of 30\,GHz AME in external galaxies, providing an effective means of target selection for follow-up studies.

Excess emission has been detected in the Magellanic clouds. Detections of excess in the Large Magellanic Cloud (LMC) claimed by \cite{Israel2010} and \cite{Bot2010} may be largely or completely caused by a CMB (hotspot) fluctuation \mbox{\citep{PEP_XVII}}. \mbox{\cite{Planck2015_XXV}} show that the integrated spectrum of the LMC shows a marginal (few $\%$) excess at $\approx 30$\,GHz with an emissivity comparable to the Milky Way. On the other hand, \mbox{\cite{PEP_XVII}} did find a significant excess at sub-mm wavelengths ($\sim100$--500\,GHz) in the SMC, which extends down to $\approx 30$\,GHz \citep{Israel2010,Bot2010,PEP_XVII}. \cite{Bot2010} suggested that the 50--300\,GHz excess in the SMC could be explained by spinning dust, but \cite{Draine2012} argued that this was inconsistent with physical conditions in the SMC, and that magnetic dipole radiation from magnetic nanoparticles (either inclusions or free-flying) could provide a more natural explanation. 

\mbox{\cite{PIP_XXV}} found a marginal detection of AME from the integrated spectrum of the Andromeda galaxy (M31), at an amplitude of $0.7\pm0.3$\,Jy ($2.3\,\sigma$). The amplitude is in line with expectations from the emissivity found in our Galaxy. M31 is also the subject of high resolution microwave observations performed with the recently commissioned 64-m Sardinia Radio Telescope \citep[SRT;][]{Prandoni2017}; detailed C-band (5.7--7.7\,GHz) intensity, spectroscopic, and polarization observations have been conducted over $\approx 7$\,deg$^2$ around M31 (Battistelli \& Fatigoni, in prep.) and K-band (18--26.5\,GHz) observations are expected in 2019.

The Local group dwarf-spiral galaxy M33 has a well-defined radio to near-infrared spectrum. An in-depth analysis of its flux densities and overall spectrum (Tibbs et al., in prep.) concludes that any AME contribution at 30\,GHz is less
than $\approx 10\%$ of the total flux density at that frequency, and almost an order of magnitude below the AME contribution expected by scaling the Milky Way results. For a few other dwarf galaxies of Magellanic type, spectra extend into
the centimetre range thanks to their compactness. The spectra of Hen\,2-10 and NGC4194 are still too poorly sampled to allow a conclusion, but the spectra of 2\,Zw\,40 and NGC5253, although needing further
accurate flux density measurements in the 40--100\,GHz range, at least do not yet rule out the presence of AME (Israel, in prep.).

%%%%%%%%%%%%%%%%%%%%%%%%%%%%%%%%%%%%%%%%%%%%%%%%%%%%%%%

\subsection{AME observations in polarization} 
\label{sec:obs_polarization}

Extensive effort has been dedicated to the theoretical modelling of the polarization spectrum of spinning and magnetic dust emission (see Sect.\,\ref{sec:theory} and references therein). Different physical conditions, including magnetic field strength, grain temperature, grain shape, size distribution, and alignment between the grain angular momentum and magnetic field direction, lead to different polarization levels and spectra, or even practically null polarization in cases that the quantization of energy levels produces a dramatic decrease of the alignment efficiency, as predicted by \cite{Draine2016}. For this reason, a measurement of the polarization of AME, or the determination of stringent upper limits, is potentially a key tool for discriminating between different models, and for obtaining information about the physical parameters associated with AME environments. Note that magnetic dust models predict much higher polarization fractions ($\gtrsim 10\,\%$) than those based on spinning dust ($\lesssim 1\,\%$). Thus, polarization can in principle be crucial to determine the relative contributions of these two mechanisms, something that is complicated in intensity data due to the presence of multiple components.

Polarization is of course more difficult to observe, because the polarization signal is very weak, and also because AME must be separated in intensity to measure a polarization fraction for the AME component i.e., $\Pi_{\rm AME}=P_{\rm AME}/I_{\rm AME} \ne P_{\rm AME}/I_{\rm total}$. For this reason, observational studies of AME polarization are scarce, and so far they have only led to upper limits. Table\,\ref{tab:pol}, which is an update of table~1 of \cite{Rubino-Martin2012a}, summarises the current constraints. Except for the value from \cite{Casassus2008}, here we list constraints on the fractional polarization relative to the AME flux density in intensity (rather than total flux density), which is derived by modelling and subtracting the other components. Note that in \cite{Battistelli2006} and in \cite{Lopez-Caraballo2011} they referenced their polarization fractions relative to the total intensity, while here we quote the values relative to the AME intensity that were calculated by \cite{Genova-Santos2015b}.

Better and more reliable limits on AME polarization come from specific Galactic clouds, which have a bright AME signal without significant contamination from other emission mechanisms. The best targets are the Perseus (specifically the dust shell G159.6--18.5) and $\rho$~Oph molecular clouds, where AME clearly dominates the intensity spectrum at frequencies $\approx 10$--50\,GHz. Another advantage is that the emission at low frequencies is dominated by optically thin free-free emission, which is much lower than AME at frequencies near 30\,GHz and is unpolarized. Also, these clouds are relatively well isolated and away from the Galactic plane, therefore avoiding contamination from diffuse Galactic emission or from nearby compact objects. 

\cite{Lopez-Caraballo2011} and \cite{Dickinson2011} obtained $\lesssim 1$\,\% limits on these regions at 23\,GHz using WMAP observations. Previously \cite{Battistelli2006} reported in G159.6--18.5 a marginal detection of $\Pi=3.4^{+1.5}_{-1.9}\%$ (or $\Pi_{\rm AME}=5.3^{+2.5}_{-3.1}\%$) using data from the COSMOSOMAS experiment at 11\,GHz, which has not yet been confirmed nor ruled out. \cite{Genova-Santos2015b} obtained upper limits in the same frequency range using QUIJOTE data, but they have a larger uncertainty. \cite{Reich2009} proposed that G159.6--18.5 may be acting as a Faraday Screen, which rotates the polarization angle of background radiation, and this could be potentially contributing to this 11\,GHz signal. The other reported detection by \cite{Battistelli2015} towards the H{\sc ii} region RCW175, at a level of $2.2\pm0.4\,\%$. However, they claim that there could be a significant contribution to the measured polarization from synchrotron emission along the line-of-sight, making this effectively an upper limit of $<2.6\,\%$. Other upper limits come from well-studied regions in total intensity, like the Lynds dark cloud LDN1622. Although they are more difficult to obtain due to the weakness of the signal, some upper limits associated with the diffuse AME emission have also been derived, using either the full sky \citep{Kogut2007,Macellari2011} or extended regions \citep{Planck2015_XXV}. Note that \cite{Planck2015_XXV} obtained a detection in the Pegasus region ($\Pi=9\pm2\,\%$), but suggested that it was likely contaminated by polarized synchrotron emission. 

More stringent upper limits on the fractional polarization require more sensitive data, or alternatively AME regions that are brighter in total intensity. The sample of 98 AME sources analysed in \citet{PIP_XXV} provides a guide to search for good candidates. One of them is the molecular complex W43, which has an AME peak flux eight times higher than G159.6--18.5, although in this case the level of free-free emission is as high as the AME in the relevant frequency range. Using a multi-frequency analysis combining QUIJOTE, WMAP and {\it Planck}, \cite{Genova-Santos2017} have provided the best upper limits obtained so far: $<0.22$\,\% at 41\,GHz, from WMAP data, and $<0.39$\,\% at 17\,GHz from QUIJOTE data. \cite{Genova-Santos2017} claim that the free-free level has been determined with an accuracy of around $20$\,\% thanks to the use of low-frequency continuum and radio-recombination line data. 

\begin{figure*}
\begin{center}
\includegraphics[width=1.0\textwidth,angle=0]{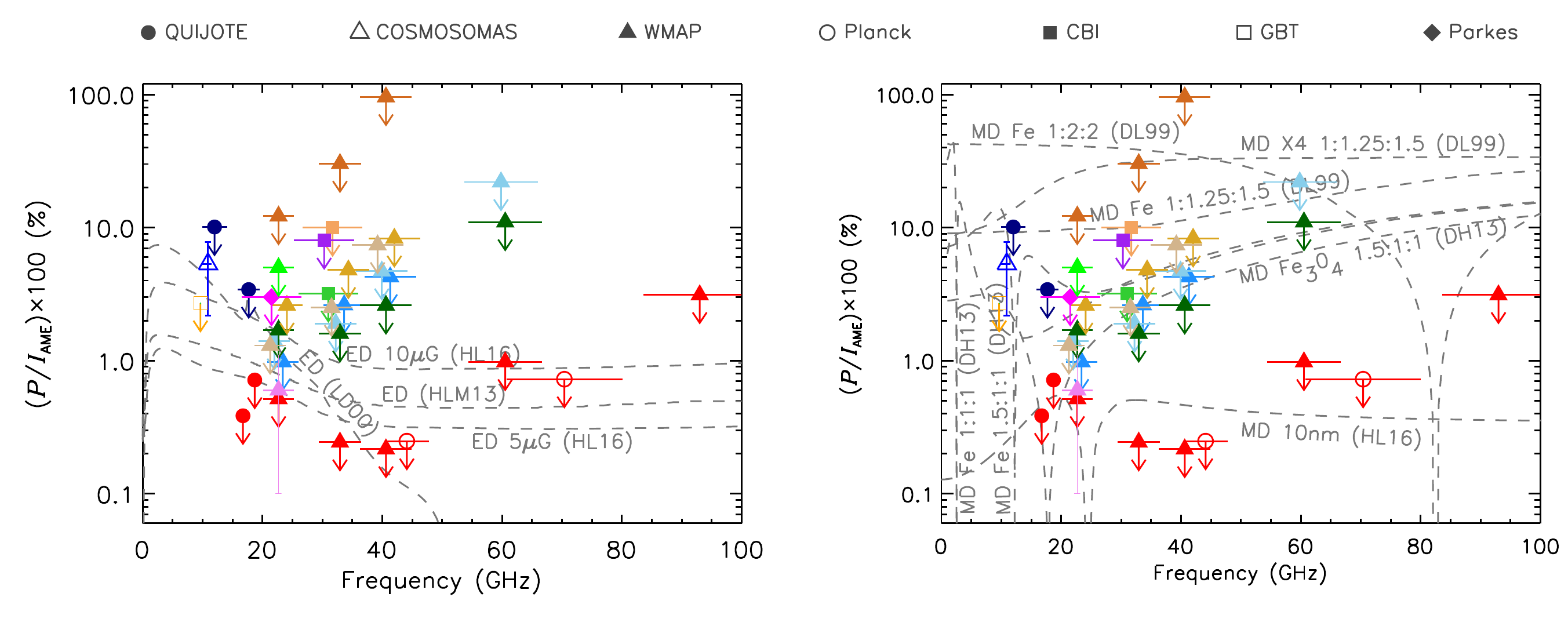}
\caption{Constraints on AME polarization fraction at different frequencies, from different experiments and in different regions, compared with the predictions from different models ({\it grey dashed lines}) based on electric dipole (ED; {\it left} panel) and magnetic dipole (MD; {\it right} panel) emissions. The horizontal lines around each data point represent the bandwidth of the corresponding detector. We have used different colours for each object and reference (see Table\,\ref{tab:pol}), and different symbols for each experiment, as indicated in the legend on top. In some cases, where there are many points at the same frequency (as for WMAP and CBI), for the sake of clarity, we have shifted the central frequencies in the plot. If the quantum suppression mechanism described by \protect\cite{Draine2016} is in effect then the polarization fraction would be $\lesssim 0.0001\,\%$.}
\label{fig:pol}
\end{center}
\end{figure*}

The constraints summarized in Table\,\ref{tab:pol} are compared in Fig.\,\ref{fig:pol} with various theoretical predictions for the polarization fraction of the spinning dust \mbox{\citep{Lazarian2000,Hoang2013,Hoang2016b}} and magnetic dust emissions \citep{Draine1999,Draine2013,Hoang2016b}. Fig.\,\ref{fig:pol} follows the same organization as figure~8 of \mbox{\cite{Genova-Santos2015b}} and of \cite{Genova-Santos2017}, and depicts the same theoretical models (in the previous references predictions for different physical conditions of the AME environment than those adopted in Fig.\,\ref{fig:pol} are also provided). 

It can be seen that many of the upper limits fall well below early predictions of polarization for the spinning dust models. However, it must be taken into account that the models of \cite{Lazarian2000} and \cite{Hoang2013} give, in practice, upper limits on the real spinning dust polarization due to various depolarization effects \citep[see e.g.,][]{Genova-Santos2015b}. Most of the previous constraints have been obtained at angular resolutions of $\sim 1^{\circ}$, so a detection of polarization would require coherence of the magnetic field over these angular scales. Similarly, a decrease of the observed polarization could be produced by the combination of different emission mechanisms along the same line-of-sight with different polarization directions. Furthermore, the observations could be affected by instrumental systematic effects, or artificially lower by depolarization, such as multiple components within a single beam (``beam depolarization"). 

\cite{Draine2016} concluded that quantization of energy levels in very small grains effectively suppresses alignment in nanoparticles that are small enough to spin at $\approx 20$--40\,GHz, leading to negligible polarization levels. In that case, only grains $>30$\,\AA~could produce polarized emission at a level comparable to the best upper limits represented in Fig.\,\ref{fig:pol}. Models predict the magnetic dust emission to have even higher polarization degrees, and therefore those models are also inconsistent with the measurements. However, several assumptions have been adopted while deriving some of these theoretical predictions, in particular: perfect alignment between the grain angular momentum and the magnetic field, magnetic field parallel to the plane of the sky, or dust grains ordered in a single magnetic domain \mbox{\citep{Draine1999}}. If any of these hypotheses does not hold the real polarization degree may be lower. 
For certain extreme models -- e.g., perfectly aligned grains consisting of a single ferromagnetic domain -- the magnetic dipole thermal emission can be strongly polarized \citep{Draine2013}, but these extreme assumptions seem unlikely to be realized even if some or all of the AME is thermal emission from ferromagnetic materials.  Thermal emission from silicate grains containing randomly-oriented magnetic inclusions is predicted to have only a small degree of polarization in the 20--40\,GHz region \citep[see fig.~10 of][]{Draine2013}.

Given the abundance of free parameters (mainly related with the geometry of the grains and magnetic field, and with the physical conditions of the environment) and possible unaccounted effects (for instance, quantum suppression of the dissipation required to produce alignment) in the models, there may be multiple models consistent with any upper limit derived from the observations. However, \mbox{\cite{Draine2016}} have made the strong prediction that spinning dust emission should be unpolarized, so if AME polarization is detected, this would either require a non-spinning dust origin, or invalidate the physical arguments for quantum suppression of dissipation. For this reason, it is important to increase the angular resolution of the observations, to improve the quality of the data, and the analysis techniques, in order to try to eventually reach a detection of the polarization of AME at different frequencies. This would be important not only for constraining spinning/magnetic dust models, but also for characterising foregrounds for future high-sensitivity measurements of the CMB.

\setlength{\tabcolsep}{3pt}
\begin{table*}
\caption{Summary of AME polarization fraction constraints, $(P/I_{\rm AME})\times 100$. The horizontal line separates measurements on individual non-resolved Galactic objects and on large regions of the sky. Columns 1 to 3 indicate the name of the object or region, the experiment from which the data were taken and its angular resolution, respectively. The next four columns show the constraints in different frequency ranges. Here, upper limits are referred to the 95\% confidence level. The last column provides the reference. 
This is an updated version of table~1 of \protect\cite{Rubino-Martin2012a}.}
\begin{tabular}{lccccccl}
\hline
\noalign{\smallskip}
Target 	       &Experiment		  &Angular        &\multicolumn{4}{c}{Polarization fraction $\Pi~[\%]$}				       &Reference  \\
	       &	             &Res. [arcmin]  &9--19\,GHz	    &20--23\,GHz     &30--33\,GHz       &41\,GHz 	       & \\ 
\noalign{\smallskip}
\hline 
\noalign{\smallskip}
 & \multicolumn{7}{c}{Galactic AME regions}\\
\noalign{\smallskip}
G159.6--18.5            &COSMOSOMAS		  &$60$	       &$5.3^{+2.5}_{-3.1}$  &\ldots  	       &\ldots            &\ldots         &\cite{Battistelli2006} \\      
---"---	                 &WMAP-7  	          &$60$	       &\ldots  	     &$<1.0$  	       &$<2.6$		  &$<4.2$	  &\cite{Lopez-Caraballo2011} \\
---"---	                 &WMAP-7  	          &$60$	       &\ldots  	     &$<1.4$  	       &$<1.9$		  &$<4.7$	  &\cite{Dickinson2011} \\
---"---	                 &QUIJOTE                 &$\approx 60$&$<3.4$  	     &\ldots 	       &\ldots  	  &\ldots 	  &\cite{Genova-Santos2015b} \\
$\rho$~Oph~W          &CBI	                	&$\approx 9$ 	&\ldots  	     &\ldots  	       &$<3.2$  	  &\ldots         &\cite{Casassus2008} \\
---"---                         &WMAP-7  		  &$60$	       &\ldots  	     &$<1.7$	       &$<1.6$		  &$<2.6$	  &\cite{Dickinson2011} \\
LDN1622	                 &GBT		          &$1.3$       &$<2.7$  	     &\ldots  	       &\ldots  	  &\ldots         &\cite{Mason2009} \\
---"---	                 &WMAP-7  	          &$60$	       &\ldots  	     &$<2.6$  	       &$<4.8$  	  &$<8.3$         &\cite{Rubino-Martin2012a} \\
RCW175	                 &{\it Parkes} 64-m	&$\approx 1$ &\ldots  	     &$<2.6$     &\ldots  	  &\ldots         &\cite{Battistelli2015} \\
W43	                 &QUIJOTE/WMAP            &$\approx 60$&$<0.39$  	     &$<0.52$ 	       &$<0.24$ 	  &$<0.22$ 	  &\cite{Genova-Santos2017} \\
Pleiades                 	&WMAP-7  		  &$60$	       &\ldots  	     &$<12$ 	       &$<32$             &$<96$	  &\cite{Genova-Santos2011} \\
\big[LPH96\big]201.663+1.643	 &CBI	&$\approx 7$ &\ldots  	     &\ldots  	       &$<10$		  &\ldots         &\cite{Dickinson2006} \\
---"---	                 &WMAP-7  	          &$60$	       &\ldots  	     &$<1.3$  	       &$<2.5$  	  &$<7.4$         &\cite{Rubino-Martin2012a} \\
Helix nebula             &CBI		          &$\approx 9$ &\ldots  	     &\ldots  	       &$<8$		  &\ldots         &\cite{Casassus2007} \\
\noalign{\smallskip}
\hline       
\noalign{\smallskip}
& \multicolumn{7}{c}{Diffuse AME}\\
\noalign{\smallskip}
All-sky	                 &WMAP-3  	          &60	       &\ldots  	     &$<1$	       &$<1$		  &$<1$	          &\cite{Kogut2007} \\
All-sky	                 &WMAP-5  	          &60	       &\ldots  	     &$<5$	       &\ldots  	  &\ldots         &\cite{Macellari2011} \\ 		
Perseus	                 &WMAP-9/{\it Planck}	  &$60$	       &\ldots  	     &$0.6\pm 0.5$     &\ldots  	  &\ldots         &\cite{Planck2015_XXV} \\
\noalign{\smallskip}
\hline
\end{tabular}
\label{tab:pol}
\end{table*}
\setlength{\tabcolsep}{6pt}

%%%%%%%%%%%%%%%%%%%%%%%%%%%%%%%%%%%%%%%%%%%%%%%%%%%%%%%

\subsection{Summary of Observational Constraints and Discussion}
\label{sec:obs_summary}

The first topic of discussion is simply, is AME real? i.e., is AME really a new component of emission such as spinning or magnetic dust, or is it our lack of understanding of synchrotron/free-free/thermal dust/CMB emissions?

For several years after the initial detection \mbox{\citep{Leitch1997}}, there was much debate whether AME was in fact real. Indeed in the first WMAP data release, the WMAP team discussed foregrounds extensively \citep{Bennett2003b}. The possibility of spinning/magnetic dust was considered, yet the discussion largely rejected such alternatives. Instead, they believed that a harder (flatter spectrum) component of synchrotron emission, which was not traced by low-frequency radio surveys, was the most likely explanation for the excess. The correlation with dust was explained by the fact that hard synchrotron would be correlated with star-formation due to energy injection, which would be close to molecular clouds, and therefore dust emission. This model would also explain the flatter spectrum ($\beta \approx -2.2$) at frequencies $\approx 15$\,GHz but steepening to $\approx -3.0$ at 30\,GHz and above. At high Galactic latitudes, there appears to be no evidence for a significantly flatter component of synchrotron emission \citep[e.g.,][]{Miville-Deschenes2008,Vidal2015}, except possibly from the Galactic centre region \citep{PIP_IX} and at low Galactic latitude \citep{Fuskeland2014}; a template-fitting analysis of WMAP data using the 2.3\,GHz southern sky survey \citep{Peel2012} showed that the AME was robust to using different radio templates. 

In the early AME detections, there was concern that free-free emission could be responsible. The optical recombination line H$\alpha$ is a tracer of warm ionized gas and therefore radio free-free emission, with little dependence on electron temperature for the temperatures $3000 < T_e < 15000$\,K expected for photoionized gas \citep{Dickinson2003}. Earlier observations of H$\alpha$ had already indicated that the free-free emission could only account for a small fraction of the total foreground signal \citep{Gaustad1996,Leitch1997}. Full-sky H$\alpha$ maps \citep{Dickinson2003,Finkbeiner2003} became available and quickly ruled out traditional free-free emission \mbox{\citep[e.g.,][]{Banday2003,Davies2006}}. Optically thick free-free emission could be contributing to the AME signal along some sight-lines, but, in general, is not a major contributor at frequencies $\approx 30$\,GHz and on scales of $\approx 1^{\circ}$ \citep[e.g.,][]{Planck2015_XXV}.

The major break-through came with the detection of AME from Galactic molecular clouds - the Perseus G159.6--18.5 region {\citep{Watson2005} and the $\rho$~Oph~W \citep{Casassus2008}. Both these sources contained strong dust-correlated emission that was clearly well in excess of the expected synchrotron and free-free intensity levels. There was still some doubt about the extrapolation of the thermal dust tail from higher frequencies. However, the early results paper from {\it Planck} \mbox{\citep{PEP_XX}} showed the peaked spectrum expected from spinning dust (on both sides), providing definitive evidence for the reality of AME, and consistent with predictions for spinning dust (see Sect.\,\ref{sec:theory_spinningdust}).

The large body of evidence for excess emission at frequencies $\approx 10$--60\,GHz is now difficult to refute. The recent {\it Planck} component separation analysis included spinning dust, which used a spectral parametric fitting code to fit each pixel seperately \citep{Planck2015_IX}. They found a strong component that was correlated with dust even though this was not assumed in the analysis. Fig.\,\ref{fig:planck_sed} presents a summary of the separation on large angular scales ($81$--93\,\% coverage). In this particular model, the AME is represented by a 2-component spinning dust model, which dominates the foreground emission near 20\,GHz. Although the details of the separation could be in doubt (e.g., due to the fixed synchrotron spectrum) it appears to be clear that AME (modelled as spinning dust in this case) is a significant fraction (of order half) of the total foreground signal at frequencies $\approx 20$--40\,GHz.

One potential issue relates to the absolute calibration of large-scale radio maps. Large-scale radio data, up to now, have often been taken with large radio dishes working as a total-power radiometer. The illumination of the dish is such that a significant amount of the power response of the telescope is outside of the main beam, resulting in a scale-dependent calibration that is difficult to account for \citep[see][for a detailed discussion]{Du2016}. For many of the large-scale radio surveys, this has resulted in a calibration scale that can be incorrect by up to tens of per cent when comparing compact sources with large-scale emission over many degrees \citep[e.g.,][]{Jonas1998}. The end result is that large-scale emission on scales of $1^{\circ}$ and larger will be enhanced relative to compact sources (which are typically used for calibration). A simple extrapolation from low to high frequencies could then yield excess emission at the higher frequencies by a similar fraction i.e. a few tens of \%, which is indeed what is observed. Given this, it is difficult to account for the ubiquity of AME, even when not relying on low frequency radio maps. A number of studies have relied on such data while others have not. Furthermore, some AME sources show effectively no signal at all at low frequencies, such as $\rho$~Oph~W.

The origin of the diffuse AME found at high Galactic latitudes is still not completely settled. Although spinning dust can readily account for the bulk of the AME \mbox{\citep[e.g.,][]{Planck2015_X}}, other emission mechanisms could be contributing at some level (Sect.\,\ref{sec:theory_other}). Magnetic dipole radiation from fluctuations in dust grain magnetization could be significant \citep{Draine1999,Draine2013,Hensley2016}, although upper limits on AME polarization \citep{Lopez-Caraballo2011,Dickinson2011,Macellari2011,Rubino-Martin2012a} appear to indicate that this does not account for the bulk of the signal. Similarly, a harder (flatter spectrum) component of synchrotron radiation may also be responsible for part of the AME at high latitudes, as proposed by \mbox{\cite{Bennett2003b}} in the original {\it WMAP} data release. The harder spectrum would naturally explain the correlation with dust, since both are related to the process of star-formation. Furthermore, we already know that there are regions that have synchrotron spectral indices that are at $\beta \approx -2.5$ or flatter, both from supernova remnants \citep{Onic2013} as well as more diffuse regions such as the {\it WMAP}/{\it Planck} haze \mbox{\citep{Finkbeiner2004,PIP_IX}}.

A contribution of harder synchrotron component may have been missed when applying component separation methods to microwave data. The majority of AME detections from fluctuations at high Galactic latitudes have been made using the ``template fitting" technique, i.e., fitting multiple templates for each foreground component to CMB data, accounting for CMB fluctuations and noise \mbox{\citep{Kogut1996,Banday2003}}. The synchrotron template is traditionally the 408\,MHz all-sky map \citep{Haslam1982}, or other low frequency template. However, data at these frequencies will naturally be sensitive to the softer (steeper spectrum) synchrotron emission, which has a temperature spectral index $\beta \approx -3.0$ \citep{Davies2006,Kogut2007,Dunkley2009b,Gold2011}. This leads to a significant AME signal at $\approx$10--60\,GHz that is correlated with FIR templates, which cannot be accounted for by the R-J tail of dust emission. \cite{Peel2012} used the 2.3\,GHz Southern-sky survey of \cite{Jonas1998} as a synchrotron template for the {\it WMAP} data and found that the dust-correlated AME component changed by only $\approx 7$\,\%, compared to using the 408\,MHz template. This suggests that the bulk of the diffuse high latitude synchrotron emission is indeed steep ($\beta \approx -3.0$), resulting in little change to the AME at 20--40\,GHz. New data from C-BASS at 5\,GHz (Dickinson et al., in prep.) have shown that synchrotron emission in the NCP region is not a major contributor to AME at 20--40\,GHz and that the synchrotron emission follows a power-law with $\beta=-2.9$, from 408\,MHz all the way to tens of GHz. If this is true for the entire sky, then synchrotron emission cannot explain the bulk of AME at high latitudes.

Recent works have also cast doubt on the spinning dust interpretation. A number of works have found that the small-scale morphology within AME regions does not always correlate with tracers of PAHs and/or the smallest grains \citep[e.g., ][]{Tibbs2011,Tibbs2012,Tibbs2013} and, in some cases, it even tends to correlate better with far-IR/sub-mm wavelengths \citep[e.g., ][]{Planck2015_XXV}. More striking is that spatial variations in the PAH abundance appear to have no correlation with variations in the ratio of AME/${\cal R}$ \citep{Hensley2016}, whereas a correlation was expected if AME is produced by spinning PAHs.  This may indicate that the PAHs have small dipole moments, so that AME is dominated by non-PAH nanoparticles, which could be as abundant as the PAHs, but which might have suitable dipole moments.  \cite{Hoang2016b} showed that small silicate grains could reproduce the observed AME, and \cite{Hensley2016} demonstrated that nanosilicates could reproduce the AME without violating observational constraints on the mid-IR emission spectrum of diffuse clouds.

The lack of correlation with PAH abundance is surprising, especially because variations in PAH abundance might be expected to correlate with variations in the abundance of other nanoparticles such as nanosilicates. Nevertheless, spinning dust still appears to be the most likely of the proposed emission mechanisms, particularly given the increasingly stringent upper limits on AME polarization.

\cite{Bernstein2017} have suggested that each of the unidentified infrared (UIR) bands may be due to rotational structure from one or two relatively small molecules (i.e., less than 30\,C atoms).  In contrast, the PAH paradigm attributes the UIR bands to the blending of narrow spectral profiles from many (30 or more), much larger molecules (50 or more C atoms). \cite{Bernstein2017} demonstrated that the $11.2\,\mu$m UIR band profiles from multiple sources can be plausibly modelled as a vibration-rotation band of a small fullerene, C$_{24}$.  If valid, this model offers a possible explanation for the lack of observed correlation between AME and some of the UIR features \citep[e.g.,][]{Hensley2016}.  C$_{24}$ has no dipole moment and therefore cannot produce pure rotational emission (i.e., no AME).  Furthermore, it is not clear that the abundance of C$_{24}$ should be proportional to (i.e., a tracer for) the PAH abundance.  The implication is that the observed lack of correlation does not exclude PAHs as a major source of AME.

%%%%%%%%%%%%%%%%%%%%%%%%%%%%%%%%%%%%%%%%%%%%%%%%%%%%%%%
%%%%%%%%%%%%%%%%%%%%%%%%%%%%%%%%%%%%%%%%%%%%%%%%%%%%%%%

\begin{figure}
\begin{center}
\includegraphics[width=0.45\textwidth,angle=0]{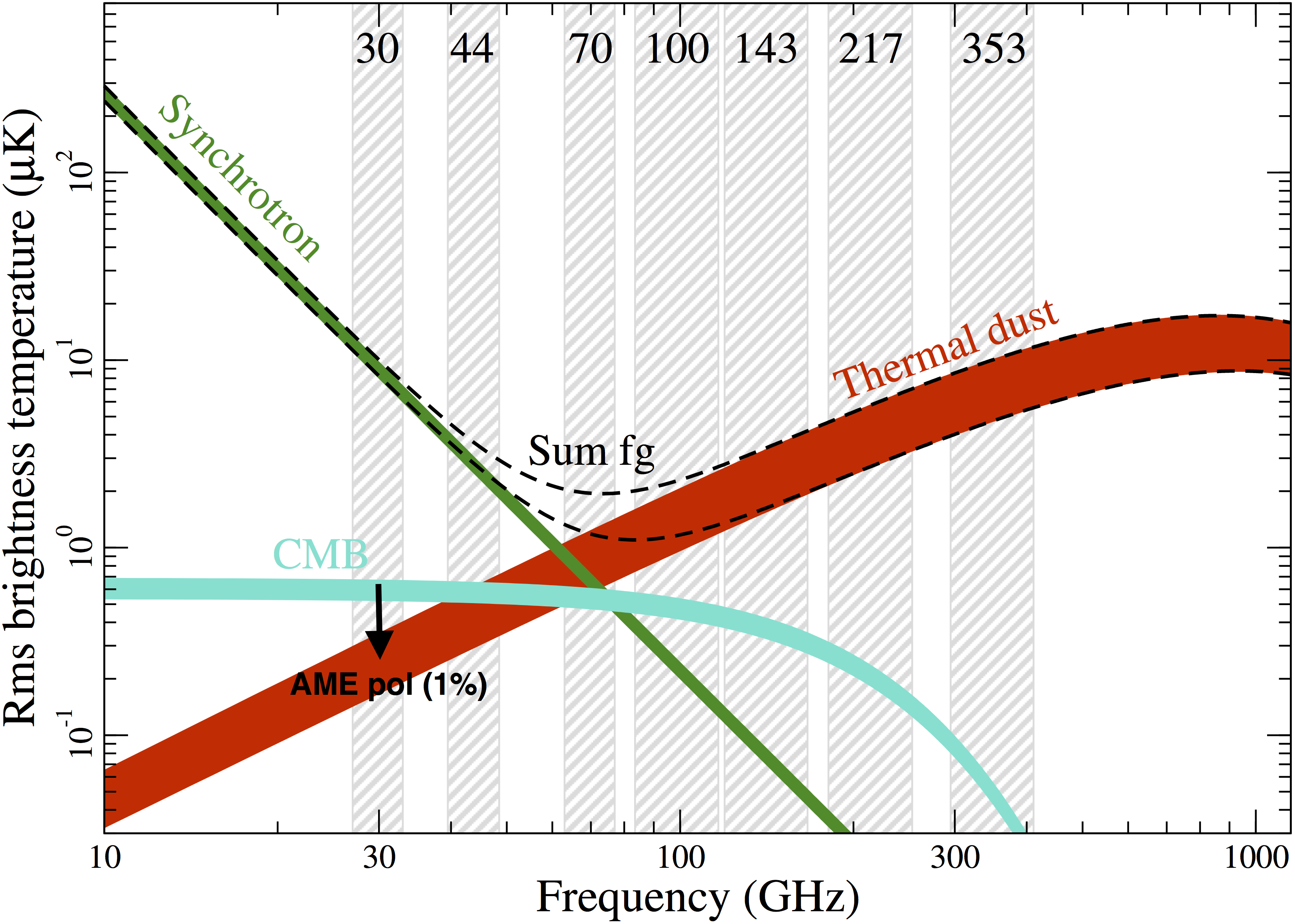}
\caption{Summary of the amplitude of polarized foregrounds from the {\it Planck} component separation 2015 results; figure taken from \protect\cite{Planck2015_X}. The brightness temperature r.m.s. against frequency, on angular scales of 40\,arcmin, is plotted for each component; this figure is the same as Fig.\,\protect\ref{fig:planck_sed} but for polarized emission. For this case, the width of the curves represents the spread when using $73\,\%$ and $93\,\%$ of the sky. The approximate maximum amplitude from AME polarization (assuming $1\%$ upper limit) at 30\,GHz is indicated by the downward arrow.}
\label{fig:planck_pol}
\end{center}
\end{figure}

\section{AME as a CMB Foreground} 
\label{sec:cmb}

AME has been demonstrated to be a significant component of the diffuse foreground emission at frequencies $\approx 10$--60\,GHz. Some analyses put AME as the dominant foreground in this frequency range, accounting for perhaps half of the total emission \mbox{\citep[e.g.,][]{Davies2006,Planck2015_X}}. It is therefore clear that AME is a major foreground for studying the CMB. Yet, CMB studies in intensity appear to not be limited by foregrounds \citep[e.g.,][]{Dunkley2009a,Bennett2013,Planck2015_X}, even though we know so little about AME. The question is how do we know that AME is not an issue?

There are two main reasons why AME does not appear to be a major problem for CMB studies. The first is that the spectrum of AME is so different from that of the CMB. AME has a steeply falling spectrum above $\approx 20$\,GHz, while the CMB is flat (in brightness temperature units). The high frequency part of the AME spectrum may be similar to that of synchrotron radiation and thus fitting for a single power-law in the WMAP data accounts for the bulk of the foreground emission \citep[e.g.,][]{Dickinson2009b,Planck2013_XII}. The second reason is that because AME is so closely correlated with far-IR data, spatial templates can be used to account for most of the foreground emission, such as those used in producing the foreground-reduced maps from WMAP \citep[e.g.,][]{Bennett2003b,Bennett2013}. Thus, both spatially and spectrally, AME can be relatively easily separated from the CMB, down to levels set by cosmic variance \citep{Planck2015_X}. Nevertheless, as CMB data become more sensitive on small scales (large scales are limited by cosmic variance), dust-correlated emission will need to be removed with higher precision. 

In polarization and on large angular scales ($\gtrsim 1^{\circ}$), the highly polarized synchrotron and thermal dust foregrounds are brighter than the CMB anisotropies over much of the sky and all frequencies. Fig.\,\ref{fig:planck_pol} depicts the average amplitude of polarized foregrounds relative to that of the total CMB polarization. The strongest polarized foregrounds are synchrotron radiation at low frequencies and thermal dust emission at high ($\gtrsim 100$\,GHz) frequencies. Fortunately, AME appears to be weakly (possibly negligibly) polarized (Sect.\,\ref{sec:theory_spinningdust} \& Sect.\,\ref{sec:obs_polarization}) and may not be a major foreground for future CMB studies; Fig.\,\ref{fig:pol} shows the latest observational constraints, which nearly all are consistent with a polarization fraction of $\lesssim 1\%$. In Fig.\,\ref{fig:planck_pol} we indicate the approximate upper limit on AME polarization assuming 1\% polarization fraction. It can be seen that AME is below the level of the CMB at 30\,GHz and at higher frequencies will likely be much lower because of the falling spinning dust spectrum. 

However, one of the major aims of current and future CMB experiments is to measure the stochastic background of gravitational waves that are a prediction of many models of inflation. This is possible with CMB polarization because these fluctuations are the only natural way to create ``B-modes" on scales of a degree and larger, with an amplitude known as the tensor-to-scalar ratio, $r$ \citep{Seljak1997,Kamionkowski1997}. Current constraints are at the level of $r<0.07$ \citep{BICEP2016} and are continuing to improve. The B-mode fluctuations at this level correspond to $\sim 0.1\,\mu$K in brightness temperature but could be much lower than this. For {\it Planck} sensitivity levels, a $1\,\%$ polarization fraction for AME has negligible impact on the recovery of the tensor-to-scalar ratio, $r$ \citep{Armitage-Caplan2012}. Nevertheless, even a small amount of AME polarization could be problematic for component separation with future ultra-high sensitivity data that aim to constrain the tensor-to-scalar ratio at the level $r \sim 10^{-3}$ or even lower \citep{Remazeilles2016,Remazeilles2017,Alonso2017}. Careful consideration of all foregrounds will be critical to achieve this level of sensitivity, including any potential spinning and magnetic dust components.

Although AME has been modelled as a single (or double) component of spinning dust, most analyses tend to reveal a residual excess at higher ($>50$\,GHz) frequencies  \citep{PIP_XXII}. This may well be a consequence of incorrect modelling, especially since a typical spectral model for spinning dust is computed for one set of parameters; an example would be a single run of the {\sc spdust}2 code, corresponding to a single component of spinning dust grains. In reality, there will be a distribution of dust grains and environments along the line-of-sight, which will inevitably broaden the spectrum compared to that of a single component. Indeed, this was the reasoning behind fitting for a 2-component model in the {\it Planck} 2015 results \citep{Planck2015_X}. It is possible that the higher frequency residuals are due to another component such as MDE, which could be more strongly polarized. Indeed, there are already hints in the {\it Planck} data that MDE might be contributing at frequencies $\sim 100$\,GHz \citep{PIP_XXII}.

In summary, AME does not appear to be a major limitation for CMB studies so far. In many ways, nature has been kind to cosmologists. If all AME can be explained by spinning dust, it will probably not be a major foreground even for future CMB polarization missions. However, a highly polarized component (e.g. MDE) contributing to higher frequencies ($>50$\,GHz) could be problematic. It is also worth noting that on degree scales, the frequency at which foreground fluctuations are at a minimum is $\approx 70$\,GHz, which is not far from the peak of where spinning dust emits. For cosmologists wishing to make simulations to test their analyses, we suggest using maps of the thermal dust optical depth at 353\,GHz ($\tau_{353}$) at an angular resolution of 5\,arcmin from {\it Planck} \citep[e.g.,][]{PIP_XLVIII} as a simple model for AME. At 30\,GHz, a multiplying factor of $\approx 8 \times 10^6\,\mu$K/$\tau_{353}$  can be used and a {\sc spdust2} spectrum \citep[e.g.,][]{Planck2015_X} to give a reasonably realistic sky model. Other alternative templates include the {\it Planck} 545\,GHz map or the total dust radiance also from {\it Planck} using the coefficients given in Table~\ref{tab:emissivities}. For polarization simulations, assuming a $\approx 1\,\%$ average polarization fraction should provide conservative estimates of any potential AME contamination.

%%%%%%%%%%%%%%%%%%%%%%%%%%%%%%%%%%%%%%%%%%%%%%%%%%%%%%
%%%%%%%%%%%%%%%%%%%%%%%%%%%%%%%%%%%%%%%%%%%%%%%%%%%%%%

\section{Methodology for future AME research} 
\label{sec:discussion}

AME research has been progressing steadily since its initial detections \citep{Kogut1996,Leitch1997}. Nevertheless, after 2 decades, we are still not sure about the exact emission mechanism responsible, and with only a handful of reliable detections. We are even further away from using the spinning dust intensity and spectrum as a tool for studying the interstellar dust distribution and environment. So the question is where do we go from here?

There are several possible approaches. Certainly more accurate multi-frequency data (see Sect.\,\ref{sec:future}) are required to make progress in AME research. In particular, the frequency coverage and quality of low frequency (few GHz up to $\approx 20$\,GHz)  data is a major limitation. This will allow accurate spectra to be produced to more precisely discern the physics of the emission. High angular resolution ($\sim$arcmin or better) data are required to study Galactic clouds and external galaxies in detail. More sophisticated analysis techniques are also needed. Photometric techniques are often susceptible to systematic errors, particularly for diffuse clouds. For example, aperture photometry is a well established technique, but can be problematic for diffuse sources where the background level is difficult to define and measure \mbox{\citep{PIP_XV}}. Component separation techniques can also be improved upon, to better remove non-AME emission such as synchrotron, free-free, thermal dust and CMB. More high quality data would provide more AME detections for deriving statistics (e.g., emissivity, peak frequency, correlations), which would inform us about the properties of AME.

To date, most of the detections of AME have been made focusing on individual regions, with only a handful of studies focusing on larger samples of AME regions including H{\sc ii} regions \citep{Dickinson2007,Scaife2008}, Lynds dark clouds \citep{Scaife2009b}, diffuse clouds \citep{PIP_XV}, and cold cores \citep{Tibbs2015}. Going forward, this approach would ideally continue as a homogeneous sample of AME detections observed with the same telescope of the same environmental conditions enables us to make accurate comparisons of the AME. Although practically this may not be possible, a more homogeneous approach would mitigate against some systematic errors and potentially provide information on the effect of environmental conditions.    Even if unbiased and complete surveys are not possible, selection effects should be considered when drawing statistical conclusions.

A more focused approach is required if we are to answer specific scientific questions and provide a ``smoking gun" for spinning dust and/or magnetic dust emissions. Specifically, we should have testable predictions that observations can either prove or disprove. These would ideally begin with a theoretical prediction that can be tested by observations (see Sect.\,\ref{sec:targets} and Sect.\,\ref{sec:future_theory}). 

Although most observations of AME have been successfully fitted with a spinning dust model, the large number of parameters in the model mean that it is very difficult to test it in this manner. Therefore, to actually test the spinning dust model, we need to look at relative changes. For example, the spinning dust model predicts that, when other parameters are fixed, the amplitude and peak frequency of the spinning dust emission will increase with increasing gas density. Therefore, by observing a sample of regions covering a range of densities (e.g., cold cores), we can test this simple prediction. However, other parameters, such as the size distribution, will also be density-dependent, so it will not be easy to find ways to clearly test the predictions of spinning dust theory. 

In summary, the optimal strategy that will allow moving forward with AME research is a combination of the following: 

\begin{enumerate}
\item{A set of observables and clear predictions from models, allowing to bridge the gap between theory and observations;}
\item{Systematic observations of statistically representative samples of astrophysical classes of sources (e.g., PDRs, cold cores, nearby galaxies) that allow investigating correlations between AME (e.g., peak intensity, peak frequency) and quantities that regulate the physics of the observed sources (e.g., radiation field intensity, density, abundance of dust grains, gas species abundance); }
\item{Within a given astrophysical class of objects, targeted observations of sources whose properties are well known from prior investigations and for which a plethora of ancillary data exist, thus allowing careful modelling and interpretation of the observations.}
\end{enumerate}

More details will be provided in the following subsections.

%%%%%%%%%%%%%%%%%%%%%%%%%%%%%%%%%%%%%%%%%%%%%%%%%%%%%%%

\subsection{Future Directions for Galactic AME Studies}
\label{sec:targets}

Observationally, high-density PDRs have proven to be the best targets for the search of AME. As we discussed in Sect.\,\ref{sec:obs_smallscales}, to date the most significant AME detections are in the Perseus and $\rho$~Oph molecular clouds, which either host PDRs or are PDR-like in nature (e.g., LDN1622, RCW175). What makes these environments ideal is the combination of high gas densities, charged PAHs, and a moderate radiation field that allows PAHs to survive, i.e., all factors that are conducive to producing spinning dust. Due to the small statistics of confirmed AME sources, the range of physical conditions in PDRs that have been probed so far is still very limited. Future observational work should focus towards expanding the currently available set of PDRs. These types of sources can be regarded as a testbed for verifying a predicted correlation between the AME peak frequency and the total gas ion column density. In particular, high-density PDRs are expected to have a high peak frequency \citep{Draine1998b,Ali-Hamoud2009}. So far, we have only found mild evidence that the peak frequency can go higher than $\sim 30$\,GHz, with some regions peaking at $\sim 50$\,GHz \mbox{\citep{PIP_XV,Planck2015_XXV}}. An experiment that would allow testing the model prediction would entail executing pointed observation of a face-on PDR in which each pointing position would probe increasingly higher densities within the photodissociation front. If the spectral energy distributions at each pointing position showed a separation in peak frequencies, this would be a strong confirmation of the spinning dust model; modelling of the observed spatial variations of AME with frequency in the $\rho$~Oph~W PDR (Casassus et al., in prep.) might yield such a result. 

Similar to PDRs are Reflection Nebulae (RN), although, in this case, the UV flux generated by nearby stars is generally less intense. An example of an AME detection in this category is represented by M78 (Sect.\,\ref{sec:obs_smallscales}). \cite{Hoang2010} estimate that in RNs, due to fast internal relaxation, the peak emissivity and peak frequency of spinning dust are increased, respectively, by a factor $\sim$ 4 and 2. These predictions still require adequate observational vetting. 

Another class of astrophysical sources that appear to be promising for studies of AME are cold molecular cores. In general terms, cores are interesting environments to explore as high densities (typically n$_{H} >$ 10$^{3}$ cm$^{-3}$) make spinning dust more likely. In addition, as noted in Sect.\,\ref{sec:obs_smallscales}, coagulation of dust particles is known to occur inside cold ($T \lesssim 15$\,K) cores, as a consequence of high density, and for such a scenario clear theoretical  predictions have been formulated, which can be tested. One caveat to consider, though, is the fact that many molecular cores are forming stars, and thus are found to be associated with Young Stellar Objects (YSOs). Depending on the evolutionary stage and inclination with respect to the observer's line-of-sight, YSOs can be relatively strong sources of cm radiation. Observing cores is facilitated by the large inventories that were recently made available, for the Galactic population, such as the surveys by {\it Planck} \citep{PEP_XXIII,Planck2015_XXVIII} and {\it Herschel} \citep{Juvela2010, Andre2010}. \cite{Tibbs2015,Tibbs2016} observed a sample of cold cores with a range of densities ($5 \times 10^3 < n_{\rm H} < 120 \times 10^3$\, H\,cm$^{-3}$) and found a lack of AME, which could be explained if AME is produced by the smallest dust grains, which are depleted due to coagulation in these dense environments. As in the case of PDRs, a more complete systematic investigation of the presence of AME in cores has yet to be conducted, which is an issue that should be addressed by future observational work (larger sample, multi-frequency observations ideally including polarization to rule out MDE). Indeed new observations, such as VLA observations of high-mass star forming region at 5 and 23\,GHz \citep[e.g.,][]{Rosero2016} that has detected rising emission with frequency (5--25\,GHz) in some sources (thought to be from ionized jets), are likely able to be useful for constraining AME, even if they were not originally for that purpose.

While Galactic cold cores are showing very low levels of AME that may be indicative of very small dust grains depletion, still nothing is known about the level of AME associated to YSOs. Radio observations obtained with the Giant Metre Radio Telescope \citep[GMRT;][]{Ainsworth2016,Coughlan2017} are probing the free-free emission associated with the outflow of T\,Tau. Any detection of AME associated with such objects in the frequency range 10--60\,GHz would be indicative of dust grain evolution and possibly could be used as a star-formation evolution tracer.

While the Warm Ionized Medium (WIM) is included by \cite{Draine1998a} as one of the environments where one might expect to find spinning dust, hot ionized objects with compact support, such as HII regions, may not represent the best targets for shedding light on the nature of this emission. The main reason is that the interior of HII regions is notoriously devoid of dust, especially PAHs and large grains, likely due to radiation pressure drift that causes dust grains to move outwards from the location of the ionizing sources and in the direction of the HII region PDR, or, because they have been destroyed \citep{Draine_book}. This paradigm, which is supported by a wealth of IR data \citep{Povich2007,Tibbs2012}, holds in particular for very bright HII regions, powered by large OB associations. In this case, the excess of cm emission that is observed could also result from the combined effect of stellar winds and shocks that are generated by the stellar cluster, rather than by spinning dust \citep{Paladini2015}. An exception with respect to this picture is represented by older, more diffuse HII regions, for which dust depletion in the ionization zone is typically less dramatic and which are not characterized by strong wind nor shock activity. 

If the carriers producing AME are magnetically aligned, with angular momenta tending to be parallel to the magnetic field, one could test the predictions for the expected range of polarization fractions provided by some models. Table\,\ref{tab:pol_prediction} gives the expected maximum polarization for a small sample of molecular clouds for which averaged magnetic field orientations with respect to the line-of-sight, $\alpha$, are known \citep[][and references therein]{Poidevin2013}. In this example, magnetic dust grains from \cite{Draine1999} of Fe with axial ratios $1\!:\!1.25\!:\!1.5$ are considered. In the case where this dust material was perfectly aligned and the magnetic field direction parallel to the plane-of-sky, Fig.\,\ref{fig:pol} shows that one would expect maximum polarization $P_{\rm MAX}$ varying from 0.09 to 0.12 in the frequency range 10--60\,GHz. Once the inclination angle of the mean field is considered one would expect the polarization degree of order the values given in columns 3 and 4 of Table\,\ref{tab:pol_prediction}, i.e., $P_{\rm EXP} \approx 100 \times [1-{\rm cos}(\alpha)^{2}] \times \epsilon$. Measured values of the degree of polarization associated to AME in these molecular clouds are lower than expected and this would therefore rule out the \cite{Draine1999} magnetic dust model while measured values greater than about 6$\%$ would rule out almost all of the ED models displayed in Fig.\,\ref{fig:pol}.  Precise polarization measurements will also test the prediction that quantum suppression of dissipation requires that 20--50\,GHz spinning dust emission be essentially unpolarized.

The possibility of detecting AME in circumstellar disks around T\,Tauri (M$_{\ast}< 2\,$M$_{\odot}$) and Herbig Ae/Be (2\,M$_{\odot} <$M$_{\ast} <10$\,M$_{\odot}$) stars has  also emerged, as discussed by \cite{Rafikov2006}. Indeed, while the existence in circumstellar disks of very large, cm-size grains has long been recognized  and explained through coagulation, that of PAHs is a discovery of the last ten years or so \citep{Visser2007}. At the temperatures characteristic of these environments, spinning dust is expected to peak around 30--50\,GHz and, if at least 5$\%$ of C is locked up in small grains, the AME will be  dominant with respect to thermal dust emission for $\nu <$ 50\,GHz. Broad PAH features have been detected in the pre-transitional disk around Herbig Ae star HD169142 \citep{Seok2016}, which incidentally appears to have an excess of emission at a wavelength of 7\,mm (43\,GHz) relative to their thermal-dust only model, possibly due to spinning dust.

Future observational investigations of AME may also target dust-producing evolved stars, which could be used to discern the origin of this emission. For example, while there has been no clear detection of magnetised grains around evolved stars, there is some suggestion that they may exist around metal-poor evolved stars \citep{McDonald2010,McDonald2011}. Alternatively, if PAHs are the carrier, they may be observed as a very strong feature of certain carbon-rich evolved objects but be entirely absent in comparable oxygen-rich stars. However, detecting AME around carbon-rich evolved stars is no simple matter, as strong excitation of PAHs requires an ultra-violet radiation field, and in fact the prototypical carbon-rich AGB star, IRC+10216, shows no obvious PAH features \citep{Sloan2003} nor obvious AME feature \citep{Wendker1995}. For this reason, PAH features are typically best seen in post-AGB stars or planetary nebulae \citep{Woods2011,Sloan2014,Matsuura2014}. In such objects, though, the AME signal may be masked by a high background flux resulting from the high emissivity from hot dust and strong free-free emission from the ionised central cavity. In addition, the intermediate size of this type of source (often marginally resolved by mm-wavelength observatories), together with strong variation in physical conditions across each nebulae, and sometimes strong temporal variations in emission, make the detection and characterization difficult. PAH excitation will be critically dependent on UV radiation field (produced either from the interstellar medium or companion star) which is known to vary substantially in such systems \citep[e.g.,][]{McDonald2015}.

A single, weak, tentative detection of AME has been made using a combination of \emph{Planck} and ancillary observations of NGC\,40 \citep{PIP_XVIII}, a nearby carbon-rich planetary nebula \citep{Ramos-Larios2011}. While dust-producing evolved stars are naturally of interest, the overall formation site of PAHs is still quite controversial: indeed it could be stellar origin, as now described, or either molecular origin \citep{Paradis2009,Sandstrom2010} or shattering in grain-grain collisions \citep{Draine1990,Draine2009}. We also note that PAH emission has been seen in reflection nebulae excited by cool stars, and this is consistent with theoretical models \cite{Li2002}.  If PAH emission is absent in outflows from regions excited by stars with $T_{\mathrm{eff}} > 3500$\,K, then it is because small PAHs are not abundant there.  A survey of such stars with a range of effective temperatures would elucidate the role of PAHs for AME. 

Last of all, if AME is indeed produced by relatively small interstellar PAHs, a smoking-gun confirmation would be the detection of their line emission, as discussed in Sect.\,\ref{sec:lines}. Testing this hypothesis would require more observations, generalizing those of \cite{Ali-Haimoud2015} to different environments and lines-of-sight.

\begin{table}
\begin{center}
\caption{Expected degrees of polarization for four molecular clouds for which the averaged magnetic field orientations with respect to the line-of-sight, $\alpha$, are known. As an example the material considered here is magnetic dust of Fe with axial ratios $1\!:\!1.25\!:\!1.5$ and with a polarization efficiency that varies between 0.9 and 1.2 between 10 and 60\,GHz.}
\label{tab:pol_prediction}
\begin{tabular}{cccc}
\hline
\noalign{\smallskip}
Molecular             & $\alpha$ [deg.]      & $P_{\rm EXP} [\%]$                 &$P_{\rm EXP} [\%]$\\
Cloud                   &                               & $P_{\rm MAX} = 0.09$             &$P_{\rm MAX} =0.12$ \\
Region                &                               &Fe 1:1.25:1.5                            & Fe 1:1.25:1.5\\
\noalign{\smallskip}
\hline 
\noalign{\smallskip}
  S106            &   [50--55] 	& $\approx$ 5.7 & $\approx$ 7.6 \\
  OMC-2/3      &   [72--80] 	& $\approx$ 8.5 & $\approx$ 11.3 \\
  W49             &   [50--60] 	&$ \approx$ 6.0 & $\approx$ 8.1 \\
  DR21           &   [60--70] 	& $\approx$ 7.4 & $\approx$ 9.9 \\
\noalign{\smallskip}
\hline
\end{tabular}
\end{center}
\end{table}

%%%%%%%%%%%%%%%%%%%%%%%%%%%%%%%%%%%%%%%%%%%%%%%%%%%%%%%

\subsection{Extragalactic AME Studies} 
\label{sec:extragalactic}

Since AME has only been detected in one or two external galaxies, it is vital that we pursue AME in extragalactic sources. It is interesting to note that in our Galaxy, the fraction of AME to total emission at frequencies near $\sim 30$\,GHz is as much as a half. Thus, one might expect to see a similar ratio in other galaxies. There is little radio data at frequencies above $\sim 10$\,GHz, nevertheless, AME does not appear to be this strong in extragalactic sources. For instance, \citet{Murphy2012} observed 103 galaxy nuclei and extranuclear star-forming complexes taken with the GBT as part of the Star Formation in Radio Survey (SFRS).  Among the 53 sources also having ancillary far infrared and 1.7\,GHz data, $\approx$10\% exhibited radio spectral indices and 33\,GHz to IR flux density ratios consistent with that of the source in NGC\,6946 showing a strong AME signal.  It appears that external star-forming galaxies emit less than a few per cent of AME at frequencies $\sim 30$\,GHz.

If the AME intensity (or emissivity) is similar to the levels we believe in our Galaxy, we would expect AME to be a significant contributor to the radio/microwave flux at rest frequencies near 30\,GHz. This does not appear to be the case (at least in the few objects that have been studied) and the reason why is not clear. Possibilities include (i) the environment in the Milky Way is conducive for producing strong AME but is relatively rare among star-forming galaxies, (ii) AME is primarily a local phenomenon in the vicinity of the Sun, or (iii) we have significantly over-estimated AME in our own Galaxy (or a combination of these).

If AME is significant in external galaxies, it could have an impact on several important areas of astrophysics. For example, free-free emission has been proposed as a reliable estimator of the star-formation rate \citep[SFR; e.g.,][]{Murphy2009, Murphy2011, Murphy2017}, and 30\,GHz has been proposed as the ideal frequency, with it being least contaminated by synchrotron or thermal dust emission. However, this is exactly the frequency at which AME appears to be strongest (at least in our own Galaxy). Not accounting for AME would result in over-estimates of the SFR and could bias SFR estimates for a large sample. 

Identifying AME candidate sources based on a coarse sampling of the radio spectrum, even with a large lever arm spanning 1.7, and 33\,GHz as was done for the GBT survey of \cite{Murphy2012}, is inconclusive. A much finer (i.e., better than a factor of two) sampling in frequency space, spanning $\approx 3$--90\,GHz would be ideal to measure the peak for conclusive detections.  Consequently, to further investigate the impact on AME studies of star formation both at low and high redshifts, the SFR analysis has been extended to include $\approx 2^{\prime\prime}$ resolution ($\approx 30$--300\,pc) JVLA 3, 15, and 33\,GHz data of 112 galaxy nuclei and extranuclear star-forming complexes to both better characterize thermal radio fractions at 33\,GHz and search for discrete AME candidates in external galaxies.  Any robust AME candidates can then be confirmed by more focused follow-up observations at 90\,GHz.  

A morphological comparison between the JVLA 33\,GHz radio, H$\alpha$ nebular line, and 24\,$\mu$m warm dust emission shows remarkably tight similarities in their distributions suggesting that each of these emission components are indeed powered by a common source expected to be massive star-forming regions, and that the 33\,GHz emission is primarily powered by free-free emission (Murphy et al., 2017a, submitted).  The full multi-frequency spectral analysis to robustly measure thermal fractions and characterize AME will be presented in a future paper (Murphy et al. 2017b, in prep.). High precision and fidelity data will be essential to obtain AME detections or to place meaningful constraints.

We make a final remark that current observations may already be seeing evidence of AME, but may have been interpreted as something else. A recent example would be the 33\,GHz measurements of 22 local luminous galaxies \citep{Barcos-Munoz2017}, where some sources showed a remarkably flat ($\alpha \approx 0$) flux density spectral index. Although this may be due to self-absorption in the densest regions, it could also be indicating a (small) component of AME. Accurate multi-frequency data, with careful consideration of all contributing components, will be essential for disentangling the radio-microwave spectrum.

We note that AME for more distant, high redshift, sources will be redshifted to lower frequencies. Future high redshift observations with, for example, the SKA at frequencies $\lesssim 20$\,GHz, may need to take AME into account.

%%%%%%%%%%%%%%%%%%%%%%%%%%%%%%%%%%%%%%%%%%%%%%%%%%%%%%%

\subsection{Current and Future Instruments and Surveys}
\label{sec:future}

Looking to the future, what new instruments and surveys are needed to make further progress in understanding AME? Future studies will encompass detailed modelling of compact regions of the Galaxy (e.g., PDR, dark clouds, cold cores); observations of large numbers of such regions in order to gather statistics; large-scale surveys of the diffuse emission in the Galaxy; and searches for AME in other galaxies. Continuing to constrain the polarization of AME will be an important, if increasingly difficult, task. Resolving the spatial distribution of the  AME on increasingly smaller scales will be limited by the maximum available resolution of the synchrotron and free-free templates available for subtracting these contributions.

Table\,\ref{tab:data} summarises some of the main current and planned astronomical facilities and surveys that will be important for AME research. To date,  AME studies have relied on radio data from  a disparate collection of single-dish telescopes and interferometers. Since AME is typically extended and diffuse, in order to make accurate, multi-frequency comparisons, the angular content of a map needs to be known, and ideally, contain the complete range of angular scales. For example, single-dish total power instruments typically require switching of the signal against a reference source or other form of high-pass filtering, which can reduce the sky signal on large angular scales. On the other hand, maps made using interferometers are sensitive to specific (and missing the largest) angular scales. Care must therefore be taken when constructing SEDs combining multi-frequency data from different interferometers and single-dish telescopes. It would be advantageous for AME studies if a single future instrument were able to cover the frequencies relevant to the AME itself, as well as synchrotron and free-free. The proposed next-generation VLA (ngVLA) and SKA interferometric arrays are promising in this regard \citep[e.g.,][]{Dickinson2015}, particularly if they can be supplemented with single-dish observations to fill in the missing large-scale information. 

The C-Band All-Sky Survey (C-BASS; \citealt{King2010}) is a full-sky 5\,GHz survey of intensity and polarization, at an angular resolution of $\approx 45$\,arcmin. With high sensitivity ($\lesssim 0.1$\,mK) and accurate calibration, it will provide a vital data point at 5\,GHz. C-BASS data will be important for any component separation and large-scale studies with WMAP/{\it Planck} data \citep[e.g.,][]{Irfan2015}. 

The Q-U-I JOint ExperimenT (QUIJOTE) is a dedicated CMB and foregrounds experiment, operating at Tenerife \citep{Genova-Santos2015a}. Although limited to $\approx 1^{\circ}$ resolution, and only northern sky,\!\footnote{There are ongoing discussions about deploying the QUIJOTE instrument in South Africa, to allow a full-sky survey to be made.} QUIJOTE has a unique frequency coverage of 11, 13, 17 and 19\,GHz, which covers the rising part of the spinning dust spectrum. QUIJOTE is primarily a polarization experiment but retains sensitivity to total-intensity, which is readily detectable at low Galactic latitudes. In combination with C-BASS, WMAP and {\it Planck} data, will be a powerful dataset for AME studies.

On small scales, the AMI \citep{Zwart2008} telescope, is well-suited for studying AME. The AMI small array (AMI-SA) is a compact interferometer operating at 12--18\,GHz with sensitivity to angular scales $\approx 2$--10\,arcmin while the large array (AMI-LA) has a resolution of $\approx 0.5$\,arcmin. The telescope has recently been upgraded with better receivers and a digital spectral backend \citep{Hickish2017}, resulting in better dynamic range allowing observations in the presence of bright radio frequency interference (RFI). 

The available frequency range, dish size and baseline lengths makes the next-generation ngVLA, the Atacama Large Millimeter Array (ALMA), and the Square Kilometre Array (SKA), suitable for high resolution  studies  of compact and extended AME regions, but not for large-area surveys. ALMA is a large interferometric array located in Chile that can operate from 30\,GHz up to $\sim 1000$\,GHz and can achieve sub-arcsec angular resolution. The current capability is limited to frequencies above 84\,GHz (band 3) while bands 1 (35--50\,GHz) and 2 (67--90\,GHz) are expected to come online around 2020. 

A number of CMB experiments continue to operate from the ground with frequency channels in the tens of GHz. For example, the Cosmology Large Angular Scale Surveyor \citep[CLASS,][]{essenger-hileman_etal_14} is focused on CMB (at multipoles $2 \lesssim \ell \lesssim 150$) and foreground characterization in intensity and in polarization. It will map around 70\% of the sky from the Cerro Toco in the Atacama Desert (Chile), in frequency bands centred at 40, 90, 150 and 220\,GHz with angular resolution 90, 40, 24 and 18\,arcmin, respectively. 

The CO Mapping Array Pathfinder \citep[COMAP,][]{li_etal_16} is an intensity mapping experiment aiming to constrain the  carbon monoxide (CO) power spectrum from  the epoch of reionization, but initially targeting intermediate redshifts of $z=2.4$--3.4 with a 19-pixel focal plane array operating at 26--34\,GHz. As ancillary science, COMAP will provide maps of extended  AME regions with 4\,arcmin angular resolution and frequency resolution of around 8\,MHz.

It is clear that for all of the above, accurate multi-frequency data are critical. However in order to be useful, the data need to be well-calibrated in terms of amplitude scale and quantification of the telescope beam. The use of existing large-scale radio maps, such as the Haslam 408\,MHz \citep{Haslam1982,Remazeilles2015}, Reich 1.4\,GHz \mbox{\citep{Reich1986,Reich2001}} and HartRAO 2.3\,GHz maps \citep{Jonas1998} can be problematic. These surveys were made using general-purpose large radio telescopes that have relatively low beam efficiencies (i.e., the integrated sidelobe response represents a significant fraction of the response compared to the main beam). This means that the amplitude scale varies as a function of angular scale, often by tens of per cent \citep[see][for a discussion]{Du2016}. For example, the full-beam to main-beam conversion factor for the Reich 1.4\,GHz map is as much as a factor of 1.55. Correcting for this afterwards is not trivial. New data, such as C-BASS, should improve on this situation with improved calibration and knowledge of the full beam. 

Since AME is dust-correlated, the availability of infra-red observations  plays a critical role  in AME studies. All-sky surveys from IRAS/IRIS and WISE are complemented by large-area surveys and observations of compact regions using the {\it Spitzer} Space Telescope. The AKARI satellite \citep{Ishihara2010} made an all-sky survey with the $9\,\mu$m filter, which covers the entire PAH emission bands in the mid-IR, i.e., the bands at 6.2, 7.7, 8.6, and 11.3$\,\mu$m, while the 12\,$\mu$m band of IRAS and WISE do not completely
cover the strong 6.2 and 7.7\,$\mu$m bands and may suffer contribution from warm dust emission.  
Since the 6.2 and 7.7\,$\mu$m bands (and the weaker 8.6\,$\mu$m band) are thought to arise from ionized PAHs, and thus the
11\,$\mu$m band comes from neutral PAHs, the correlation of AME with AKARI 9\,$\mu$m
could potentially be a better test for the spinning dust model, in which ionized PAHs
contribute more significantly than neutral.   However, the latest analysis
of the correlation of AME with the AKARI 9\,$\mu$m data do not show better correlation
than with the far-IR emission (Bell et al., in prep.) in agreement with previous analyses 
with IRAS and WISE data \citep{Hensley2016,Planck2015_XXV}.

There is currently no active mid-far infra-red space telescope capability, although the airborne Sofia observatory is available. The launch of the James Webb Telescope (currently scheduled for Oct 2018) will provide space-based mid-infrared (5--28$\,\mu$m) imaging and spectroscopic capability at sub-arcsec  resolution, supporting detailed modelling of compact regions, but studies of  larger areas will rely on existing data from IRAS/IRIS, {\it Spitzer} and WISE for the foreseeable future.

\setlength{\tabcolsep}{2pt}
\small
\begin{table*}
\caption{Current and future astronomical facilities and datasets that can be used for AME research. The table is separated in to different types of observational facilities (radio surveys, interferometers, IR surveys, and future). Values in parentheses are expected to be available in the future.}
\begin{tabular}{lcccl}
\hline
\noalign{\smallskip}
Telescope/ 			&Frequencies  			&Angular Res.    				&Type				&Status \& Notes   	\\
facility				&[GHz]				&[arcmin] 				&                            		&				\\
\noalign{\smallskip}
\hline 
\noalign{\smallskip}
C-BASS				&5					&45						&Full-sky survey		&Ongoing. Data to be made available \\
Effelsberg				&$<34$				&$\approx 0.3\,@30$\,GHz	&Single-dish			&Available by request \\
GBT 100-m			&$<115$				&$\approx 0.3\,@30$\,GHz	&Single-dish			&Available \\
Parkes				&$<24$				&$\approx 1\,@21$\,GHz		&Single-dish			&Available by request \\
{\it Planck}			&28,44,70,100+		&$\approx 33\,@28.4$\,GHz	&Full-sky surveys		&Final data published \\
QUIJOTE-MFI			&11,13,17,19			&$\approx 55$--36			&Single-dish			&Ongoing. Data to be made available \\
S-PASS				&2.3					&$\approx 9$				&Southern sky survey	&Ongoing \\
WMAP				&23,33,41,61,94		&$\approx 50\,@22.8$GHz	&Full-sky surveys		&Final data published \\
\noalign{\smallskip}
\hline
\noalign{\smallskip}
ALMA				&$\approx 30$--50 (Band 1) 	&$\approx 0.1$--4		&Interferometer 	&Online and in development \\
AMI					&13--18				&$\approx 0.5$--10			&Interferometer		&Available by request \\
ATCA				&$<105$				&$<0.1$					&Interferometer		&Available \\
VLA					&$<100$				&$<0.1$					&Interferometer		&Available and in development \\
\noalign{\smallskip}
\hline
\noalign{\smallskip}
AKARI				&9--180$\,\mu$m		&$\approx 0.1$--1.5			&Full-sky surveys		&Completed. Data to be made available \\
IRIS (IRAS/DIRBE)		&12,25,60,100$\,\mu$m	&$\approx 2$--4			&Full-sky surveys		&Final data published \\	
{\it Spitzer} 			&3.6-160$\,\mu$m  &$\approx 0.02-0.8$ 				&IR 					 &Data available \\
WISE				&3--25$\,\mu$m		&0.1--0.2					&Full-sky surveys		&Completed. Data available \\
\noalign{\smallskip}
\hline
\noalign{\smallskip}
CLASS                              &40,90,150,220                 &90--18                                      & Southern sky survey      &Data  will be made publicly available \\
COMAP				&26--34				&$\approx 4$				&Single-dish			&19 detectors, 0.2--8\,MHz resolution (in prep.) \\
James Webb		 	&5--28$\,\mu$m (MIRI)	&$<0.02$					&Optical/IR 			&2018 launch \\
ngVLA				&$\approx$1--116		&$\approx 9$mas--1 arcmin\,$@30$\,GHz &Interferometer &Concept stage\\
Sardinia Radio Telescope 	&$<26$ 				&1\,@19.7\,GHz			&Single-dish			&7-beam dual-pol array. Operational 2019\\
SKA					&($<14+$)			&$<0.1$					&Interferometer			&Phase 1 from $\approx 2023$ \\
\noalign{\smallskip}
\hline
\end{tabular}
\label{tab:data}
\end{table*}
\normalsize
\setlength{\tabcolsep}{6pt}

%%%%%%%%%%%%%%%%%%%%%%%%%%%%%%%%%%%%%%%%%%%%%%%%%%%%%%%

\subsection{Laboratory studies}
\label{sec:lab}

Laboratory studies are needed to help constrain the magnetic relaxation of small
grains, which is one of the key problems relating to alignment of spinning dust grains \citep{Lazarian2000}. Measurements of the spin-lattice relaxation time will test
theoretical models of both paramagnetic relaxation, which aligns a grain
with an external magnetic field, and the alignment of the angular
momentum vector with the principal axes of a wobbling grain.

For a spin to flip in a rotating grain, it is necessary for the total energy in lattice vibrations to change by $\hbar \omega$. Because the density of states is finite, it may not be possible for the lattice vibrations to absorb such energy. Indeed, the minimal bending mode frequency for grains smaller than $10^{-7}$\,cm can be estimated to be orders of magnitude larger than $kT/\hbar$. According to  \cite{Lazarian2000} this does not mean that the spin relaxation time becomes infinite. To show this the authors considered the Raman scattering of phonons process \citep[see][]{Waller1932,Pake1962} where the change of the spin happens due to the scattering of phonons with energies much larger than $\hbar \omega$.

The calculations in \cite{Lazarian2000} using the accepted approaches \citep[see][]{Altshuler1964} provided the spin-lattice relaxation rates for the PAHs that may interfere with the resonance paramagnetic relaxation of smallest grains and can also affect the internal Barnett relaxation of energy within these particles. However, the Waller theory is known to overestimate the spin-lattice relaxation time by many orders of magnitude. Therefore, reliable estimates of the spin-lattice relaxation should be obtained through the laboratory studies of the magnetic response of suspended nano-particles. 

Even more straightforward laboratory studies are required to test the expected magneto-dipole emission of magnetic grains as suggested in \cite{Lazarian2000} and \cite{Draine2013}. The two studies used different magnetic response expressions and none of the expressions seem to fit well to the limited available experimental data. Therefore it is important to do the corresponding laboratory studies in order to establish the magnetic response of the candidate materials at AME frequencies. For large grains the studies of the magnetic response of the bulk samples are adequate. Such studies can carried out using existing laboratory equipment.

%%%%%%%%%%%%%%%%%%%%%%%%%%%%%%%%%%%%%%%%%%%%%%%%%%%%%%%

\subsection{Next Steps for AME Modelling} 
\label{sec:future_theory}

As reviewed in Sect.\,\ref{sec:theory}, theoretical models of spinning dust emission are quite detailed in their consideration of various effects, which determine the rotational velocities and consequent microwave emission of ultrasmall interstellar grains. However, theoretical modelling faces two primary hurdles. First is that the identity of the AME carrier, and thus its material properties, are unknown. Second is that, even with perfect knowledge of the grain physics, predicting the emission from a particular interstellar environment is challenging due to the number of poorly constrained or completely unconstrained environmental parameters, such as the grain size distribution or the radiation spectrum and intensity. We address each of these hurdles in turn in this section.

Detailed calculations of rotational emission from a population of
spinning grains require knowledge of: (1) the composition of the
nanoparticles; (2) the distribution of sizes and shapes; (3) the
electric (and perhaps magnetic) dipole moments of these particles; (4)
the orientation of the dipole moment relative to the principal axes of
the moment-of-inertia tensor of each particle; (5) the optical
properties of the particles for absorption of optical-UV photons, and
subsequent emission of infrared photons; (6) the vibrational mode
spectrum of the particles, which determines the specific heat and the
``temperature" of the particle as a function of internal energy; (7)
the frequency-dependent magnetic polarizability of the nanoparticles,
which will determine the importance of the Barnett effect,
paramagnetic dissipation and ``Barnett relaxation"
\citep{Purcell1979,Lazarian1999}; (8) the photoelectric emission
properties and (9) electron capture cross sections of
the nanoparticles, which affect their electric charge; and (10) the
spin-lattice relaxation times, which affect the 
rotational dynamics and alignment of grains.  All of these physical properties
are uncertain, and current theoretical modelling is based on
assumptions whose validity is not always clear.  Progress on these
fundamental issues will be based on a combination of laboratory
experiments and improved theoretical modelling of the structure of the
particles, based on density functional theory, etc.

Even with perfect knowledge of the above properties, an {\it a priori} prediction of an AME spectrum of a specific interstellar region would be daunting due to the sensitivity of the calculations to various environmental parameters, which are typically poorly constrained. In light of this, future theoretical efforts should be directed toward what observational predictions can be made {\it irrespective of these modelling uncertainties}. A simple example is that the spinning dust emissivity per H atom should decrease in very high density gas where ultrasmall grains are expected to be depleted due to coagulation. Future observational efforts will be directed toward understanding systematic changes of the AME spectrum (e.g., emissivity per H atom, peak frequency, frequency width, shape) with environment. Assessing the magnitude of these changes in various idealized environments and identifying the environmental parameters inducing the largest changes will be invaluable in both selecting observational targets and for interpreting the data.

One step in this direction is continued development of the {\sc SpDust} code. In particular, porting the code to a language more amenable to integration in large simulations (e.g., Python, C, Fortran) than its native IDL would enable detailed exploration of the high-dimensional parameter space in which current models lie. Integration of such a code with models of interstellar clouds, as pioneered by \citet{Ysard2011}, could identify generic predictions of spinning dust theory robust to variations in the grain properties or the specifics of the local interstellar environment.

In addition to improving the modelling of spinning dust emission and articulating generic observational predictions of how the AME spectrum should vary, theoretical efforts should also be directed toward connecting the AME to observational probes at other wavelengths. In particular, an abundance of AME carriers may have observable impact on the UV extinction, photoelectric heating, diffuse interstellar bands \citep[see][]{Bernstein2015}, and/or MIR emission. Self consistent modelling of the AME and these phenomena may yield feasible observational tests of AME models, particularly the identity of the AME carrier(s).

Finally, while the hypothesis that the AME is predominantly rotational emission from rapidly-spinning nanoparticles is favoured at this time, the possibility remains that the AME is ``thermal" emission from the ``big" grains that dominate the grain mass, with the spectrum due to unusual frequency dependence of the material opacity (from either electric dipole or magnetic dipole absorption).  Laboratory studies of magnetic absorption in candidate magnetic materials (e.g., metallic iron, magnetite, maghemite) are feasible and will be able to tell us whether currently-favoured phenomenological models based on the Gilbert equation \citep[see][]{Draine2013} provide a good approximation to the actual magnetization dynamics. Laboratory studies on candidate amorphous materials at low temperatures can test whether any plausible materials exhibit a strong opacity peak in the 20--40\,GHz region, possibly arising from conformational changes, as suggested by \cite{Jones2009}.

%%%%%%%%%%%%%%%%%%%%%%%%%%%%%%%%%%%%%%%%%%%%%%%%%%%%%%%
%%%%%%%%%%%%%%%%%%%%%%%%%%%%%%%%%%%%%%%%%%%%%%%%%%%%%%%

\section{Concluding remarks}
\label{sec:conclusion}

AME has become an important topic in astrophysics, both as a foreground for CMB observations and as a new component of the ISM. A strong, FIR-correlated component of emission has been detected at frequencies $\sim 10$--100\,GHz, which cannot be easily explained by CMB, synchrotron, free-free or thermal dust radiation. The lack of significant AME polarization appears to rule out magnetic dipole emission as the main source. The most favoured explanation is electric dipole radiation from tiny spinning dust grains. The underlying theory of spinning dust emission is well known, even if the detailed physics is not fully understood. There are at least two clear examples of high S/N detections, from the Perseus and $\rho$~Ophiuchus molecular clouds, where the data are easily explained by a simple model of spinning dust emission.

Nevertheless, the picture is still far from settled on a number of aspects. There is a dearth of detections on very small angular scales; attempts to detect AME from compact Galactic objects (e.g., H{\sc ii} regions) has resulted in no clear detections. Indeed, in nearly all cases, AME is observed to have a diffuse morphology. The lack of correlation with PAHs appears to rule out PAHs as the main carriers of AME, at least along some sight-lines, even though theory would suggest that PAHs (being even smaller than the small dust ``grains") are sufficiently numerous and spinning rapidly enough to naturally produce the majority of spinning dust emission, given a suitable electric dipole moment. On the other hand, the lack of AME in dense cores appears to rule out ``big" dust grains as the carriers of AME. Therefore, the AME appears to be due to small spinning nano-particles containing an electric dipole moment, but the nature of these nano-particles is still undetermined.

Future AME research should continue on a number of independent fronts, including theory, laboratory measurements, as well as high-fidelity multi-frequency observations. New facilities have the capabilities to observe over a wide frequency range and at high angular resolution. New data should focus on testing specific aspects of the spinning (or magnetic) dust model predictions.

AME appears to not be a major limitation as a foreground in CMB temperature measurements. Careful component separation yields results that are not limited by foregrounds. However, CMB polarization data may be impacted. Current upper limits on AME polarization are at the level of $\approx 1$\,\% and this could be non-negligible for future ultra-deep CMB polarization surveys.

%% Get figure permissions - Parthansarathy Vasu from journal will help (v.parthasarathy.1@elsevier.com) - email when you submit!

%%%%%%%%%%%%%%%%%%%%%%%%%%%%%%%%%%%%%%%%%%%%%%%%%%%%%%%
%%%%%%%%%%%%%%%%%%%%%%%%%%%%%%%%%%%%%%%%%%%%%%%%%%%%%%%

\begin{acknowledgements}
We thank Chris Tibbs and colleagues at ESTEC for organising the 3rd AME workshop, held 23--24 June 2016 at ESTEC, Noordwijk, The Netherlands. CD acknowledges support from an STFC Consolidated Grants (ST/P000649/1) and an ERC Starting (Consolidator) Grant (no.\,307209). BTD was supported in part by NSF grant AST-1408723. TH acknowledges the support by the Basic Science Research Program through the National Research Foundation of Korea (NRF), funded by the Ministry of Education (2017R1D1A1B03035359). CTT acknowledges an ESA Research Fellowship. ESB acknowledges support from Sapienza Ateneo projects 2016. AB and TO are supported by JSPS and CNRS under the Japan--France Research Cooperative Program. CHLC thanks CONICYT for grant Anillo ACT-1417. MP acknowledges grant \#2015/19936-1, S\~{a}o Paulo Research Foundation (FAPESP). YCP acknowledges support from a Trinity College JRF. FP thanks the European Commission under the Marie Sklodowska-Curie Actions within the H2020 program, Grant Agreement number: 658499-PolAME-H2020-MSCA-IF-2014. JARM acknowledges the funding from the European Union's Horizon 2020 research and innovation programme under grant agreement number 687312 (RADIOFOREGROUNDS). This work has been partially funded by the Spanish Ministry of Economy and Competitiveness (MINECO) under the project AYA2014-60438-P. MV acknowledges  support from FONDECYT through grant 3160750. We thank George Bendo and Tim Pearson for helpful discussions while preparing the draft. This research was carried out in part at the Jet Propulsion Laboratory, California Institute of Technology, under a contract with the National Aeronautics and Space Administration.
\end{acknowledgements}

%%%%%%%%%%%%%%%%%%%%%%%%%%%%%%%%%%%%%%%%%%%%%%%%%%%%%%%
%%%%%%%%%%%%%%%%%%%%%%%%%%%%%%%%%%%%%%%%%%%%%%%%%%%%%%%

%\paragraph{Paragraph headings} Use paragraph headings as needed.

\bibliographystyle{abbrvnat}
\bibliography{refs}

\end{document}